\documentclass[longauth]{aa}

\usepackage{euclid}
\usepackage{graphicx}
\usepackage{natbib}
\usepackage{scalerel}
\usepackage{multirow}
\usepackage{booktabs}
\usepackage{enumitem}
\usepackage[table]{xcolor}
\usepackage{fontawesome5}
\usepackage{tikz}

\usepackage{subcaption} 

\usepackage{rotating}
\usepackage[table]{xcolor}

\usepackage{glossaries}
\glsdisablehyper

\usepackage{txfonts}
\usepackage[pdfencoding=auto,psdextra]{hyperref}
\hypersetup{
    colorlinks=true,
    linkcolor=blue,
    filecolor=magenta,      
    urlcolor=blue,
    citecolor=blue
}
\makeatletter
\renewcommand*\aa@pageof{, page \thepage{} of \pageref*{LastPage}}
\makeatother

\usepackage[utf8]{inputenc}

\renewcommand{\linenumbers}[0]{}

\newcommand{\Plin}{P_\mathrm{lin}}

\newcommand{\chired}{\chi^2_{\mathrm{red}}}

\newcommand{\Mpc}{\, h^{-1} \, {\rm Mpc}}

\newcommand{\cGpc}{\, h^{-3} \, {\rm Gpc}^3}

\newcommand{\kcMpc}{\, h^3 \, {\rm Mpc}^{-3}}
\newcommand{\CPUh}{\, \mathrm{CPU} \, \mathrm{h}}

\newcommand*{\NDS}{\ensuremath{N_\sfont{DS}}}
\newcommand*{\NSS}{\ensuremath{N_\sfont{SS}}}

\newacronym{2pcf}{2PCF}{2-point correlation function}
\newacronym{3pcf}{3PCF}{3-point correlation function}
\newacronym{spt}{SPT}{standard perturbation theory}
\newacronym{rpt}{RPT}{renormalised perturbation theory}
\newacronym{eft}{EFT}{effective field theory of large-scale structure}
\newacronym{bao}{BAO}{baryon acoustic oscillation}
\newacronym{rsd}{RSD}{redshift-space distortions}
\newacronym{hod}{HOD}{halo occupation distribution}
\newacronym{los}{LOS}{line of sight}
\newacronym{fob}{FoB}{figure of bias}
\newacronym{fom}{FoM}{figure of merit}
\newacronym{mgf}{MGF}{moment generating function}
\newacronym{left}{LEFT}{Lagrangian effective field theory}
\newacronym{cdm}{CDM}{cold dark matter}
\newacronym{lpt}{LPT}{Lagrangian perturbation theory}
\newacronym{dof}{dof}{degrees of freedom}
\newacronym{pdfabbrev}{PDF}{probability distribution function}
\newacronym{DRone}{DR1}{Data Release 1}
\newacronym{DRthree}{DR3}{Data Release 3}
\newacronym{flagshipone}{FS1}{Flagship 1}
\newacronym{hmc}{HMC}{Hamiltonian Monte Carlo}
\newacronym{mcmc}{MCMC}{Markov chain Monte Carlo}
\newacronym{apdist}{AP}{Alcock--Paczynski}

\newcommand{\tpcf}{\gls{2pcf}\xspace}

\newcommand{\bao}{\gls{bao}\xspace}
\newcommand{\rsd}{\gls{rsd}\xspace}

\newcommand{\hod}{\gls{hod}\xspace}
\newcommand{\los}{\gls{los}\xspace}

\newcommand{\fom}{\gls{fom}\xspace}

\newcommand{\cdm}{\gls{cdm}\xspace}

\newcommand{\DRone}{\gls{DRone}\xspace}

\newcommand{\flagshipone}{\gls{flagshipone}\xspace}
\newcommand{\hmc}{\gls{hmc}\xspace}
\newcommand{\mcmc}{\gls{mcmc}\xspace}
\newcommand{\apdist}{\gls{apdist}\xspace}

\begin{document}
%
%
    \title{\Euclid preparation. Baryon acoustic oscillations extraction techniques: comparison and optimisation}

\newcommand{\orcid}[1]{} 
\author{Euclid Collaboration: E.~Sarpa\orcid{0000-0002-1256-655X}\thanks{\email{elena.sarpa@inaf.it}}\inst{\ref{aff1}}
\and A.~Veropalumbo\orcid{0000-0003-2387-1194}\inst{\ref{aff2},\ref{aff3},\ref{aff4}}
\and M.~Bonici\orcid{0000-0002-8430-126X}\inst{\ref{aff5},\ref{aff6}}
\and M.~K{\"a}rcher\orcid{0000-0001-5868-647X}\inst{\ref{aff7}}
\and M.~Crocce\orcid{0000-0002-9745-6228}\inst{\ref{aff8},\ref{aff9}}
\and E.~Sefusatti\orcid{0000-0003-0473-1567}\inst{\ref{aff1},\ref{aff10},\ref{aff11}}
\and E.~Maragliano\orcid{0009-0009-6004-4156}\inst{\ref{aff4},\ref{aff3}}
\and E.~Branchini\orcid{0000-0002-0808-6908}\inst{\ref{aff4},\ref{aff3},\ref{aff2}}
\and C.~Oliveri\orcid{0009-0001-0346-6530}\inst{\ref{aff12},\ref{aff10},\ref{aff11}}
\and G.~Gambardella\orcid{0009-0001-1281-1746}\inst{\ref{aff8},\ref{aff9}}
\and B.~Camacho~Quevedo\orcid{0000-0002-8789-4232}\inst{\ref{aff10},\ref{aff12},\ref{aff1}}
\and C.~Moretti\orcid{0000-0003-3314-8936}\inst{\ref{aff1},\ref{aff10},\ref{aff11}}
\and P.~Monaco\orcid{0000-0003-2083-7564}\inst{\ref{aff13},\ref{aff1},\ref{aff11},\ref{aff10}}
\and J.~Bautista\orcid{0000-0002-9885-3989}\inst{\ref{aff14}}
\and M.~Viel\orcid{0000-0002-2642-5707}\inst{\ref{aff10},\ref{aff1},\ref{aff12},\ref{aff11},\ref{aff15}}
\and W.~J.~Percival\orcid{0000-0002-0644-5727}\inst{\ref{aff5},\ref{aff16},\ref{aff17}}
\and S.~Nadathur\orcid{0000-0001-9070-3102}\inst{\ref{aff18}}
\and A.~Pezzotta\orcid{0000-0003-0726-2268}\inst{\ref{aff2}}
\and A.~Eggemeier\orcid{0000-0002-1841-8910}\inst{\ref{aff19}}
\and A.~G.~S\'anchez\orcid{0000-0003-1198-831X}\inst{\ref{aff20}}
\and J.~Bel\inst{\ref{aff21}}
\and C.~Carbone\orcid{0000-0003-0125-3563}\inst{\ref{aff6}}
\and A.~Crespi\orcid{0009-0009-9437-0232}\inst{\ref{aff16}}
\and S.~Radinovi\'c\orcid{0009-0002-1535-7953}\inst{\ref{aff8},\ref{aff22}}
\and G.~Parimbelli\orcid{0000-0002-2539-2472}\inst{\ref{aff8},\ref{aff12}}
\and A.~Farina\orcid{0009-0000-3420-929X}\inst{\ref{aff2},\ref{aff3}}
\and I.~Risso\orcid{0000-0003-2525-7761}\inst{\ref{aff2},\ref{aff3},\ref{aff4}}
\and M.~Guidi\orcid{0000-0001-9408-1101}\inst{\ref{aff23},\ref{aff24}}
\and G.~Degni\orcid{0009-0001-4912-1087}\inst{\ref{aff14}}
\and D.~Eisenstein\orcid{0000-0002-2929-3121}\inst{\ref{aff25}}
\and F.~Beutler\orcid{0000-0003-0467-5438}\inst{\ref{aff26}}
\and C.~Garc\'ia-Garc\'ia\orcid{0000-0001-6394-7494}\inst{\ref{aff5},\ref{aff16}}
\and G.~Piccirilli\orcid{0000-0002-3341-1872}\inst{\ref{aff27},\ref{aff28}}
\and J.~G.~Sorce\orcid{0000-0002-2307-2432}\inst{\ref{aff29},\ref{aff30}}
\and B.~Altieri\orcid{0000-0003-3936-0284}\inst{\ref{aff31}}
\and S.~Andreon\orcid{0000-0002-2041-8784}\inst{\ref{aff2}}
\and C.~Baccigalupi\orcid{0000-0002-8211-1630}\inst{\ref{aff10},\ref{aff1},\ref{aff11},\ref{aff12}}
\and M.~Baldi\orcid{0000-0003-4145-1943}\inst{\ref{aff23},\ref{aff24},\ref{aff32}}
\and S.~Bardelli\orcid{0000-0002-8900-0298}\inst{\ref{aff24}}
\and P.~Battaglia\orcid{0000-0002-7337-5909}\inst{\ref{aff24}}
\and A.~Biviano\orcid{0000-0002-0857-0732}\inst{\ref{aff1},\ref{aff10}}
\and M.~Brescia\orcid{0000-0001-9506-5680}\inst{\ref{aff33},\ref{aff34}}
\and S.~Camera\orcid{0000-0003-3399-3574}\inst{\ref{aff35},\ref{aff36},\ref{aff37}}
\and G.~Ca\~nas-Herrera\orcid{0000-0003-2796-2149}\inst{\ref{aff38}}
\and V.~Capobianco\orcid{0000-0002-3309-7692}\inst{\ref{aff37}}
\and J.~Carretero\orcid{0000-0002-3130-0204}\inst{\ref{aff39},\ref{aff40}}
\and F.~J.~Castander\orcid{0000-0001-7316-4573}\inst{\ref{aff8},\ref{aff9}}
\and M.~Castellano\orcid{0000-0001-9875-8263}\inst{\ref{aff41}}
\and G.~Castignani\orcid{0000-0001-6831-0687}\inst{\ref{aff24}}
\and S.~Cavuoti\orcid{0000-0002-3787-4196}\inst{\ref{aff34},\ref{aff42}}
\and K.~C.~Chambers\orcid{0000-0001-6965-7789}\inst{\ref{aff43}}
\and A.~Cimatti\inst{\ref{aff44}}
\and C.~Colodro-Conde\inst{\ref{aff45}}
\and G.~Congedo\orcid{0000-0003-2508-0046}\inst{\ref{aff26}}
\and L.~Conversi\orcid{0000-0002-6710-8476}\inst{\ref{aff46},\ref{aff31}}
\and Y.~Copin\orcid{0000-0002-5317-7518}\inst{\ref{aff47}}
\and F.~Courbin\orcid{0000-0003-0758-6510}\inst{\ref{aff48},\ref{aff49},\ref{aff50}}
\and H.~M.~Courtois\orcid{0000-0003-0509-1776}\inst{\ref{aff51}}
\and H.~Degaudenzi\orcid{0000-0002-5887-6799}\inst{\ref{aff52}}
\and S.~de~la~Torre\inst{\ref{aff53}}
\and G.~De~Lucia\orcid{0000-0002-6220-9104}\inst{\ref{aff1}}
\and F.~Dubath\orcid{0000-0002-6533-2810}\inst{\ref{aff52}}
\and X.~Dupac\inst{\ref{aff31}}
\and S.~Escoffier\orcid{0000-0002-2847-7498}\inst{\ref{aff14}}
\and M.~Farina\orcid{0000-0002-3089-7846}\inst{\ref{aff54}}
\and R.~Farinelli\inst{\ref{aff24}}
\and F.~Faustini\orcid{0000-0001-6274-5145}\inst{\ref{aff41}}
\and S.~Ferriol\inst{\ref{aff47}}
\and F.~Finelli\orcid{0000-0002-6694-3269}\inst{\ref{aff24},\ref{aff55}}
\and P.~Fosalba\orcid{0000-0002-1510-5214}\inst{\ref{aff9},\ref{aff8}}
\and N.~Fourmanoit\orcid{0009-0005-6816-6925}\inst{\ref{aff14}}
\and M.~Frailis\orcid{0000-0002-7400-2135}\inst{\ref{aff1}}
\and E.~Franceschi\orcid{0000-0002-0585-6591}\inst{\ref{aff24}}
\and M.~Fumana\orcid{0000-0001-6787-5950}\inst{\ref{aff6}}
\and S.~Galeotta\orcid{0000-0002-3748-5115}\inst{\ref{aff1}}
\and K.~George\orcid{0000-0002-1734-8455}\inst{\ref{aff56}}
\and W.~Gillard\orcid{0000-0003-4744-9748}\inst{\ref{aff14}}
\and B.~Gillis\orcid{0000-0002-4478-1270}\inst{\ref{aff26}}
\and C.~Giocoli\orcid{0000-0002-9590-7961}\inst{\ref{aff24},\ref{aff32}}
\and J.~Gracia-Carpio\orcid{0000-0003-4689-3134}\inst{\ref{aff20}}
\and A.~Grazian\orcid{0000-0002-5688-0663}\inst{\ref{aff57}}
\and F.~Grupp\inst{\ref{aff20},\ref{aff58}}
\and L.~Guzzo\orcid{0000-0001-8264-5192}\inst{\ref{aff7},\ref{aff2},\ref{aff59}}
\and S.~V.~H.~Haugan\orcid{0000-0001-9648-7260}\inst{\ref{aff22}}
\and W.~Holmes\orcid{0009-0007-8554-4646}\inst{\ref{aff60}}
\and F.~Hormuth\inst{\ref{aff61}}
\and A.~Hornstrup\orcid{0000-0002-3363-0936}\inst{\ref{aff62},\ref{aff63}}
\and K.~Jahnke\orcid{0000-0003-3804-2137}\inst{\ref{aff64}}
\and M.~Jhabvala\inst{\ref{aff65}}
\and B.~Joachimi\orcid{0000-0001-7494-1303}\inst{\ref{aff66}}
\and S.~Kermiche\orcid{0000-0002-0302-5735}\inst{\ref{aff14}}
\and A.~Kiessling\orcid{0000-0002-2590-1273}\inst{\ref{aff60}}
\and B.~Kubik\orcid{0009-0006-5823-4880}\inst{\ref{aff47}}
\and M.~K\"ummel\orcid{0000-0003-2791-2117}\inst{\ref{aff58}}
\and M.~Kunz\orcid{0000-0002-3052-7394}\inst{\ref{aff67}}
\and H.~Kurki-Suonio\orcid{0000-0002-4618-3063}\inst{\ref{aff68},\ref{aff69}}
\and A.~M.~C.~Le~Brun\orcid{0000-0002-0936-4594}\inst{\ref{aff70}}
\and S.~Ligori\orcid{0000-0003-4172-4606}\inst{\ref{aff37}}
\and P.~B.~Lilje\orcid{0000-0003-4324-7794}\inst{\ref{aff22}}
\and V.~Lindholm\orcid{0000-0003-2317-5471}\inst{\ref{aff68},\ref{aff69}}
\and I.~Lloro\orcid{0000-0001-5966-1434}\inst{\ref{aff71}}
\and G.~Mainetti\orcid{0000-0003-2384-2377}\inst{\ref{aff72}}
\and O.~Mansutti\orcid{0000-0001-5758-4658}\inst{\ref{aff1}}
\and O.~Marggraf\orcid{0000-0001-7242-3852}\inst{\ref{aff19}}
\and M.~Martinelli\orcid{0000-0002-6943-7732}\inst{\ref{aff41},\ref{aff73}}
\and N.~Martinet\orcid{0000-0003-2786-7790}\inst{\ref{aff53}}
\and F.~Marulli\orcid{0000-0002-8850-0303}\inst{\ref{aff74},\ref{aff24},\ref{aff32}}
\and R.~J.~Massey\orcid{0000-0002-6085-3780}\inst{\ref{aff75}}
\and E.~Medinaceli\orcid{0000-0002-4040-7783}\inst{\ref{aff24}}
\and S.~Mei\orcid{0000-0002-2849-559X}\inst{\ref{aff76},\ref{aff77}}
\and M.~Melchior\inst{\ref{aff78}}
\and M.~Meneghetti\orcid{0000-0003-1225-7084}\inst{\ref{aff24},\ref{aff32}}
\and E.~Merlin\orcid{0000-0001-6870-8900}\inst{\ref{aff41}}
\and G.~Meylan\orcid{0000-0001-6503-0209}\inst{\ref{aff79}}
\and A.~Mora\orcid{0000-0002-1922-8529}\inst{\ref{aff80}}
\and M.~Moresco\orcid{0000-0002-7616-7136}\inst{\ref{aff74},\ref{aff24}}
\and L.~Moscardini\orcid{0000-0002-3473-6716}\inst{\ref{aff74},\ref{aff24},\ref{aff32}}
\and C.~Neissner\orcid{0000-0001-8524-4968}\inst{\ref{aff81},\ref{aff40}}
\and S.-M.~Niemi\orcid{0009-0005-0247-0086}\inst{\ref{aff82}}
\and C.~Padilla\orcid{0000-0001-7951-0166}\inst{\ref{aff81}}
\and S.~Paltani\orcid{0000-0002-8108-9179}\inst{\ref{aff52}}
\and F.~Pasian\orcid{0000-0002-4869-3227}\inst{\ref{aff1}}
\and K.~Pedersen\inst{\ref{aff83}}
\and V.~Pettorino\orcid{0000-0002-4203-9320}\inst{\ref{aff82}}
\and S.~Pires\orcid{0000-0002-0249-2104}\inst{\ref{aff84}}
\and G.~Polenta\orcid{0000-0003-4067-9196}\inst{\ref{aff85}}
\and M.~Poncet\inst{\ref{aff86}}
\and L.~A.~Popa\inst{\ref{aff87}}
\and F.~Raison\orcid{0000-0002-7819-6918}\inst{\ref{aff20}}
\and J.~Rhodes\orcid{0000-0002-4485-8549}\inst{\ref{aff60}}
\and G.~Riccio\inst{\ref{aff34}}
\and F.~Rizzo\orcid{0000-0002-9407-585X}\inst{\ref{aff1}}
\and E.~Romelli\orcid{0000-0003-3069-9222}\inst{\ref{aff1}}
\and M.~Roncarelli\orcid{0000-0001-9587-7822}\inst{\ref{aff24}}
\and R.~Saglia\orcid{0000-0003-0378-7032}\inst{\ref{aff58},\ref{aff20}}
\and Z.~Sakr\orcid{0000-0002-4823-3757}\inst{\ref{aff88},\ref{aff89},\ref{aff90}}
\and D.~Sapone\orcid{0000-0001-7089-4503}\inst{\ref{aff91}}
\and M.~Schirmer\orcid{0000-0003-2568-9994}\inst{\ref{aff64}}
\and P.~Schneider\orcid{0000-0001-8561-2679}\inst{\ref{aff19}}
\and T.~Schrabback\orcid{0000-0002-6987-7834}\inst{\ref{aff92}}
\and M.~Scodeggio\inst{\ref{aff6}}
\and A.~Secroun\orcid{0000-0003-0505-3710}\inst{\ref{aff14}}
\and E.~Sihvola\orcid{0000-0003-1804-7715}\inst{\ref{aff93}}
\and C.~Sirignano\orcid{0000-0002-0995-7146}\inst{\ref{aff94},\ref{aff95}}
\and G.~Sirri\orcid{0000-0003-2626-2853}\inst{\ref{aff32}}
\and L.~Stanco\orcid{0000-0002-9706-5104}\inst{\ref{aff95}}
\and P.~Tallada-Cresp\'{i}\orcid{0000-0002-1336-8328}\inst{\ref{aff39},\ref{aff40}}
\and D.~Tavagnacco\orcid{0000-0001-7475-9894}\inst{\ref{aff1}}
\and A.~N.~Taylor\inst{\ref{aff26}}
\and I.~Tereno\orcid{0000-0002-4537-6218}\inst{\ref{aff96},\ref{aff97}}
\and N.~Tessore\orcid{0000-0002-9696-7931}\inst{\ref{aff98}}
\and S.~Toft\orcid{0000-0003-3631-7176}\inst{\ref{aff99},\ref{aff100}}
\and R.~Toledo-Moreo\orcid{0000-0002-2997-4859}\inst{\ref{aff101},\ref{aff102}}
\and F.~Torradeflot\orcid{0000-0003-1160-1517}\inst{\ref{aff40},\ref{aff39}}
\and I.~Tutusaus\orcid{0000-0002-3199-0399}\inst{\ref{aff8},\ref{aff9},\ref{aff89}}
\and L.~Valenziano\orcid{0000-0002-1170-0104}\inst{\ref{aff24},\ref{aff55}}
\and J.~Valiviita\orcid{0000-0001-6225-3693}\inst{\ref{aff68},\ref{aff69}}
\and T.~Vassallo\orcid{0000-0001-6512-6358}\inst{\ref{aff1},\ref{aff56}}
\and G.~Verdoes~Kleijn\orcid{0000-0001-5803-2580}\inst{\ref{aff103}}
\and Y.~Wang\orcid{0000-0002-4749-2984}\inst{\ref{aff104}}
\and J.~Weller\orcid{0000-0002-8282-2010}\inst{\ref{aff58},\ref{aff20}}
\and A.~Zacchei\orcid{0000-0003-0396-1192}\inst{\ref{aff1},\ref{aff10}}
\and G.~Zamorani\orcid{0000-0002-2318-301X}\inst{\ref{aff24}}
\and F.~M.~Zerbi\orcid{0000-0002-9996-973X}\inst{\ref{aff2}}
\and E.~Zucca\orcid{0000-0002-5845-8132}\inst{\ref{aff24}}
\and M.~Ballardini\orcid{0000-0003-4481-3559}\inst{\ref{aff105},\ref{aff106},\ref{aff24}}
\and A.~Boucaud\orcid{0000-0001-7387-2633}\inst{\ref{aff76}}
\and E.~Bozzo\orcid{0000-0002-8201-1525}\inst{\ref{aff52}}
\and C.~Burigana\orcid{0000-0002-3005-5796}\inst{\ref{aff107},\ref{aff55}}
\and R.~Cabanac\orcid{0000-0001-6679-2600}\inst{\ref{aff89}}
\and M.~Calabrese\orcid{0000-0002-2637-2422}\inst{\ref{aff108},\ref{aff6}}
\and A.~Cappi\inst{\ref{aff109},\ref{aff24}}
\and T.~Castro\orcid{0000-0002-6292-3228}\inst{\ref{aff1},\ref{aff11},\ref{aff10},\ref{aff15}}
\and J.~A.~Escartin~Vigo\inst{\ref{aff20}}
\and G.~Fabbian\orcid{0000-0002-3255-4695}\inst{\ref{aff30}}
\and J.~Garc\'ia-Bellido\orcid{0000-0002-9370-8360}\inst{\ref{aff88}}
\and J.~Macias-Perez\orcid{0000-0002-5385-2763}\inst{\ref{aff110}}
\and R.~Maoli\orcid{0000-0002-6065-3025}\inst{\ref{aff111},\ref{aff41}}
\and J.~Mart\'{i}n-Fleitas\orcid{0000-0002-8594-569X}\inst{\ref{aff112}}
\and N.~Mauri\orcid{0000-0001-8196-1548}\inst{\ref{aff44},\ref{aff32}}
\and R.~B.~Metcalf\orcid{0000-0003-3167-2574}\inst{\ref{aff74},\ref{aff24}}
\and M.~P\"ontinen\orcid{0000-0001-5442-2530}\inst{\ref{aff68}}
\and V.~Scottez\orcid{0009-0008-3864-940X}\inst{\ref{aff113},\ref{aff114}}
\and M.~Sereno\orcid{0000-0003-0302-0325}\inst{\ref{aff24},\ref{aff32}}
\and M.~Tenti\orcid{0000-0002-4254-5901}\inst{\ref{aff32}}
\and M.~Tucci\inst{\ref{aff52}}
\and M.~Wiesmann\orcid{0009-0000-8199-5860}\inst{\ref{aff22}}
\and Y.~Akrami\orcid{0000-0002-2407-7956}\inst{\ref{aff88},\ref{aff115}}
\and I.~T.~Andika\orcid{0000-0001-6102-9526}\inst{\ref{aff58}}
\and M.~Archidiacono\orcid{0000-0003-4952-9012}\inst{\ref{aff7},\ref{aff59}}
\and F.~Atrio-Barandela\orcid{0000-0002-2130-2513}\inst{\ref{aff116}}
\and E.~Aubourg\orcid{0000-0002-5592-023X}\inst{\ref{aff76},\ref{aff117}}
\and L.~Bazzanini\orcid{0000-0003-0727-0137}\inst{\ref{aff105},\ref{aff24}}
\and D.~Bertacca\orcid{0000-0002-2490-7139}\inst{\ref{aff94},\ref{aff57},\ref{aff95}}
\and M.~Bethermin\orcid{0000-0002-3915-2015}\inst{\ref{aff118}}
\and A.~Blanchard\orcid{0000-0001-8555-9003}\inst{\ref{aff89}}
\and L.~Blot\orcid{0000-0002-9622-7167}\inst{\ref{aff119},\ref{aff70}}
\and S.~Borgani\orcid{0000-0001-6151-6439}\inst{\ref{aff13},\ref{aff10},\ref{aff1},\ref{aff11},\ref{aff15}}
\and M.~L.~Brown\orcid{0000-0002-0370-8077}\inst{\ref{aff120}}
\and S.~Bruton\orcid{0000-0002-6503-5218}\inst{\ref{aff121}}
\and A.~Calabro\orcid{0000-0003-2536-1614}\inst{\ref{aff41}}
\and F.~Caro\orcid{0009-0003-1053-0507}\inst{\ref{aff41}}
\and C.~S.~Carvalho\inst{\ref{aff97}}
\and F.~Cogato\orcid{0000-0003-4632-6113}\inst{\ref{aff74},\ref{aff24}}
\and S.~Contarini\orcid{0000-0002-9843-723X}\inst{\ref{aff20}}
\and A.~R.~Cooray\orcid{0000-0002-3892-0190}\inst{\ref{aff122}}
\and O.~Cucciati\orcid{0000-0002-9336-7551}\inst{\ref{aff24}}
\and S.~Davini\orcid{0000-0003-3269-1718}\inst{\ref{aff3}}
\and T.~de~Boer\orcid{0000-0001-5486-2747}\inst{\ref{aff43}}
\and F.~De~Paolis\orcid{0000-0001-6460-7563}\inst{\ref{aff123},\ref{aff124},\ref{aff125}}
\and G.~Desprez\orcid{0000-0001-8325-1742}\inst{\ref{aff103}}
\and A.~D\'iaz-S\'anchez\orcid{0000-0003-0748-4768}\inst{\ref{aff126}}
\and S.~Di~Domizio\orcid{0000-0003-2863-5895}\inst{\ref{aff4},\ref{aff3}}
\and J.~M.~Diego\orcid{0000-0001-9065-3926}\inst{\ref{aff127}}
\and V.~Duret\orcid{0009-0009-0383-4960}\inst{\ref{aff14}}
\and M.~Y.~Elkhashab\orcid{0000-0001-9306-2603}\inst{\ref{aff1},\ref{aff11},\ref{aff13},\ref{aff10}}
\and Y.~Fang\orcid{0000-0002-0334-6950}\inst{\ref{aff58}}
\and P.~G.~Ferreira\orcid{0000-0002-3021-2851}\inst{\ref{aff128}}
\and A.~Finoguenov\orcid{0000-0002-4606-5403}\inst{\ref{aff68}}
\and A.~Franco\orcid{0000-0002-4761-366X}\inst{\ref{aff124},\ref{aff123},\ref{aff125}}
\and K.~Ganga\orcid{0000-0001-8159-8208}\inst{\ref{aff76}}
\and T.~Gasparetto\orcid{0000-0002-7913-4866}\inst{\ref{aff41}}
\and E.~Gaztanaga\orcid{0000-0001-9632-0815}\inst{\ref{aff8},\ref{aff9},\ref{aff18}}
\and Z.~Ghaffari\orcid{0000-0002-6467-8078}\inst{\ref{aff1},\ref{aff10}}
\and F.~Giacomini\orcid{0000-0002-3129-2814}\inst{\ref{aff32}}
\and F.~Gianotti\orcid{0000-0003-4666-119X}\inst{\ref{aff24}}
\and E.~J.~Gonzalez\orcid{0000-0002-0226-9893}\inst{\ref{aff129},\ref{aff130}}
\and G.~Gozaliasl\orcid{0000-0002-0236-919X}\inst{\ref{aff131},\ref{aff68}}
\and A.~Gruppuso\orcid{0000-0001-9272-5292}\inst{\ref{aff24},\ref{aff32}}
\and C.~M.~Gutierrez\orcid{0000-0001-7854-783X}\inst{\ref{aff45},\ref{aff132}}
\and A.~Hall\orcid{0000-0002-3139-8651}\inst{\ref{aff26}}
\and H.~Hildebrandt\orcid{0000-0002-9814-3338}\inst{\ref{aff133}}
\and J.~Hjorth\orcid{0000-0002-4571-2306}\inst{\ref{aff83}}
\and J.~J.~E.~Kajava\orcid{0000-0002-3010-8333}\inst{\ref{aff134},\ref{aff135},\ref{aff136}}
\and Y.~Kang\orcid{0009-0000-8588-7250}\inst{\ref{aff52}}
\and V.~Kansal\orcid{0000-0002-4008-6078}\inst{\ref{aff137},\ref{aff138}}
\and D.~Karagiannis\orcid{0000-0002-4927-0816}\inst{\ref{aff105},\ref{aff139}}
\and K.~Kiiveri\orcid{0000-0002-3711-3346}\inst{\ref{aff93}}
\and J.~Kim\orcid{0000-0003-2776-2761}\inst{\ref{aff128}}
\and C.~C.~Kirkpatrick\inst{\ref{aff93}}
\and K.~Koyama\orcid{0000-0001-6727-6915}\inst{\ref{aff18}}
\and S.~Kruk\orcid{0000-0001-8010-8879}\inst{\ref{aff31}}
\and M.~C.~Lam\orcid{0000-0002-9347-2298}\inst{\ref{aff26}}
\and F.~Leclercq\orcid{0000-0002-9339-1404}\inst{\ref{aff140}}
\and L.~Legrand\orcid{0000-0003-0610-5252}\inst{\ref{aff141},\ref{aff142}}
\and M.~Lembo\orcid{0000-0002-5271-5070}\inst{\ref{aff140}}
\and F.~Lepori\orcid{0009-0000-5061-7138}\inst{\ref{aff143}}
\and G.~Leroy\orcid{0009-0004-2523-4425}\inst{\ref{aff144},\ref{aff75}}
\and G.~F.~Lesci\orcid{0000-0002-4607-2830}\inst{\ref{aff74},\ref{aff24}}
\and J.~Lesgourgues\orcid{0000-0001-7627-353X}\inst{\ref{aff145}}
\and T.~I.~Liaudat\orcid{0000-0002-9104-314X}\inst{\ref{aff117}}
\and S.~J.~Liu\orcid{0000-0001-7680-2139}\inst{\ref{aff54}}
\and M.~Magliocchetti\orcid{0000-0001-9158-4838}\inst{\ref{aff54}}
\and C.~J.~A.~P.~Martins\orcid{0000-0002-4886-9261}\inst{\ref{aff146},\ref{aff147}}
\and L.~Maurin\orcid{0000-0002-8406-0857}\inst{\ref{aff30}}
\and M.~Migliaccio\inst{\ref{aff27},\ref{aff28}}
\and M.~Miluzio\inst{\ref{aff31},\ref{aff148}}
\and G.~Morgante\inst{\ref{aff24}}
\and K.~Naidoo\orcid{0000-0002-9182-1802}\inst{\ref{aff18},\ref{aff64}}
\and A.~Navarro-Alsina\orcid{0000-0002-3173-2592}\inst{\ref{aff19}}
\and S.~Nesseris\orcid{0000-0002-0567-0324}\inst{\ref{aff88}}
\and F.~Pace\orcid{0000-0001-8039-0480}\inst{\ref{aff35},\ref{aff36},\ref{aff37}}
\and D.~Paoletti\orcid{0000-0003-4761-6147}\inst{\ref{aff24},\ref{aff55}}
\and K.~Paterson\orcid{0000-0001-8340-3486}\inst{\ref{aff64}}
\and L.~Patrizii\inst{\ref{aff32}}
\and C.~Pattison\orcid{0000-0003-3272-2617}\inst{\ref{aff18}}
\and A.~Pisani\orcid{0000-0002-6146-4437}\inst{\ref{aff14}}
\and D.~Potter\orcid{0000-0002-0757-5195}\inst{\ref{aff149}}
\and A.~Pourtsidou\orcid{0000-0001-9110-5550}\inst{\ref{aff26},\ref{aff150}}
\and G.~W.~Pratt\inst{\ref{aff84}}
\and S.~Quai\orcid{0000-0002-0449-8163}\inst{\ref{aff74},\ref{aff24}}
\and M.~Radovich\orcid{0000-0002-3585-866X}\inst{\ref{aff57}}
\and G.~Rodighiero\orcid{0000-0002-9415-2296}\inst{\ref{aff94},\ref{aff57}}
\and W.~Roster\orcid{0000-0002-9149-6528}\inst{\ref{aff20}}
\and S.~Sacquegna\orcid{0000-0002-8433-6630}\inst{\ref{aff151}}
\and M.~Sahl\'en\orcid{0000-0003-0973-4804}\inst{\ref{aff152}}
\and D.~B.~Sanders\orcid{0000-0002-1233-9998}\inst{\ref{aff43}}
\and A.~Schneider\orcid{0000-0001-7055-8104}\inst{\ref{aff149}}
\and D.~Sciotti\orcid{0009-0008-4519-2620}\inst{\ref{aff41},\ref{aff73}}
\and E.~Sellentin\inst{\ref{aff153},\ref{aff38}}
\and L.~C.~Smith\orcid{0000-0002-3259-2771}\inst{\ref{aff154}}
\and I.~Szapudi\orcid{0000-0003-2274-0301}\inst{\ref{aff43}}
\and K.~Tanidis\orcid{0000-0001-9843-5130}\inst{\ref{aff155}}
\and C.~Tao\orcid{0000-0001-7961-8177}\inst{\ref{aff14}}
\and F.~Tarsitano\orcid{0000-0002-5919-0238}\inst{\ref{aff156},\ref{aff52}}
\and G.~Testera\orcid{0000-0003-2970-766X}\inst{\ref{aff3}}
\and R.~Teyssier\orcid{0000-0001-7689-0933}\inst{\ref{aff157}}
\and S.~Tosi\orcid{0000-0002-7275-9193}\inst{\ref{aff4},\ref{aff2},\ref{aff3}}
\and A.~Troja\orcid{0000-0003-0239-4595}\inst{\ref{aff1}}
\and C.~Uhlemann\orcid{0000-0001-7831-1579}\inst{\ref{aff158},\ref{aff159}}
\and C.~Valieri\inst{\ref{aff32}}
\and F.~Vernizzi\orcid{0000-0003-3426-2802}\inst{\ref{aff160}}
\and G.~Verza\orcid{0000-0002-1886-8348}\inst{\ref{aff161},\ref{aff162}}
\and S.~Vinciguerra\orcid{0009-0005-4018-3184}\inst{\ref{aff53}}
\and M.~von~Wietersheim-Kramsta\orcid{0000-0003-4986-5091}\inst{\ref{aff75},\ref{aff144}}
\and N.~A.~Walton\orcid{0000-0003-3983-8778}\inst{\ref{aff154}}
\and A.~H.~Wright\orcid{0000-0001-7363-7932}\inst{\ref{aff133}}
\and H.~W.~Yeung\orcid{0000-0002-4993-9014}\inst{\ref{aff26}}}
										   
\institute{INAF-Osservatorio Astronomico di Trieste, Via G. B. Tiepolo 11, 34143 Trieste, Italy\label{aff1}
\and
INAF-Osservatorio Astronomico di Brera, Via Brera 28, 20122 Milano, Italy\label{aff2}
\and
INFN-Sezione di Genova, Via Dodecaneso 33, 16146, Genova, Italy\label{aff3}
\and
Dipartimento di Fisica, Universit\`a di Genova, Via Dodecaneso 33, 16146, Genova, Italy\label{aff4}
\and
Waterloo Centre for Astrophysics, University of Waterloo, Waterloo, Ontario N2L 3G1, Canada\label{aff5}
\and
INAF-IASF Milano, Via Alfonso Corti 12, 20133 Milano, Italy\label{aff6}
\and
Dipartimento di Fisica "Aldo Pontremoli", Universit\`a degli Studi di Milano, Via Celoria 16, 20133 Milano, Italy\label{aff7}
\and
Institute of Space Sciences (ICE, CSIC), Campus UAB, Carrer de Can Magrans, s/n, 08193 Barcelona, Spain\label{aff8}
\and
Institut d'Estudis Espacials de Catalunya (IEEC),  Edifici RDIT, Campus UPC, 08860 Castelldefels, Barcelona, Spain\label{aff9}
\and
IFPU, Institute for Fundamental Physics of the Universe, via Beirut 2, 34151 Trieste, Italy\label{aff10}
\and
INFN, Sezione di Trieste, Via Valerio 2, 34127 Trieste TS, Italy\label{aff11}
\and
SISSA, International School for Advanced Studies, Via Bonomea 265, 34136 Trieste TS, Italy\label{aff12}
\and
Dipartimento di Fisica - Sezione di Astronomia, Universit\`a di Trieste, Via Tiepolo 11, 34131 Trieste, Italy\label{aff13}
\and
Aix-Marseille Universit\'e, CNRS/IN2P3, CPPM, Marseille, France\label{aff14}
\and
ICSC - Centro Nazionale di Ricerca in High Performance Computing, Big Data e Quantum Computing, Via Magnanelli 2, Bologna, Italy\label{aff15}
\and
Department of Physics and Astronomy, University of Waterloo, Waterloo, Ontario N2L 3G1, Canada\label{aff16}
\and
Perimeter Institute for Theoretical Physics, Waterloo, Ontario N2L 2Y5, Canada\label{aff17}
\and
Institute of Cosmology and Gravitation, University of Portsmouth, Portsmouth PO1 3FX, UK\label{aff18}
\and
Universit\"at Bonn, Argelander-Institut f\"ur Astronomie, Auf dem H\"ugel 71, 53121 Bonn, Germany\label{aff19}
\and
Max Planck Institute for Extraterrestrial Physics, Giessenbachstr. 1, 85748 Garching, Germany\label{aff20}
\and
Aix-Marseille Universit\'e, Universit\'e de Toulon, CNRS, CPT, Marseille, France\label{aff21}
\and
Institute of Theoretical Astrophysics, University of Oslo, P.O. Box 1029 Blindern, 0315 Oslo, Norway\label{aff22}
\and
Dipartimento di Fisica e Astronomia, Universit\`a di Bologna, Via Gobetti 93/2, 40129 Bologna, Italy\label{aff23}
\and
INAF-Osservatorio di Astrofisica e Scienza dello Spazio di Bologna, Via Piero Gobetti 93/3, 40129 Bologna, Italy\label{aff24}
\and
Center for Astrophysics | Harvard \& Smithsonian, 60 Garden St., Cambridge, MA 02138, USA\label{aff25}
\and
Institute for Astronomy, University of Edinburgh, Royal Observatory, Blackford Hill, Edinburgh EH9 3HJ, UK\label{aff26}
\and
Dipartimento di Fisica, Universit\`a di Roma Tor Vergata, Via della Ricerca Scientifica 1, Roma, Italy\label{aff27}
\and
INFN, Sezione di Roma 2, Via della Ricerca Scientifica 1, Roma, Italy\label{aff28}
\and
Univ. Lille, CNRS, Centrale Lille, UMR 9189 CRIStAL, 59000 Lille, France\label{aff29}
\and
Universit\'e Paris-Saclay, CNRS, Institut d'astrophysique spatiale, 91405, Orsay, France\label{aff30}
\and
ESAC/ESA, Camino Bajo del Castillo, s/n., Urb. Villafranca del Castillo, 28692 Villanueva de la Ca\~nada, Madrid, Spain\label{aff31}
\and
INFN-Sezione di Bologna, Viale Berti Pichat 6/2, 40127 Bologna, Italy\label{aff32}
\and
Department of Physics "E. Pancini", University Federico II, Via Cinthia 6, 80126, Napoli, Italy\label{aff33}
\and
INAF-Osservatorio Astronomico di Capodimonte, Via Moiariello 16, 80131 Napoli, Italy\label{aff34}
\and
Dipartimento di Fisica, Universit\`a degli Studi di Torino, Via P. Giuria 1, 10125 Torino, Italy\label{aff35}
\and
INFN-Sezione di Torino, Via P. Giuria 1, 10125 Torino, Italy\label{aff36}
\and
INAF-Osservatorio Astrofisico di Torino, Via Osservatorio 20, 10025 Pino Torinese (TO), Italy\label{aff37}
\and
Leiden Observatory, Leiden University, Einsteinweg 55, 2333 CC Leiden, The Netherlands\label{aff38}
\and
Centro de Investigaciones Energ\'eticas, Medioambientales y Tecnol\'ogicas (CIEMAT), Avenida Complutense 40, 28040 Madrid, Spain\label{aff39}
\and
Port d'Informaci\'{o} Cient\'{i}fica, Campus UAB, C. Albareda s/n, 08193 Bellaterra (Barcelona), Spain\label{aff40}
\and
INAF-Osservatorio Astronomico di Roma, Via Frascati 33, 00078 Monteporzio Catone, Italy\label{aff41}
\and
INFN section of Naples, Via Cinthia 6, 80126, Napoli, Italy\label{aff42}
\and
Institute for Astronomy, University of Hawaii, 2680 Woodlawn Drive, Honolulu, HI 96822, USA\label{aff43}
\and
Dipartimento di Fisica e Astronomia "Augusto Righi" - Alma Mater Studiorum Universit\`a di Bologna, Viale Berti Pichat 6/2, 40127 Bologna, Italy\label{aff44}
\and
Instituto de Astrof\'{\i}sica de Canarias, E-38205 La Laguna, Tenerife, Spain\label{aff45}
\and
European Space Agency/ESRIN, Largo Galileo Galilei 1, 00044 Frascati, Roma, Italy\label{aff46}
\and
Universit\'e Claude Bernard Lyon 1, CNRS/IN2P3, IP2I Lyon, UMR 5822, Villeurbanne, F-69100, France\label{aff47}
\and
Institut de Ci\`{e}ncies del Cosmos (ICCUB), Universitat de Barcelona (IEEC-UB), Mart\'{i} i Franqu\`{e}s 1, 08028 Barcelona, Spain\label{aff48}
\and
Instituci\'o Catalana de Recerca i Estudis Avan\c{c}ats (ICREA), Passeig de Llu\'{\i}s Companys 23, 08010 Barcelona, Spain\label{aff49}
\and
Institut de Ciencies de l'Espai (IEEC-CSIC), Campus UAB, Carrer de Can Magrans, s/n Cerdanyola del Vall\'es, 08193 Barcelona, Spain\label{aff50}
\and
UCB Lyon 1, CNRS/IN2P3, IUF, IP2I Lyon, 4 rue Enrico Fermi, 69622 Villeurbanne, France\label{aff51}
\and
Department of Astronomy, University of Geneva, ch. d'Ecogia 16, 1290 Versoix, Switzerland\label{aff52}
\and
Aix-Marseille Universit\'e, CNRS, CNES, LAM, Marseille, France\label{aff53}
\and
INAF-Istituto di Astrofisica e Planetologia Spaziali, via del Fosso del Cavaliere, 100, 00100 Roma, Italy\label{aff54}
\and
INFN-Bologna, Via Irnerio 46, 40126 Bologna, Italy\label{aff55}
\and
University Observatory, LMU Faculty of Physics, Scheinerstr.~1, 81679 Munich, Germany\label{aff56}
\and
INAF-Osservatorio Astronomico di Padova, Via dell'Osservatorio 5, 35122 Padova, Italy\label{aff57}
\and
Universit\"ats-Sternwarte M\"unchen, Fakult\"at f\"ur Physik, Ludwig-Maximilians-Universit\"at M\"unchen, Scheinerstr.~1, 81679 M\"unchen, Germany\label{aff58}
\and
INFN-Sezione di Milano, Via Celoria 16, 20133 Milano, Italy\label{aff59}
\and
Jet Propulsion Laboratory, California Institute of Technology, 4800 Oak Grove Drive, Pasadena, CA, 91109, USA\label{aff60}
\and
Felix Hormuth Engineering, Goethestr. 17, 69181 Leimen, Germany\label{aff61}
\and
Technical University of Denmark, Elektrovej 327, 2800 Kgs. Lyngby, Denmark\label{aff62}
\and
Cosmic Dawn Center (DAWN), Denmark\label{aff63}
\and
Max-Planck-Institut f\"ur Astronomie, K\"onigstuhl 17, 69117 Heidelberg, Germany\label{aff64}
\and
NASA Goddard Space Flight Center, Greenbelt, MD 20771, USA\label{aff65}
\and
Department of Physics and Astronomy, University College London, Gower Street, London WC1E 6BT, UK\label{aff66}
\and
Universit\'e de Gen\`eve, D\'epartement de Physique Th\'eorique and Centre for Astroparticle Physics, 24 quai Ernest-Ansermet, CH-1211 Gen\`eve 4, Switzerland\label{aff67}
\and
Department of Physics, P.O. Box 64, University of Helsinki, 00014 Helsinki, Finland\label{aff68}
\and
Helsinki Institute of Physics, Gustaf H{\"a}llstr{\"o}min katu 2, University of Helsinki, 00014 Helsinki, Finland\label{aff69}
\and
Laboratoire d'etude de l'Univers et des phenomenes eXtremes, Observatoire de Paris, Universit\'e PSL, Sorbonne Universit\'e, CNRS, 92190 Meudon, France\label{aff70}
\and
SKAO, Jodrell Bank, Lower Withington, Macclesfield SK11 9FT, UK\label{aff71}
\and
Centre de Calcul de l'IN2P3/CNRS, 21 avenue Pierre de Coubertin 69627 Villeurbanne Cedex, France\label{aff72}
\and
INFN-Sezione di Roma, Piazzale Aldo Moro, 2 - c/o Dipartimento di Fisica, Edificio G. Marconi, 00185 Roma, Italy\label{aff73}
\and
Dipartimento di Fisica e Astronomia "Augusto Righi" - Alma Mater Studiorum Universit\`a di Bologna, via Piero Gobetti 93/2, 40129 Bologna, Italy\label{aff74}
\and
Department of Physics, Institute for Computational Cosmology, Durham University, South Road, Durham, DH1 3LE, UK\label{aff75}
\and
Universit\'e Paris Cit\'e, CNRS, Astroparticule et Cosmologie, 75013 Paris, France\label{aff76}
\and
CNRS-UCB International Research Laboratory, Centre Pierre Bin\'etruy, IRL2007, CPB-IN2P3, Berkeley, USA\label{aff77}
\and
University of Applied Sciences and Arts of Northwestern Switzerland, School of Engineering, 5210 Windisch, Switzerland\label{aff78}
\and
Institute of Physics, Laboratory of Astrophysics, Ecole Polytechnique F\'ed\'erale de Lausanne (EPFL), Observatoire de Sauverny, 1290 Versoix, Switzerland\label{aff79}
\and
Telespazio UK S.L. for European Space Agency (ESA), Camino bajo del Castillo, s/n, Urbanizacion Villafranca del Castillo, Villanueva de la Ca\~nada, 28692 Madrid, Spain\label{aff80}
\and
Institut de F\'{i}sica d'Altes Energies (IFAE), The Barcelona Institute of Science and Technology, Campus UAB, 08193 Bellaterra (Barcelona), Spain\label{aff81}
\and
European Space Agency/ESTEC, Keplerlaan 1, 2201 AZ Noordwijk, The Netherlands\label{aff82}
\and
DARK, Niels Bohr Institute, University of Copenhagen, Jagtvej 155, 2200 Copenhagen, Denmark\label{aff83}
\and
Universit\'e Paris-Saclay, Universit\'e Paris Cit\'e, CEA, CNRS, AIM, 91191, Gif-sur-Yvette, France\label{aff84}
\and
Space Science Data Center, Italian Space Agency, via del Politecnico snc, 00133 Roma, Italy\label{aff85}
\and
Centre National d'Etudes Spatiales -- Centre spatial de Toulouse, 18 avenue Edouard Belin, 31401 Toulouse Cedex 9, France\label{aff86}
\and
Institute of Space Science, Str. Atomistilor, nr. 409 M\u{a}gurele, Ilfov, 077125, Romania\label{aff87}
\and
Instituto de F\'isica Te\'orica UAM-CSIC, Campus de Cantoblanco, 28049 Madrid, Spain\label{aff88}
\and
Institut de Recherche en Astrophysique et Plan\'etologie (IRAP), Universit\'e de Toulouse, CNRS, UPS, CNES, 14 Av. Edouard Belin, 31400 Toulouse, France\label{aff89}
\and
Universit\'e St Joseph; Faculty of Sciences, Beirut, Lebanon\label{aff90}
\and
Departamento de F\'isica, FCFM, Universidad de Chile, Blanco Encalada 2008, Santiago, Chile\label{aff91}
\and
Universit\"at Innsbruck, Institut f\"ur Astro- und Teilchenphysik, Technikerstr. 25/8, 6020 Innsbruck, Austria\label{aff92}
\and
Department of Physics and Helsinki Institute of Physics, Gustaf H\"allstr\"omin katu 2, University of Helsinki, 00014 Helsinki, Finland\label{aff93}
\and
Dipartimento di Fisica e Astronomia "G. Galilei", Universit\`a di Padova, Via Marzolo 8, 35131 Padova, Italy\label{aff94}
\and
INFN-Padova, Via Marzolo 8, 35131 Padova, Italy\label{aff95}
\and
Departamento de F\'isica, Faculdade de Ci\^encias, Universidade de Lisboa, Edif\'icio C8, Campo Grande, PT1749-016 Lisboa, Portugal\label{aff96}
\and
Instituto de Astrof\'isica e Ci\^encias do Espa\c{c}o, Faculdade de Ci\^encias, Universidade de Lisboa, Tapada da Ajuda, 1349-018 Lisboa, Portugal\label{aff97}
\and
Mullard Space Science Laboratory, University College London, Holmbury St Mary, Dorking, Surrey RH5 6NT, UK\label{aff98}
\and
Cosmic Dawn Center (DAWN)\label{aff99}
\and
Niels Bohr Institute, University of Copenhagen, Jagtvej 128, 2200 Copenhagen, Denmark\label{aff100}
\and
Universidad Polit\'ecnica de Cartagena, Departamento de Electr\'onica y Tecnolog\'ia de Computadoras,  Plaza del Hospital 1, 30202 Cartagena, Spain\label{aff101}
\and
European University of Technology EUt+, European Union\label{aff102}
\and
Kapteyn Astronomical Institute, University of Groningen, PO Box 800, 9700 AV Groningen, The Netherlands\label{aff103}
\and
Caltech/IPAC, 1200 E. California Blvd., Pasadena, CA 91125, USA\label{aff104}
\and
Dipartimento di Fisica e Scienze della Terra, Universit\`a degli Studi di Ferrara, Via Giuseppe Saragat 1, 44122 Ferrara, Italy\label{aff105}
\and
Istituto Nazionale di Fisica Nucleare, Sezione di Ferrara, Via Giuseppe Saragat 1, 44122 Ferrara, Italy\label{aff106}
\and
INAF, Istituto di Radioastronomia, Via Piero Gobetti 101, 40129 Bologna, Italy\label{aff107}
\and
Astronomical Observatory of the Autonomous Region of the Aosta Valley (OAVdA), Loc. Lignan 39, I-11020, Nus (Aosta Valley), Italy\label{aff108}
\and
Universit\'e C\^{o}te d'Azur, Observatoire de la C\^{o}te d'Azur, CNRS, Laboratoire Lagrange, Bd de l'Observatoire, CS 34229, 06304 Nice cedex 4, France\label{aff109}
\and
Univ. Grenoble Alpes, CNRS, Grenoble INP, LPSC-IN2P3, 53, Avenue des Martyrs, 38000, Grenoble, France\label{aff110}
\and
Dipartimento di Fisica, Sapienza Universit\`a di Roma, Piazzale Aldo Moro 2, 00185 Roma, Italy\label{aff111}
\and
Aurora Technology for European Space Agency (ESA), Camino bajo del Castillo, s/n, Urbanizacion Villafranca del Castillo, Villanueva de la Ca\~nada, 28692 Madrid, Spain\label{aff112}
\and
Institut d'Astrophysique de Paris, 98bis Boulevard Arago, 75014, Paris, France\label{aff113}
\and
ICL, Junia, Universit\'e Catholique de Lille, LITL, 59000 Lille, France\label{aff114}
\and
CERCA/ISO, Department of Physics, Case Western Reserve University, 10900 Euclid Avenue, Cleveland, OH 44106, USA\label{aff115}
\and
Departamento de F{\'\i}sica Fundamental. Universidad de Salamanca. Plaza de la Merced s/n. 37008 Salamanca, Spain\label{aff116}
\and
IRFU, CEA, Universit\'e Paris-Saclay 91191 Gif-sur-Yvette Cedex, France\label{aff117}
\and
Universit\'e de Strasbourg, CNRS, Observatoire astronomique de Strasbourg, UMR 7550, 67000 Strasbourg, France\label{aff118}
\and
Center for Data-Driven Discovery, Kavli IPMU (WPI), UTIAS, The University of Tokyo, Kashiwa, Chiba 277-8583, Japan\label{aff119}
\and
Jodrell Bank Centre for Astrophysics, Department of Physics and Astronomy, University of Manchester, Oxford Road, Manchester M13 9PL, UK\label{aff120}
\and
California Institute of Technology, 1200 E California Blvd, Pasadena, CA 91125, USA\label{aff121}
\and
Department of Physics \& Astronomy, University of California Irvine, Irvine CA 92697, USA\label{aff122}
\and
Department of Mathematics and Physics E. De Giorgi, University of Salento, Via per Arnesano, CP-I93, 73100, Lecce, Italy\label{aff123}
\and
INFN, Sezione di Lecce, Via per Arnesano, CP-193, 73100, Lecce, Italy\label{aff124}
\and
INAF-Sezione di Lecce, c/o Dipartimento Matematica e Fisica, Via per Arnesano, 73100, Lecce, Italy\label{aff125}
\and
Departamento F\'isica Aplicada, Universidad Polit\'ecnica de Cartagena, Campus Muralla del Mar, 30202 Cartagena, Murcia, Spain\label{aff126}
\and
Instituto de F\'isica de Cantabria, Edificio Juan Jord\'a, Avenida de los Castros, 39005 Santander, Spain\label{aff127}
\and
Department of Physics, Oxford University, Keble Road, Oxford OX1 3RH, UK\label{aff128}
\and
Departament de F\'{\i}sica, Universitat Aut\`onoma de Barcelona, 08193 Bellaterra (Barcelona), Spain\label{aff129}
\and
Instituto de Astronomia Teorica y Experimental (IATE-CONICET), Laprida 854, X5000BGR, C\'ordoba, Argentina\label{aff130}
\and
Department of Computer Science, Aalto University, PO Box 15400, Espoo, FI-00 076, Finland\label{aff131}
\and
Universidad de La Laguna, Dpto. Astrof\'\i sica, E-38206 La Laguna, Tenerife, Spain\label{aff132}
\and
Ruhr University Bochum, Faculty of Physics and Astronomy, Astronomical Institute (AIRUB), German Centre for Cosmological Lensing (GCCL), 44780 Bochum, Germany\label{aff133}
\and
Department of Physics and Astronomy, Vesilinnantie 5, University of Turku, 20014 Turku, Finland\label{aff134}
\and
Finnish Centre for Astronomy with ESO (FINCA), Quantum, Vesilinnantie 5, University of Turku, 20014 Turku, Finland\label{aff135}
\and
Serco for European Space Agency (ESA), Camino bajo del Castillo, s/n, Urbanizacion Villafranca del Castillo, Villanueva de la Ca\~nada, 28692 Madrid, Spain\label{aff136}
\and
ARC Centre of Excellence for Dark Matter Particle Physics, Melbourne, Australia\label{aff137}
\and
Centre for Astrophysics \& Supercomputing, Swinburne University of Technology,  Hawthorn, Victoria 3122, Australia\label{aff138}
\and
Department of Physics and Astronomy, University of the Western Cape, Bellville, Cape Town, 7535, South Africa\label{aff139}
\and
Institut d'Astrophysique de Paris, UMR 7095, CNRS, and Sorbonne Universit\'e, 98 bis boulevard Arago, 75014 Paris, France\label{aff140}
\and
DAMTP, Centre for Mathematical Sciences, Wilberforce Road, Cambridge CB3 0WA, UK\label{aff141}
\and
Kavli Institute for Cosmology Cambridge, Madingley Road, Cambridge, CB3 0HA, UK\label{aff142}
\and
Departement of Theoretical Physics, University of Geneva, Switzerland\label{aff143}
\and
Department of Physics, Centre for Extragalactic Astronomy, Durham University, South Road, Durham, DH1 3LE, UK\label{aff144}
\and
Institute for Theoretical Particle Physics and Cosmology (TTK), RWTH Aachen University, 52056 Aachen, Germany\label{aff145}
\and
Centro de Astrof\'{\i}sica da Universidade do Porto, Rua das Estrelas, 4150-762 Porto, Portugal\label{aff146}
\and
Instituto de Astrof\'isica e Ci\^encias do Espa\c{c}o, Universidade do Porto, CAUP, Rua das Estrelas, PT4150-762 Porto, Portugal\label{aff147}
\and
HE Space for European Space Agency (ESA), Camino bajo del Castillo, s/n, Urbanizacion Villafranca del Castillo, Villanueva de la Ca\~nada, 28692 Madrid, Spain\label{aff148}
\and
Department of Astrophysics, University of Zurich, Winterthurerstrasse 190, 8057 Zurich, Switzerland\label{aff149}
\and
Higgs Centre for Theoretical Physics, School of Physics and Astronomy, The University of Edinburgh, Edinburgh EH9 3FD, UK\label{aff150}
\and
INAF - Osservatorio Astronomico d'Abruzzo, Via Maggini, 64100, Teramo, Italy\label{aff151}
\and
Theoretical astrophysics, Department of Physics and Astronomy, Uppsala University, Box 516, 751 37 Uppsala, Sweden\label{aff152}
\and
Mathematical Institute, University of Leiden, Einsteinweg 55, 2333 CA Leiden, The Netherlands\label{aff153}
\and
Institute of Astronomy, University of Cambridge, Madingley Road, Cambridge CB3 0HA, UK\label{aff154}
\and
Center for Astrophysics and Cosmology, University of Nova Gorica, Nova Gorica, Slovenia\label{aff155}
\and
Institute for Particle Physics and Astrophysics, Dept. of Physics, ETH Zurich, Wolfgang-Pauli-Strasse 27, 8093 Zurich, Switzerland\label{aff156}
\and
Department of Astrophysical Sciences, Peyton Hall, Princeton University, Princeton, NJ 08544, USA\label{aff157}
\and
Fakult\"at f\"ur Physik, Universit\"at Bielefeld, Postfach 100131, 33501 Bielefeld, Germany\label{aff158}
\and
School of Mathematics, Statistics and Physics, Newcastle University, Herschel Building, Newcastle-upon-Tyne, NE1 7RU, UK\label{aff159}
\and
Institut de Physique Th\'eorique, CEA, CNRS, Universit\'e Paris-Saclay 91191 Gif-sur-Yvette Cedex, France\label{aff160}
\and
International Centre for Theoretical Physics (ICTP), Strada Costiera 11, 34151 Trieste, Italy\label{aff161}
\and
Center for Computational Astrophysics, Flatiron Institute, 162 5th Avenue, 10010, New York, NY, USA\label{aff162}}    

%
%
\abstract{
We present the first end-to-end validation of the \Euclid baryon acoustic oscillation (BAO) analysis pipeline, encompassing density-field reconstruction, 2-point correlation function (2PCF) measurement, and cosmological parameter inference. Using eight \Euclid-like mock catalogues extracted from each of the four snapshots of the \flagshipone simulation -- designed to replicate the statistical properties of the first \Euclid data release (DR1) -- we assess the performance of the two standard BAO reconstruction methods based on the Zeldovich approximation, \textsc{RecSym} and \textsc{RecIso}, across four redshift snapshots ($0.9 \leq z \leq 1.8$).
The pipeline introduces several methodological advances: an emulator-based model evaluator (\texttt{Bora.jl}) combined with a Hamiltonian Monte Carlo sampler (\texttt{NUTS}), achieving over 500-fold speed-up relative to standard Monte Carlo Markov chains, and a semi-analytical covariance estimator (\texttt{BeXiCov+WinCov}) that enables robust error estimates with only eight mock realisations, remaining stable under variations in the fiducial cosmology. Together, these components ensure computational efficiency while significantly reducing the risk of underestimating parameter uncertainties.
We find that both reconstruction schemes yield unbiased BAO measurements across all redshift and user-defined choices, including the smoothing scale and fiducial cosmology. In each snapshot, reconstruction enhances the figure of merit for $\{\Omega_{\mathrm{m}}, H_0r_{\mathrm{s}}\}$ by a factor of $\sim3$, equivalent to tripling the effective survey volume. When combining the four redshift bins, the improvement remains substantial, with BAO-only constraints reaching $\sim10\%$ precision on $\Omega_{\mathrm{m}}$ and $\sim3\%$ on $H_0 r_{\mathrm{s}}$. Results from \textsc{RecSym} and \textsc{RecIso} are consistent within uncertainties, though we recommend \textsc{RecSym} during testing due to its reduced sensitivity to covariance variations.
These findings establish the accuracy, robustness, and scalability of the \Euclid BAO pipeline for DR1, providing a solid foundation for future cosmological analyses.
 }
%
%
%
%
   \titlerunning{BAO extraction techniques}
   \authorrunning{Euclid Collaboration: E. Sarpa et al.}
   
   \maketitle
%
%
%
%
\section{Introduction}\label{sc:Intro}

The \bao is among the most robust geometric probe of cosmic expansion \citep{DETF}, with minimal dependence on astrophysical systematics \citep{EP-Risso}. The Sloan Digital Sky Survey (SDSS) Baryon Oscillation Spectroscopic Survey (BOSS) established the methodology, reaching percent-level distance measurements from the clustering of 1.2\,million galaxies \citep{Alam2017}. Stage-IV spectroscopic experiments are now extending this legacy to much larger volumes and higher redshifts. The \Euclid near-infrared spectroscopic survey will obtain redshifts for about $30$\,million H$\alpha$ emission-line galaxies with a redshift uncertainty of $\sigma_z/(1+z)\le10^{-3}$ over a redshift range of $0.9\lesssim z\lesssim1.8$, covering approximately $14\,000\,\deg^{2}$ \citep{Laureijs2011, EuclidSkyOverview}. The ground-based Dark Energy Spectroscopic Instrument (DESI) will map a comparable footprint and measure redshifts for more than $30$\,million galaxies and quasars \citep{Aghamousa2016,DESI_BAO_Y1}. Both are designed to recover the \bao scale with sub-percent precision, enabling tight constraints on the matter density $\Omega_\mathrm{m}$ and on the product $H_0 r_\mathrm{s}$, linking the present-day Hubble constant $H_0$ to the sound horizon at the baryon-drag epoch $r_\mathrm{s}$.

At this level of precision, validated \Euclid\ forecasts based on the joint analysis of spectroscopic galaxy clustering, photometric clustering, and weak lensing predict uncertainties of $\sigma(w_0) \simeq 0.025$--$0.040$ as well as $\sigma(w_a) \simeq 0.092$--$0.170$ on the dark energy equation-of-state parameters \citep[][Table~11]{Blanchard-EP7}. These ranges reflect differences between optimistic and pessimistic assumptions. The forecasts assume a flat $w_0w_a$CDM cosmology -- an extension of the standard $\Lambda$CDM model, which includes \cdm\ and a cosmological constant $\Lambda$ -- in which the dark energy equation of state is parametrised as $w(a) = w_0 + (1 - a)w_a$, 
enabling stringent tests of $\Lambda$CDM and strong constraints on deviations from a cosmological constant.

These forecasts are obtained under the baseline assumptions adopted for the \Euclid spectroscopic clustering analysis, in which the clustering signal is modelled through unreconstructed two-point statistics rather than through a reconstructed \bao\ analysis.  By estimating the large-scale displacement field and partially reversing nonlinear growth, reconstruction sharpens the acoustic peak and restores part of the linear information. Since spectroscopic clustering constrains cosmic distances largely through the \bao\ feature, quantifying the information recovered by reconstruction is essential for assessing its final contribution to dark-energy constraints of \Euclid. In the SDSS Data Release 7 Luminous Red Galaxy (LRG) sample \citep{Padmanabhan2012}, reconstruction reduced the relative distance error from $3.5\,\%$ to $1.9\,\%$, effectively tripling the survey volume. This has made reconstruction a key ingredient in spectroscopic Stage-IV surveys for constraining cosmological parameters, as highlighted by recent results from DESI \citep{DESI_BAO_Y1,DESI_BAO_DR2}. However, its performance depends sensitively on analysis choices such as the smoothing scale used to infer displacements, the grid resolution used for density interpolation, and the covariance model adopted for clustering fits \citep[e.g.,][]{Burden2014,Vargas-Maga2017Smth,Paillas2025,Chen2024}. Moreover, different reconstruction implementations can yield subtly different outcomes, motivating a systematic comparison of their efficiency and stability.

A common route to optimise reconstruction and modelling is to calibrate analysis choices on large suites of realistic mock catalogues that reproduce survey purity, completeness, and selection function. However, executing the full measurement-to-inference pipeline on thousands of realisations is computationally costly -- particularly in early survey phases when observational characteristics are still evolving and multiple configurations must be tested.

In this work, we present the first run of the \Euclid \bao reconstruction pipeline and its validation on \Euclid-like nonlinear mock catalogues, providing recommendations on input settings and methods to be used for the \Euclid data analysis. The pipeline unifies reconstruction, configuration-space estimation, analytical and semi-analytical covariance modelling, and \bao fitting. We quantify and compare the performance of the two reconstruction schemes, \textsc{RecSym} and \textsc{RecIso} \citep{Chen2019}, assessing their impact on \bao precision and robustness across redshift bins. Guided by the post-reconstruction clustering model, we derive theory-informed choices for reconstruction and covariance settings and verify them against simulations. This theory-data methodology reduces the reliance on extensive mock suites and will accelerate iterative validation in the first stages of the actual data analysis, while keeping the focus on scientifically optimal configurations. 

The paper is organised as follows. In Sect.~\ref{sec:catalogues}, we describe the \Euclid spectroscopic dataset alongside the suite of mock catalogues used for validation. Section~\ref{sec: Recon_methods} introduces the theoretical framework and numerical implementation of the reconstruction algorithms. In Sect.~\ref{sec:measurements}, we present the configuration-space estimators and the \bao fitting methodology. Section~\ref{sec:Likelihood} outlines the likelihood analysis and the semi-analytical covariance modelling. The performance of the pipeline is validated on mock data in Sect.~\ref{sec:results}, where we also assess its robustness to variations in smoothing scale and fiducial cosmology, and compare the outputs of \textsc{RecSym} and \textsc{RecIso}. This section concludes with the resulting cosmological constraints and a precision forecast for \Euclid \DRone. We summarise and discuss our main findings in Sect.~\ref{sec:Discussion_Conclusions}.

\section{Data}\label{sec:catalogues}
\begin{table}
\centering
\caption{Redshift, volume, number density, and expected linear galaxy bias of the subboxes used in the analysis.}
\renewcommand{\arraystretch}{1.2}
    \begin{tabular}{|cccc|}
    \hline
    \rowcolor{blue!5}
      & & & \\[-12pt]   
    \rowcolor{blue!5}
    $z_\mathrm{snap}$ & $V~[\!\cGpc]$ & $10^{4}\,\bar{n}\,[\!\kcMpc]$&  $b^\mathrm{f}$ \\
    \hline
    0.9 & 1.33 & 20 & 1.32 \\
    1.2 & 1.52 & 10 & 1.64 \\
    1.5 & 1.69 & 6 & 1.95 \\
    1.8 & 2.69 & 3 & 2.46 \\
    \hline
    \end{tabular}
\label{tab:snapshots}
\end{table}

For our analysis, we use four sets of mock galaxy catalogues constructed from distinct comoving snapshots of the \Euclid \flagshipone simulation. We choose \flagshipone over faster, approximate mocks, such as EZmocks \citep{EZmocks}, or Euclid Large Mocks \citep[hereafter ELM,][]{EP-Monaco1}, to ensure an accurate representation of late-time nonlinear matter clustering, which is crucial for assessing the impact of \bao reconstruction. \flagshipone is a full $N$-body simulation evolved with the \textsc{PKDGRAV3} gravity solver \citep{Potter2017}, following over two trillion dark matter particles in a periodic cube of side $L = 3780\,\hMpc$ from $z = 99$ to $z = 0$. Haloes are identified using a friends-of-friends algorithm (minimum of ten particles), achieving a mass resolution of a few 
$10^{10}\,h^{-1}\si{\solarmass}$. Galaxies are then populated via a \hod model calibrated to reproduce the abundance and clustering of \Euclid H$\alpha$ emission-line galaxies (see \citealt{EP-Pezzotta} or \citealt{EuclidSkyFlagship}, for further details).

The four simulation snapshots are centred at redshifts $z_\mathrm{snap} \in \{ 0.9, 1.2, 1.5,  1.8\}$, chosen to represent the mean redshift of the tomographic bins adopted for the \Euclid \DRone spectroscopic analysis \citep{Blanchard-EP7}. 
From each snapshot, we extract \DRone-like sub-boxes, each with a comoving volume matched to the relevant redshift bin and separated by at least $300\,\hMpc$ to ensure statistical independence \citep[see, e.g.,][]{Sarpa2019}. This carving procedure breaks the periodicity of the parent \flagshipone snapshots and defines finite-volume mock catalogues with a cubic survey window. Together, these requirements limit the number of subcatalogues that can be extracted from each snapshot to a maximum of eight.
To match the expected galaxy number density in \Euclid \DRone, the mock catalogues are randomly down-sampled to achieve the target 43\% completeness level at each redshift. The resulting volumes and number densities are summarised in Table~\ref{tab:snapshots}\footnote{We note that \flagshipone\ adopts a simplified description of the Universe, neglecting massive neutrinos and evolving dark energy. Although idealised, this is sufficient for the purpose of the present work, namely to validate the BAO pipeline in a controlled setting, compare reconstruction conventions, assess the impact of analysis choices, and establish the main methodological ingredients for the forthcoming \DRone\ analysis, whose BAO constraints remain largely robust to these assumptions \citep{ZANeutrino,Carter_2020}. A more realistic assessment of the systematic error budget, including massive neutrinos and observational systematics, will be addressed in the dedicated \DRone\ preparation studies.}. All catalogues adopt the \Euclid \flagshipone fiducial flat $\Lambda$CDM cosmology as detailed in Table~\ref{tab:cosmology}.
We slightly adjusted $n_\mathrm{s}$ from its nominal simulation value following \citet{EP-Pezzotta}. The galaxy populations are modelled as biased tracers of the matter density field, with fiducial linear bias values listed in Table~\ref{tab:snapshots}; these are estimated from full-shape fits to the real-space power spectrum measured from the full \flagshipone\ snapshots at each redshift.
\begin{table}[]
    \centering
    \caption{Fiducial cosmology of the \flagshipone snapshots.}
    \renewcommand{\arraystretch}{1.2}
    \begin{tabular}{|c|c|c|c|c|c|c|}
        \hline
        \rowcolor{blue!5}
        $\Omega_{\mathrm{m}}$ & $\Omega_{\mathrm{b}}$ & $n_{\mathrm{s}}$ & $h$ & $\sigma_8$ & $10^9\,A_{\mathrm{s}}$ & $\sum m_{\nu} \, [\mathrm{eV}]$ \\
        \hline
        0.319 & 0.049 & 0.97 & 0.67 & 0.83 & 2.09 & 0 \\
        \hline
    \end{tabular}
    \tablefoot{In order, the table presents the total matter density $\Omega_{\mathrm{m}}$, the baryon density $\Omega_{\mathrm{b}}$, the spectral index of the primordial power spectrum $n_{\mathrm{s}}$, the dimensionless Hubble parameter $h$, the variance of density perturbations in spheres of $8\Mpc$ radius $\sigma_8$, the amplitude of the primordial power spectrum $A_{\mathrm{s}}$, and the sum of neutrinos masses $\sum m_{\nu}$.}
    \label{tab:cosmology}
\end{table}

%

\subsection{Redshift-space catalogues}\label{sec:redshift_cat}

We construct the redshift-space catalogues under the distant-observer approximation, fixing the \los along one Cartesian axis. The redshift-space transformation is applied to the full \flagshipone snapshot before extracting the independent sub-boxes, ensuring that large-scale velocity correlations are consistently preserved. Each box is then shifted so that its centre lies at the comoving distance $\chi$ corresponding to its snapshot redshift, given by
\begin{equation}\label{eq:red-to-dist}
\chi^\mathrm{f}(z) = \int_0^{z} \frac{c\,\diff z'}{H^\mathrm{f}(z')} \,,
\end{equation}
where $H^\mathrm{f}(z)$ denotes the Hubble parameter, $c$ is the vacuum speed of light, and the superscript ‘f’ indicates evaluation in the fiducial cosmology.
The cosmological redshift $z_\mathrm{c}$ of each galaxy is obtained by inverting the above relation for its real-space \los coordinate, while the observed redshift includes the Doppler contribution from the \los peculiar velocity $v^\parallel$,
\begin{equation}
z_\mathrm{obs} = \left(1+z_\mathrm{c}\right)\left(1+v^\parallel/c\right)-1 \,.
\end{equation}
Galaxies are placed at a comoving distance $\chi^\mathrm{f}(z_\mathrm{obs})$ along the \los, yielding the final redshift-space positions. Unless otherwise stated, all steps are performed consistently within the fiducial cosmology. In Sect.~\ref{sec:fiducial_cosmo}, we revisit this procedure to quantify the impact of assuming an incorrect fiducial cosmology on the \bao analysis.

\subsection{Random catalogues}\label{sec:random_cat}
Each galaxy catalogue is accompanied by a random catalogue with identical geometry and containing fifty times more objects to suppress shot noise. As no observational effects or selection functions are applied to the mocks, the points in the random catalogues are distributed homogeneously within the cubic volumes. These catalogues are used throughout the analysis as inputs to both the reconstruction algorithms and the \tpcf estimator, which quantifies the galaxy clustering signal.

\section{Reconstruction algorithm}\label{sec: Recon_methods}

To enhance constraints on the \bao scale, we apply a density-field reconstruction based on the Zeldovich approximation (ZA; \citealt{Eisenstein2007,Padmanabhan2009}), shifting the positions of galaxies and random tracers to recover a configuration with a sharpened acoustic peak. We consider two widely used implementations -- \textsc{RecSym} and \textsc{RecIso} -- which differ in how \rsd are modelled for the random catalogue. This section introduces the ZA formalism, examines how key analysis choices shape the reconstructed clustering signal, and motivates the theoretical models used for \bao fitting and covariance estimation in our \Euclid{} analysis.

We begin in Sect.~\ref{sec:ZA_nonlinear} by formulating the ZA in real-space, including the definition of the smoothed displacement estimator. Section~\ref{sec:ZA_redshift} extends this framework to redshift space, central for observational applications, and outlines the \textsc{RecSym} and \textsc{RecIso} approaches. The practical implementation adopted in this work, based on the publicly available \textsc{MultiGridReconstruction}\footnote{\url{https://github.com/cosmodesi/pyrecon}} algorithm, is detailed in Sect.~\ref{sec:ZA_multigrid}.

\subsection{Real space: Mitigating nonlinearities}\label{sec:ZA_nonlinear}

In the Lagrangian description of structure formation \citep{Buchert1989,Moutarde1991,Hivon1995}, the nonlinear observed galaxy overdensity field, $\delta^\mathrm{g}_\mathrm{obs}(\vec{x})=\delta^\mathrm{g}(\vec{x},t_\mathrm{obs})$, arises from the displacement of the initial underlying DM perturbations, $\delta^\mathrm{DM}(\vec{x},t=0)$, from their Lagrangian coordinates $\vec{q}$ to the Eulerian positions $\vec{x}(t_\mathrm{obs})$. In the linear regime, the matter overdensity field evolves proportionally to the linear growth factor such that
\(
\delta^{\mathrm{DM}}(\vec{x},t) = \delta^{\mathrm{DM}}_{\mathrm{lin}}(\vec{x},t_0)\,D(t)/D(t_0)
\),
where $D(t_0)$ is the growth factor at a reference epoch and $\delta^{\mathrm{DM}}_{\mathrm{lin}}(\vec{x},t_0)$ denotes the linear, Eulerian dark matter overdensity field at that time. In this regime, each Fourier mode of the density field evolves independently, preserving the shape of the initial power spectrum. As gravitational clustering proceeds, mode coupling amplifies density fluctuations and drives $\delta^\mathrm{DM}$ beyond unity, signalling the breakdown of linear theory.

At first order, the displacement field $\vec{\psi}_\mathrm{obs}$ is described by the Zeldovich approximation \citep{Zel'dovich1970} as
\begin{equation}\label{eq:ZA_disp}
\vec{\hat{\psi}}_\mathrm{obs}(\vec{k}) = \mathrm{i}\,\frac{\vec{k}}{k^2}\,\hat{\delta}_\mathrm{lin}^\mathrm{DM}(\vec{k},t_\mathrm{obs}) \, ,
\end{equation}
where hats denote Fourier transforms and $k$ is the comoving wavenumber. Assuming a linear, scale-independent bias $b$, the galaxy and matter overdensities are related by $\delta^\mathrm{g}_\mathrm{obs}=b\,\delta^\mathrm{DM}_\mathrm{obs}$, yielding
\begin{equation}\label{eq:ZA_pk}
P^\mathrm{g}(k,t_\mathrm{obs}) = b^2\,\mathrm{e}^{-k^2\Sigma^2/2} \, \Plin(k,t_\mathrm{obs}) \, ,
\end{equation}
where $\Plin$ is the linear-theory matter power spectrum and
\begin{equation}\label{eq:ZA_sigma}
\Sigma^2 = \frac{1}{3\pi^2}\int_0^{\infty} \mathrm{d}p\,\Plin(p,t_\mathrm{obs})
\end{equation}
is the variance of large-scale displacements. The exponential factor in Eq.~\eqref{eq:ZA_pk} describes \bao damping by nonlinear bulk flows \citep{Matsubara2008a,Padmanabhan2009}, which hinders \bao detection and degrades the accuracy of the inferred characteristic scale.

Zeldovich reconstruction mitigates this damping by estimating the large-scale displacement $\vec{\psi}_\mathrm{ZA}$ and removing it from the data. Because the linear density field in Eq.~\eqref{eq:ZA_disp} is unobservable, we approximate it by low-pass filtering the observed galaxy overdensity and assuming $\mathcal{S}\hat{\delta}^{\mathrm{g}}_{\rm obs}\simeq b\,\mathcal{S}\hat{\delta}_{\rm lin}$, where $\mathcal{S}$ is a Gaussian smoothing filter
\begin{equation}
\mathcal{S}(k;R_\mathrm{s})=\mathrm{exp}\left(-k^2 R_\mathrm{s}^2/2\right) \, ,
\end{equation}
of width $R_\mathrm{s}$. To ensure this approximation holds, $R_\mathrm{s}$ must exceed $\Sigma$ from Eq.~\eqref{eq:ZA_sigma}. The reconstructed displacement is then
\begin{equation}\label{eq:ZA_rec_disp}
    \hat{\vec{\psi}}_{\rm ZA}(\vec{k}) = \mathrm{i}\,\frac{\vec{k}}{k^2}\,\frac{\mathcal{S}(k)\,\hat{\delta}^{\mathrm{g}}_{\rm obs}(\vec{k},t_{\rm obs})}{b} = \mathcal{S}(k)\,\hat{\vec{\psi}}_{\rm obs}(\vec{k}) \, ,
\end{equation}
and galaxies are shifted by $-\vec{\psi}_{\rm ZA}$ to obtain their reconstructed positions $\vec{x}_{\rm d}$. Combining this with the non--linear evolution, the reconstructed overdensity $\delta_{\rm d}$ corresponds to the mapping $\vec{q}\mapsto \vec{x}_{\rm d}$ with residual displacement $\vec{\hat{\psi}}_{\rm rec}=(1-\mathcal{S})\,\vec{\hat{\psi}}_{\rm obs}$.

Following the same reasoning as for $P^\mathrm{g}$ (see Appendix~\ref{app:recon_real}), the power spectrum of shifted galaxies is
\begin{equation}\label{eq:Pk_dd}
    P_\mathrm{dd}(k) = \mathrm{exp}\left(-k^2\Sigma_\mathrm{dd}^2/2\right) \,\left[b-\mathcal{S}(k)\right]^2 \, \Plin(k, t_\mathrm{obs}) \, ,
\end{equation}
with
\begin{equation}\label{eq:sigma_dd}
    \Sigma_\mathrm{dd}^2 = \frac{1}{3\pi^2} \int_0^{\infty} \mathrm{d}p \, \left[1-\mathcal{S}(p)\right]^2 \, \Plin(p, t_\mathrm{obs}) \, ,
\end{equation}
the residual displacement variance. For typical galaxy bias values $b\in[1,3]$, the term $1-\mathcal{S}(k)$ suppresses most of the clustering power of $P^\mathrm{g}_\mathrm{obs}$, partially erasing the \bao information in $P_\mathrm{dd}$. To restore these modes, ZA reconstruction generates a new clustered field, $\delta_\mathrm{s}$ by displacing a uniform distribution -- with the same selection function as the data -- by $-\vec{\psi}_\mathrm{ZA}$. The power spectrum associated to $\delta_\mathrm{s}$ is
\begin{equation}
    P_\mathrm{ss}(k) = \mathrm{exp}\left(-k^2\Sigma_\mathrm{ss}^2/2\right) \, \mathcal{S}^2(k) \,\Plin(k,t_\mathrm{obs}) \, ,
\end{equation}
with
\begin{equation}\label{eq:sigma_ss}
    \Sigma_\mathrm{ss}^2 = \frac{1}{3\pi^2}\int_0^{\infty} \mathrm{d}p\, \mathcal{S}^2(p) \, \Plin(p,t_\mathrm{obs}) \, .
\end{equation}
Combining the two, the reconstructed field $\delta_\mathrm{ZA}=\delta_\mathrm{d}-\delta_\mathrm{s}$ yields the power spectrum
\begin{equation}
    P_\mathrm{ZA}(k)=P_\mathrm{dd}(k)+P_\mathrm{ss}(k)-2P_\mathrm{ds}(k) \, ,
\end{equation}
where the cross term is given by
\begin{equation}\label{eq:Pk_ZA_cross}
    P_\mathrm{ds}(k)=-\mathcal{S}(k) \, [b-\mathcal{S}(k)] \, \mathrm{exp}\left(-k^2\Sigma_\mathrm{ds}^2/2\right) \, \Plin(k,t_\mathrm{obs}) \, ,
\end{equation}
and
\begin{equation}\label{eq:sigma_ds}
    \Sigma_\mathrm{ds}^2=\frac{1}{2}\left(\Sigma^2_\mathrm{dd}+\Sigma^2_\mathrm{ss}\right) \, .
\end{equation}
In this approximation, the reconstructed spectrum becomes
\begin{equation}\label{eq:Pk_ZA_rec}
    P_\mathrm{ZA}(k)= \mathcal{D}_\mathrm{ZA}(k) \, \Plin(k,t_\mathrm{obs}) \, ,
\end{equation}
where
\begin{equation}\label{eq:Dk_ZA}
\begin{split}
    \mathcal{D}_\mathrm{ZA}(k) = & \left[b-\mathcal{S}(k)\right]^2 \, \mathrm{exp}\left(-k^2\Sigma_\mathrm{dd}^2/2\right)  + \mathcal{S}^2(k)\,\mathrm{exp}\left(-k^2\Sigma_\mathrm{ss}^2/2\right) \\
    & + 2\, \mathcal{S}(k) \, \left[b-\mathcal{S}(k)\right] \, \mathrm{exp}\left(-k^2\Sigma_\mathrm{ds}^2/2\right)
\end{split}
\end{equation}
is the reconstructed damping function.

Because the ZA algorithm requires the inclusion of shifted random catalogues, the reconstructed field cannot be regarded as a `back-in-time' representation of the observed distribution. It nevertheless provides a linearised description of the clustering signal at the observed epoch, recovering a substantial fraction of the primordial \bao information. The efficiency of this process depends on the adopted smoothing filter $\mathcal{S}$; excessive smoothing erases cosmological information, whereas insufficient smoothing leaves residual nonlinear noise in the reconstructed field.

\begin{figure}
\centering
\includegraphics[width=\columnwidth]{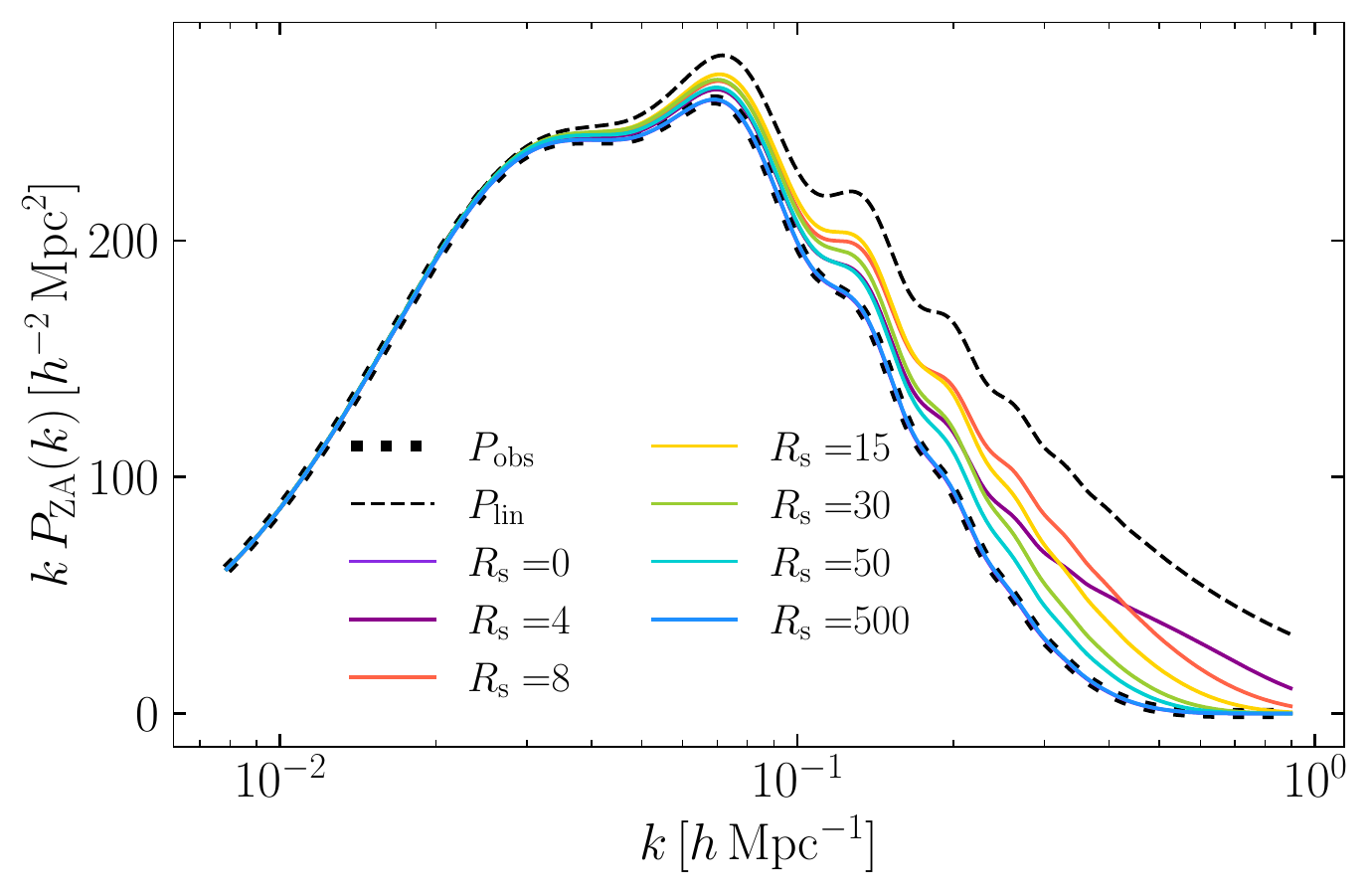}
\caption{Reconstructed power spectrum predicted by standard Zeldovich reconstruction, with coloured lines corresponding to different smoothing kernel sizes (in units of $\!\Mpc$). Black dashed and dotted lines denote the linear and nonlinear power spectra at the redshift of observation, respectively.}
\label{fig:ZA_smoothing_th}
\end{figure}


Figure~\ref{fig:ZA_smoothing_th} demonstrates the impact of the smoothing scale \( R_\mathrm{s} \) on the reconstructed power spectrum. As \( R_\mathrm{s} \) varies, the relative amplitude of the reconstruction transfer function \( \mathcal{D}_\mathrm{ZA}(k) \) changes, affecting the recovery of \bao features. An optimal value, \( R_{s,\mathrm{opt}} = 8.4\,\hMpc \), emerges that best approaches the linear prediction by balancing noise suppression with minimal loss of physical information.

However, Eq.~\eqref{eq:Dk_ZA} shows that the optimal reconstructed spectrum inherits a more complex scale dependence than the pre-reconstruction case (hereafter \textsc{PreRec}), with three independent damping terms required to model $P_\mathrm{ZA}$. A second characteristic scale, $R_{\mathrm{s,equiv}} = 17.5\,\hMpc$, marks the configuration where $\Sigma_\mathrm{dd}=\Sigma_\mathrm{ss}=\Sigma_\mathrm{ds}$, yielding a single effective damping parameter $\Sigma_\mathrm{eq}$ and the simplified form
\begin{equation}
P_\mathrm{ZA}(k,t_\mathrm{obs}) = \mathrm{exp}\left(-k^2\Sigma_\mathrm{eq}^2\right) \, \Plin(k,t_\mathrm{obs}) \, .
\end{equation}
While $R_{\mathrm{s,opt}}$ ensures the strongest \bao recovery, $R_{\mathrm{s,equiv}}$ simplifies the theoretical modelling by reducing the damping description to a single parameter. The two scales do not coincide, emphasising the trade-off between reconstruction performance and model tractability, which must be calibrated to achieve optimal \bao recovery.

If validated, these theoretical prescriptions provide direct means to select optimal reconstruction parameters and construct robust \bao models without extensive mock-based tuning. They are tested in Sect.~\ref{sec:ZA_sys_smoothing} using \Euclid \DRone mocks over $R_\mathrm{s} \in [R_{\mathrm{s, opt}}, R_{s,\mathrm{equiv}}]$ to quantify the residual damping and validate the modelling assumptions.

\subsection{Redshift space: \textsc{RecSym} and \textsc{RecIso}} \label{sec:ZA_redshift}

In redshift surveys, galaxies are observed at their apparent positions
\(\vec{s}\), displaced along the \los with respect to their
real-space coordinates \(\vec{x}\) by the effect of peculiar velocities,
\begin{equation}
    \vec{s} = \vec{x} + \int_{z_\mathrm{c}}^{z_\mathrm{obs}} \frac{c \, \mathrm{d}z'}{H^\mathrm{f}(z')}
    \simeq \vec{x} + \frac{(1+z_\mathrm{c})}{H^\mathrm{f}(z_\mathrm{c})} \vec{v}^\parallel \, .
\end{equation}
When modelling the nonlinear motions of galaxies, the Zeldovich reconstruction must therefore account for this additional \los displacement, so
that the observed redshift-space displacement field reads
\begin{equation}\label{eq:ZA_disp_rsd}
\vec{\psi}_\mathrm{obs}^{\,s}
= \vec{\psi}_\mathrm{obs}
  + \frac{(1+z_\mathrm{c}) \, \vec{v}_\parallel}{H(z_\mathrm{c})} \, .
\end{equation}
In first-order Lagrangian perturbation theory, the peculiar velocity field follows from the time derivative of the real-space displacement
(Eq.~\ref{eq:ZA_disp}), yielding
\begin{equation}\label{eq:ZA_vel}
\hat{\vec{v}}(\vec{k},t_\mathrm{obs})
= \mathrm{i} \, \frac{f(t)H(t)\big|_{t_\mathrm{obs}}}{1+z_\mathrm{c}}
  \, \frac{\vec{k}}{k^2} \, \hat{\delta}_\mathrm{lin}(\vec{k})
= \frac{f(t)H(t)}{1+z_\mathrm{c}} \, 
  \hat{\vec{\psi}}_\mathrm{obs}(\vec{k}) \, ,
\end{equation}
where \(f=\mathrm{d}\ln D/\mathrm{d}\ln a\) is the linear growth rate, $a$ the scale factor, and
\(\mu=\vec{k}\cdot\vec{z}/(k\,z)\) denotes the cosine of the angle to the \los $\vec{z}$.
Substituting Eq.~\eqref{eq:ZA_vel} into Eq.~\eqref{eq:ZA_disp_rsd} gives the
redshift-space displacement field,
\begin{equation}
\hat{\vec{\psi}}_\mathrm{obs}^{\,s}(\vec{k})
= \left(1+f\mu^2\right)\,\hat{\vec{\psi}}_\mathrm{obs}(\vec{k}) \, ,
\end{equation}
which links the redshift-space and real-space displacements through the linear
\rsd factor \((1+f\mu^2)\).  
This relation leads to the anisotropic power spectrum \citep{Matsubara2008a,Chen2019,Chudaykin2020}
\begin{equation}\label{eq:Pk_ZA_rsd}
\begin{split}
P_\mathrm{obs}^{s}(k,\mu)= & \mathrm{exp}\left\{-k^2\left[(1-\mu^2)\Sigma^2+(1+f)^2\mu^2\Sigma^2\right]/2\right\} \\
  & \times \left(b+f\mu^2\right)^2 \, \Plin(k)  \, ,
\end{split}
\end{equation}
where the \(\mu\)-dependence models both the Kaiser amplification and the anisotropic damping of the \bao feature. Accordingly, the redshift-space overdensity can be written as \(\delta_\mathrm{obs}^{\,s}=\left(1+f\mu^2\right)\,\delta_\mathrm{obs}^{\mathrm{g}}\).

When applied in redshift space, the Zeldovich reconstruction shifts the observed galaxy positions by \(-\vec{\psi}_\mathrm{ZA}^{\,s}\),
where the reconstructed displacement is derived from \(\delta_\mathrm{obs}^{\,s}\) as \citep{Chen2019}
\begin{equation}\label{eq:ZA_rec_disp_rsd}
    \hat{\vec{\psi}}_\mathrm{ZA}^{s}(\vec{k})= \mathrm{i}\,\frac{\vec{k}}{k^2} \, \frac{\mathcal{S}(k)\,\hat{\delta}_\mathrm{obs}^{\,s}(\vec{k})}{b} = \mathcal{S}(k)\,\hat{\vec{\psi}}_\mathrm{obs}^{s}(\vec{k}) \, .
\end{equation}
The reconstructed density field can thus be regarded as generated by the residual displacement
\begin{equation}\label{eq:data_disp_rec}
    \hat{\vec{\psi}}_\mathrm{rec}^{\,s} = \left(1-\mathcal{S}\right) \, \hat{\vec{\psi}}_\mathrm{obs}^{s} = \left(1-\mathcal{S}\right) \, \left(1+f\mu^2\right) \, \hat{\vec{\psi}}_\mathrm{obs} \, ,
\end{equation}
where we have omitted the $\vec{k}$ dependence for brevity.
To restore the clustering modes suppressed by the smoothing filter, we adopt
two standard approaches for shifting the random catalogue, \textsc{RecSym} and \textsc{RecIso}.  
\textsc{RecSym} follows the real-space convention by shifting the random catalogue by \(-\hat{\vec{\psi}}_\mathrm{ZA}^{\,s}\), consistently with the data.
The resulting clustered field \(\delta_{\mathrm{s},\mathrm{sym}}^{\,s}\), where the sub-script $\mathrm{s}$ refers to `shifted random', is then generated by
\begin{equation}\label{eq:rand_dis_sym}
\hat{\vec{\psi}}_\mathrm{sym}^{s}
 = \mathcal{S} \, \hat{\vec{\psi}}_\mathrm{obs}^{s}
 = \mathcal{S} \, \left(1+f\mu^2\right) \, \hat{\vec{\psi}}_\mathrm{obs} \, .
\end{equation}
In contrast, \textsc{RecIso} neglects the \los velocity contribution in
\(\hat{\vec{\psi}}_\mathrm{ZA}^{\,s}\) and shifts the random catalogue by
the real-space displacement, Eq.~\eqref{eq:ZA_disp}, producing the overdensity
field \(\delta_{\mathrm{s},\mathrm{iso}}^{\,s}\) associated with
\begin{equation}\label{eq:rand_dis_iso}
\hat{\vec{\psi}}_\mathrm{iso}^{s}
 = \mathcal{S} \, \hat{\vec{\psi}}_\mathrm{obs} \, .
\end{equation}

For both \textsc{RecSym} and \textsc{RecIso}, the reconstructed density field is
defined as
\(\delta_\mathrm{rec}^{\,s}=\delta_\mathrm{d}^{\,s}
 - \delta_{\mathrm{s},\mathrm{(sym/iso)}}^{s}\),
leading to the reconstructed power spectra (see
Appendix~\ref{app:recon_redshift})
\begin{equation}\label{eq:Pk_symiso}
P^s_{\mathrm{sym/iso}}(k,\mu)
 = \mathcal{D}_{\mathrm{sym/iso}}(k,\mu) \, \Plin(k,t_\mathrm{obs}) \, ,
\end{equation}
with damping functions
\begin{equation}\label{eq:DA_sym}
\begin{split}
    \mathcal{D}_\mathrm{sym}(k, \mu) =& \left[(b-1)+(1-\mathcal{S}) \, \left(1+f\mu^2\right)\right]^2 \, \mathcal{E}_{\mathrm{dd}} \\
    &+ 2\,\mathcal{S}\,(1+f\mu^2) \, \left[(b-1)+(1-\mathcal{S}) \, \left(1+f\mu^2\right)\right]   \\
    &\times \left[\mathcal{E}_{\mathrm{dd}}\,\mathcal{E}_{\mathrm{ss}}\right]^{1/2} + \mathcal{S}^2\,(1+f\mu^2)^2 \, \mathcal{E}_{\mathrm{ss}} 
\end{split}
\end{equation}
and
\begin{equation}\label{eq:DA_iso}
\begin{split}
    \mathcal{D}_\mathrm{iso}(k, \mu) =& \left[(b-1)+(1-\mathcal{S}) \, \left(1+f\mu^2\right)\right]^2 \, \mathcal{E}_{\mathrm{dd}} \\
    &+ 2\mathcal{S} \, \left[(b-1)+(1-\mathcal{S}) \, \left(1+f\mu^2\right)\right] \\
    & \times \left[ \mathcal{E}_{\mathrm{dd}}\,\mathcal{E}_{\mathrm{ss}}^{f=0} \right]^{1/2} + \mathcal{S}^2 \, \mathrm{exp}\left(-k^2 \Sigma_\mathrm{ss}^2 / 2\right)  \, , 
\end{split}
\end{equation}
where
\begin{equation}
\mathcal{E}_i(k,\mu)
 = \mathrm{exp}\left\{-k^2\,
   \left[\left(1-\mu^2\right) \, \Sigma_i^2 + (1+f)^2 \, \mu^2 \, \Sigma_i^2\right]/2\right\}
\end{equation}
is the anisotropic exponential damping associated with the displacement variance \(\Sigma_i\), defined in Sect.~\ref{sec:ZA_nonlinear}. We denoted with $f=0$ in the superscript the damping evaluated with the growth rate set to zero. In the expressions involving $\mathcal{E}$ we omitted the explicit dependency on $k$ and $\mu$ for better readability.

To characterise the behaviour of the reconstructed power spectrum, we consider the case \(R_\mathrm{s}=R_\mathrm{s,eq}\). In this limit, the damping functions simplify to
\begin{equation}
    \mathcal{D}_\mathrm{sym}(k,\mu;R_\mathrm{s,eq}) =\left(b+f\mu^2\right)^2 \, \mathcal{E}_{\mathrm{eq}} \, ,
\end{equation}
and
\begin{equation}\label{eq:D_iso}
\begin{split}
    \mathcal{D}_\mathrm{iso}(k,\mu;R_\mathrm{s,eq}) =&
    \left\{ \left[(b-1)+(1-\mathcal{S})\, \left(1+f\mu^2\right)\right]
    \right. \\
    & \times \left. \mathcal{E}_{\mathrm{eq}}^{1/2} +\mathcal{S}(k)\,\mathrm{exp}\left(-k^2\Sigma_\mathrm{eq}^2/4\right)\right\}^2 \, . 
\end{split}
\end{equation}

From these expressions, several key features emerge.  
First, the \textsc{RecSym} implementation preserves the overall shape of the redshift-space clustering signal. In particular, the linear Kaiser term \(\left(b+f\mu^2\right)^2\) remains
unchanged, while the damping of the \bao oscillations is reduced through
\(\Sigma_\mathrm{eq}<\Sigma\).  
Second, \textsc{RecIso} suppresses the \los velocity contribution in
the random catalogue, thereby approaching the real-space clustering pattern on
scales dominated by \((1-\mathcal{S})\).  
Both algorithms yield a similar overall improvement in the recovered \bao
contrast -- roughly \(\Sigma_\mathrm{eq}/\Sigma\) -- but their physical content
differs.  
By removing the \rsd component, \textsc{RecIso} produces a signal that is more
sensitive to nonlinear small-scale motions, which are harder to model and
cannot be described by a single damping term.  
In contrast, \textsc{RecSym} retains the anisotropic structure of the linear
redshift-space signal while simplifying its modelling through a single, well-defined damping scale.  
These distinctions motivate our comparison of the two reconstruction schemes in
Sect.~\ref{sec:ZA_sys_smoothing}.

\subsection{\textsc{MultiGridReconstruction}}
\label{sec:ZA_multigrid}
\begin{figure}
\centering
\includegraphics[width=1.\columnwidth]{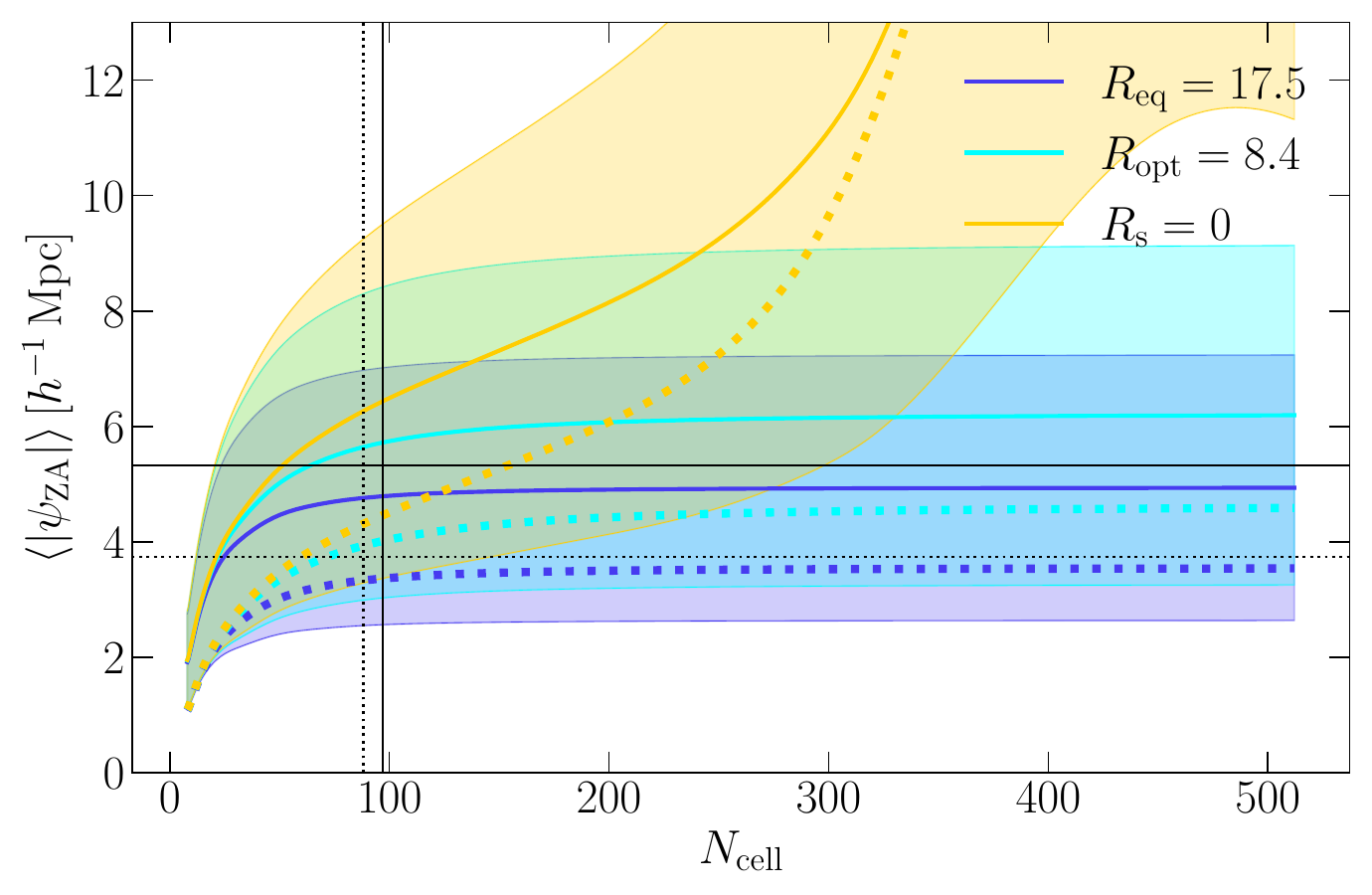}
\caption{Mean reconstructed galaxy-displacement amplitude as a function of the mesh resolution $N_\mathrm{cell}$ for three smoothing choices given by $R_\mathrm{s}=0\Mpc$ (yellow), $R_\mathrm{s}=R_\mathrm{opt}$ (light blue), and $R_\mathrm{s}=R_\mathrm{eq}$ (blue), expressed in units of $h^{-1}\,\mathrm{Mpc}$. Solid lines show results at $z=0.9$, while dotted lines correspond to $z=1.8$. Shaded bands, shown only for the case of $z=0.9$ for visual clarity, represent the sample standard deviation across galaxies. Vertical black lines mark the resolution corresponding to the mean inter-particle separation. Horizontal lines indicate the nonlinear Zeldovich displacement predicted by Eq.~\eqref{eq:ZA_sigma}.}
\label{fig:MultiGrid_massassignment}
\end{figure}
Building on the formalism developed in Sects.~\ref{sec:ZA_nonlinear} and \ref{sec:ZA_redshift}, we implement the reconstruction using the \textsc{MultiGridReconstruction} algorithm, originally developed by Martin~J.~White\footnote{\url{https://github.com/martinjameswhite/recon_code/blob/master/notes.pdf}} and integrated into the \textsc{pyrecon}\footnote{\url{https://github.com/cosmodesi/pyrecon/tree/main}} Python package. The algorithm computes the Zeldovich displacement field directly from the galaxy distribution through a multigrid relaxation scheme, specifically designed to handle non-parallel \los and optimised for multithreaded execution. It provides an efficient configuration-space alternative to Fourier-based solvers \citep[e.g.][]{iFFT, IFFTp}, ideally suited for large-scale galaxy surveys with redshift-dependent number densities and complex geometries.

The algorithm requires as input the galaxy and random catalogues (Sect.~\ref{sec:catalogues}), the fiducial values of the linear bias $b$ and growth rate $f$, the Gaussian smoothing scale $R_\mathrm{s}$, and the number of grid cells $N_\mathrm{cell}^3$ used for density-field interpolation. The random catalogue serves both to define the survey selection function and to generate the shifted random field $\delta_\mathrm{s}$ required by the reconstruction. To minimise shot noise, it must be substantially denser than the galaxy catalogue. Following White's recommendations, the grid size $L_\mathrm{cell} = L_\mathrm{box}/N_\mathrm{cell}$ should not exceed the chosen smoothing scale $R_\mathrm{s}$, ensuring that the displacement field is resolved across the smoothing kernel.\footnote{By stress-testing the pipeline, we found that the multigrid solver remains numerically stable only for grid resolutions selected among the discrete candidate values
\(
\{\,2^{n_\mathrm{bit}},\; 3\times2^{n_\mathrm{bit}-1},\; 5\times2^{n_\mathrm{bit}-2},\; 7\times2^{n_\mathrm{bit}-2},\; 2^{n_\mathrm{bit}+1}\,\}
\), where $n_\mathrm{bit}$ is a freely chosen integer.} We will test the impact of the choice for $N_\mathrm{cell}$ in Sect.~\ref{sec:mass_assignment}.

The algorithm begins by interpolating the galaxy and random populations onto a regular grid encompassing the survey volume to estimate their corresponding number density fields, $n^\mathrm{s}_\mathrm{obs}$ and $n_\mathrm{r}$, respectively. From these, the observed overdensity field is computed as
\begin{equation}
\delta^{s}_\mathrm{obs} = \frac{n^{s}_\mathrm{obs}}{n_\mathrm{r}} - 1 \, ,
\end{equation}
with cells for which $\rho_\mathrm{r} = 0$ assigned a value $\delta^s_\mathrm{obs} = 0$ to preserve the mean density.

As required by Eq.~\eqref{eq:ZA_rec_disp}, a Gaussian smoothing filter $\mathcal{S}(k)=\exp(-k^2 R_\mathrm{s}^2/2)$ is applied to the overdensity field in Fourier space using a Fast Fourier Transform (FFT), then transformed back into configuration space. The real-space Zeldovich displacement field, $\vec{\psi}_\mathrm{ZA}$, defined from Eq.~\eqref{eq:ZA_rec_disp_rsd} as
\[
\vec{\psi}_\mathrm{ZA} = \frac{\vec{\psi}^{s}_\mathrm{ZA}}{1 + f\mu^2} \, ,
\]
is inferred from the filtered overdensity by solving the following equation in configuration space
\begin{equation}
\label{eq:ZA_disp_sol}
\boldsymbol{\nabla} \cdot \vec{\psi}_\mathrm{ZA}
+ \frac{f}{b}\,\boldsymbol{\nabla} \cdot
\left[ \frac{\vec{s}}{s^2} \left(\vec{s}\cdot\vec{\psi}_\mathrm{ZA}\right) \right]
= \frac{\delta_\mathrm{obs}^s}{b} \, ,
\end{equation}
where $\vec{s}$ denotes the comoving position of the observed galaxies.

Assuming the displacement field is irrotational, that is,  $\vec{\psi}_\mathrm{ZA} = \boldsymbol{\nabla}\phi$, the equation reduces to a Poisson-like form for the scalar potential $\phi$, given by
\begin{equation}
\label{eq:Multigrd_pot}
\nabla^2\phi + \frac{f}{b}\,\boldsymbol{\nabla} \cdot
\left[ \frac{\vec{s}}{s^2} \left(\vec{s} \cdot \boldsymbol{\nabla}\phi\right) \right]
= \frac{\delta_\mathrm{obs}^s}{b} \, .
\end{equation}
This equation is solved using a multigrid relaxation scheme with a full V-cycle and damped Jacobi iterations, where the V-cycle moves through the hierarchy of increasingly coarser and finer grids, and the Jacobi steps act as a smoothing operation at each level. Once the scalar potential $\phi$ is computed, the displacement field is obtained by taking its gradient. Further technical details are provided in the solver documentation.

Finally, the shifted galaxy catalogue \smash{$\delta^s_\mathrm{d}$} is constructed by displacing each galaxy by \smash{$-\left(1+f\mu^2\right)\,\vec{\psi}_\mathrm{ZA}$}, evaluated at its observed position. The corresponding shifted random catalogue is obtained by applying either \smash{$-\left(1+f\mu^2\right)\,\vec{\psi}_\mathrm{ZA}$ or $-\vec{\psi}_\mathrm{ZA}$}, depending on whether the \textsc{RecSym} or \textsc{RecIso} implementation is adopted.

\subsubsection{Mass assignment}
\label{sec:mass_assignment}
To assess the effect of grid resolution on the \textsc{MultiGridReconstruction} output, we apply the algorithm to all mocks while systematically varying the number of grid cells $N_\mathrm{cell}$ from 8 to 512. The reconstruction efficiency is evaluated using the mean reconstructed galaxy displacement $\langle|\vec{\psi}_\mathrm{ZA}|\rangle$, which quantifies the average offset between observed and reconstructed galaxy positions within each snapshot.

Figure~\ref{fig:MultiGrid_massassignment} shows how the mean reconstructed displacement amplitude depends on mesh resolution for different smoothing scales and redshifts.
In the absence of smoothing ($R_\mathrm{s}=0$), the mean displacement increases monotonically with grid resolution, with a noticeable steepening for $N_\mathrm{cell}\gtrsim250$ (i.e.,\ $L_\mathrm{cell}\lesssim5\hMpc$). This behaviour reflects the inability of the ZA approximation to recover fully nonlinear galaxy motions. Without explicit smoothing, the accuracy of the reconstruction is governed by $L_\mathrm{cell}$, which effectively acts as a numerical smoothing scale. As $L_\mathrm{cell}$ decreases below the mean galaxy separation, the recovered displacement field approaches that of direct particle-particle (`true') calculations, capturing increasingly nonlinear motion. However, the ZA framework assumes linear trajectories and constant comoving velocities (Eq.~\ref{eq:ZA_disp}), causing it to overestimate the true displacements when applied at excessively high resolution \citep{Sarpa2022}.

Applying Gaussian filtering ($R_\mathrm{s}>0$), the displacement amplitude stabilises for $N_\mathrm{cell}\gtrsim100$, with lower values achieved for stronger smoothing. This indicates that once $L_\mathrm{cell} < R_\mathrm{s}$, the reconstruction output becomes insensitive to grid resolution. In this regime, the measured displacement amplitude agrees with the theoretical expectation to within $1\sigma$ at all redshifts, validating the accuracy and numerical stability of our implementation.

Given the algorithm's insensitivity to $N_\mathrm{cell}$ provided $L_\mathrm{cell} \lesssim R_\mathrm{s}$, we adopt $N_\mathrm{cell}=128$ in all subsequent analyses. This choice ensures that the smoothing kernel is well resolved across all snapshots while maintaining computational efficiency, as the runtime of the multigrid solver scales approximately with $N_\mathrm{cell}^3$. We investigate the reconstruction performance across the full smoothing range $R_\mathrm{s}\in[R_\mathrm{opt},\,R_\mathrm{eq}]$ in Sect.~\ref{sec:ZA_sys_smoothing}.

\subsection{Reconstructed catalogues}
\label{sec:Rec_catalogues}
For each mock galaxy catalogue, we generate the corresponding reconstructed galaxy and shifted random catalogues using the \textsc{MultiGridReconstruction} algorithm described in Sect.~\ref{sec:ZA_multigrid}. The resulting catalogues are then trimmed according to the observed survey mask. This operation, which removes fewer than 1\% of objects, is essential to ensure that the \tpcf estimator (see Sect.~\ref{sec:Estimators}) measures the clustering excess relative to the unshifted random catalogue that defines the survey geometry.

\section{2PCF: Estimators and fitting template}
\label{sec:measurements}

\subsection{Estimators}\label{sec:Estimators}
We measure the clustering signal of the pre- and post-reconstruction catalogues using the anisotropic \tpcf, \(\xi^\mathrm{data}(s,\mu)\), and its first three even multipoles, \(\curly{\xi^\mathrm{data}_\ell(s)}_{\ell=0,2,4}\). Here, \(s\) denotes the absolute value of the pair-separation vector in redshift space. For this purpose, we employ the official \Euclid implementation of the Landy--Szalay estimator \citep{Landy-Szalay1993, EP-delaTorre} in the two configurations
\begin{equation}\label{eq:LS_standard}
\xi_\mathrm{pre}(s,\nu) = \frac{\NDD(s,\nu) - 2\,\NDR(s,\nu) + \NRR(s,\nu)}{\NRR(s,\nu)}
\end{equation}
and 
\begin{equation}\label{eq:LS_shifted}
\xi_\mathrm{post}(s,\nu) = \frac{\NDD(s,\nu) - 2\,\NDS(s,\nu) + \NSS(s,\nu)}{\NRR(s,\nu)} \, .
\end{equation}
Here, $\NDD$, $\NDR$, and $\NRR$ denote the standard data-data, data-random, and random-random pair counts entering the pre-reconstruction estimator in Eq.~\eqref{eq:LS_standard} and $\nu=\vec{s}\cdot\vec{z}/(sz)$ refers to the cosine of the angle between the \los, $\vec{z}$, and the pair-separation vector, $\vec{s}$.
For the post-reconstruction estimator -- given by Eq.~\eqref{eq:LS_shifted} -- the contributions $\NDS$ and $\NSS$ correspond to pair counts computed by cross-correlating the data catalogue with the shifted random catalogue and auto-correlating the latter with itself, respectively.

The anisotropic \tpcf is measured in 40 bins of width \(5\,\hMpc\) over the range \(s \in [0,200]\,\hMpc\), and in 200 uniform bins of \(\nu \in [-1,1]\). The multipoles are obtained by projecting \(\xi^\mathrm{data}(s,\nu)\) onto the Legendre polynomials \(L_\ell(\nu)\) where the integral over $\nu$ is replaced by a Riemann sum such that
\begin{equation}\label{eq:2PCF_mult}
\xi^\mathrm{data}_\ell(s) = \sum_i \xi^\mathrm{data}(s,\nu_i) \, \mathcal{L}_\ell(\nu_i) \, \Delta \nu_i \, ,
\end{equation}
where \(\mathcal{L}_\ell\) is evaluated at the centre of each angular bin \(\nu_i\) of width $\Delta \nu_i$. Mean measurements and their mock-to-mock dispersion are shown in  Fig.~\ref{fig:2PCF_measurements} in the Appendix. A detailed validation of the estimator and its comparison with other public implementations are presented in \citet{EP-delaTorre}.

\subsection{2PCF fitting templates}\label{sec:fitting_templates}
To extract cosmological information from the \bao feature, we model the multipoles of the \tpcf, Eq.~\eqref{eq:2PCF_mult}, using the anisotropic template of \citet{Ross2017}, given by
\begin{equation}\label{eq:xi_model}
\xi_\ell^\mathrm{model}(s) = b^2\,\xi_\ell^\mathrm{ph}(s) + \mathrm{BB}_\ell(s) \, ,
\end{equation}
where the physical term \(\xi_\ell^\mathrm{ph}\) describes the reconstructed clustering signal evaluated at $\bar{z}$ and depends on the set of parameters given by $\{\alpha_\perp,\alpha_\parallel,f, \Sigma_\perp,\Sigma_\parallel,R_\mathrm{s}\}$. The broadband component \(\mathrm{BB}_\ell(s)\) marginalises over residual systematic or nonlinear effects not captured by the model.

The cosmological information is encoded in \(\xi_\ell^\mathrm{ph}\) through the \apdist parameters \citep{Alcock1979, Xu2013} given by
\begin{equation}\label{eq:alphas}
\alpha_\parallel(\bar{z}) =
\frac{H^\mathrm{f}(\bar{z}) \, r_\mathrm{s}^\mathrm{f}}
     {H^\mathrm{t}(\bar{z}) \, r_\mathrm{s}^\mathrm{t}}  \qquad \mathrm{and} \qquad
\alpha_\perp(\bar{z}) =
\frac{D_\mathrm{A}^\mathrm{t}(\bar{z}) \, r_\mathrm{s}^\mathrm{f}}
     {D_\mathrm{A}^\mathrm{f}(\bar{z}) \, r_\mathrm{s}^\mathrm{t}} \, ,
\end{equation}
which describe the apparent dilation of the \bao scale in the direction parallel and perpendicular to the \los when the fiducial cosmology differs from the true one (Eq.~\ref{eq:red-to-dist}), denoted here by the superscript `t'. Here, \(D_\mathrm{A}\) and \(H\) denote the angular diameter distance and the Hubble parameter at the mean redshift \(\bar{z}\), while \(r_\mathrm{s}\) is the sound horizon at the baryon-drag epoch. In addition to the \apdist parameters, the template depends on the linear bias \(b\), the growth rate of structures \(f\), and the damping parameters \(\Sigma_\perp\) and \(\Sigma_\parallel\), which model the exponential smoothing of the \bao feature perpendicular and parallel to the \los. The Gaussian smoothing scale \(R_\mathrm{s}\) is fixed to the value used in the reconstruction and set to zero when no reconstruction is applied.

In practice, \(\xi_\ell^\mathrm{ph}\) is obtained by Fourier-transforming the redshift-space anisotropic power spectrum \(P^s(k,\mu)\), evaluated at the fiducial cosmology of the mock catalogues and constructed following the formalism of Sect.~\ref{sec:ZA_redshift} under the assumption \(R_\mathrm{s}=R_\mathrm{eq}\) such that
\begin{equation}\label{eq:Pkmu_nl_model}
P^s(k,\mu) = \left\{1+\mu^2 \, \frac{f}{b} \, \left[1-\mathcal{S}(k)\right]\right\}^2 \, P^\mathrm{r}(k)
\end{equation}
and 
\begin{equation}\label{eq:Pk_real_nl_model}
P^\mathrm{r}(k) = \left[\Plin(k)-P_\mathrm{nw}(k)\right]
\mathrm{exp}\left(-k^2\sigma_\mathrm{v}^2/2\right) + P_\mathrm{nw}(k) \, ,
\end{equation}
with
\begin{equation} \label{eq:Preal}
\sigma_\mathrm{v}^2 = \left(1-\mu^2\right) \, \Sigma_\perp^2 + \mu^2 \, \Sigma_\parallel^2 \, .
\end{equation}
All the power spectra appearing in Eqs.~\eqref{eq:Pkmu_nl_model} and \eqref{eq:Pk_real_nl_model} are to be understood as evaluated at $\bar{z}$.
The real-space power spectrum $P^\mathrm{r}$ is built using the smooth `no-wiggle' power spectrum, \(P_\mathrm{nw}\), computed using the transfer function of \citet{EisenstainHu1998}. This weighted combination of \(\Plin\) and \(P_\mathrm{nw}\) extends the Zeldovich formalism of Sect.~\ref{sec:ZA_nonlinear} by accounting for partial mode coupling between the linear and smooth components of the density field. The Gaussian filter \(\mathcal{S}(k;R_\mathrm{s})\) modulates the residual \rsd term in \textsc{RecIso} and vanishes for \textsc{RecSym} or \textsc{PreRec} fits. Following \citet{Bautista2018}, we neglect nonlinear \rsd, whose contribution is absorbed by the broadband term.

The \apdist correction is implemented following \citet{Kazin2013}. The multipoles of \(P^s(k,\mu)\) are obtained by integrating over \(\mu\), then transformed into configuration space via a Hankel transform, and re-projected into multipoles after the coordinate transformation $\nu \rightarrow \nu'$ and  $s \rightarrow s'$ set by
\begin{equation}
\nu' = \frac{\nu\,\alpha_\perp}{\sqrt{\nu^2\alpha_\parallel^2 + \left(1-\nu^2\right)\alpha_\perp^2}}
\end{equation}
and 
\begin{equation}
    s'= s\,\sqrt{\nu^2\alpha_\parallel^2 + \left(1-\nu^2\right) \, \alpha_\perp^2} \, ,
\end{equation}
yielding the final physical template \(\xi_\ell^\mathrm{ph}(s)\).
The broadband term is modelled as a second-order polynomial,
\begin{equation}\label{eq:BB}
\mathrm{BB}_\ell(s) = A_{\ell,0}
           + A_{\ell,1}\frac{s_\mathrm{ref}}{s}
           + A_{\ell,2}\frac{s_\mathrm{ref}^2}{s^2} \, ,
\end{equation}
with coefficients normalised at \(s_\mathrm{ref} = 80\,\hMpc\) to improve numerical stability \citep{Sarpa2021}.

\section{Parameter inference}\label{sec:Likelihood}

\begin{figure*}
\centering
\includegraphics[width=1\textwidth]{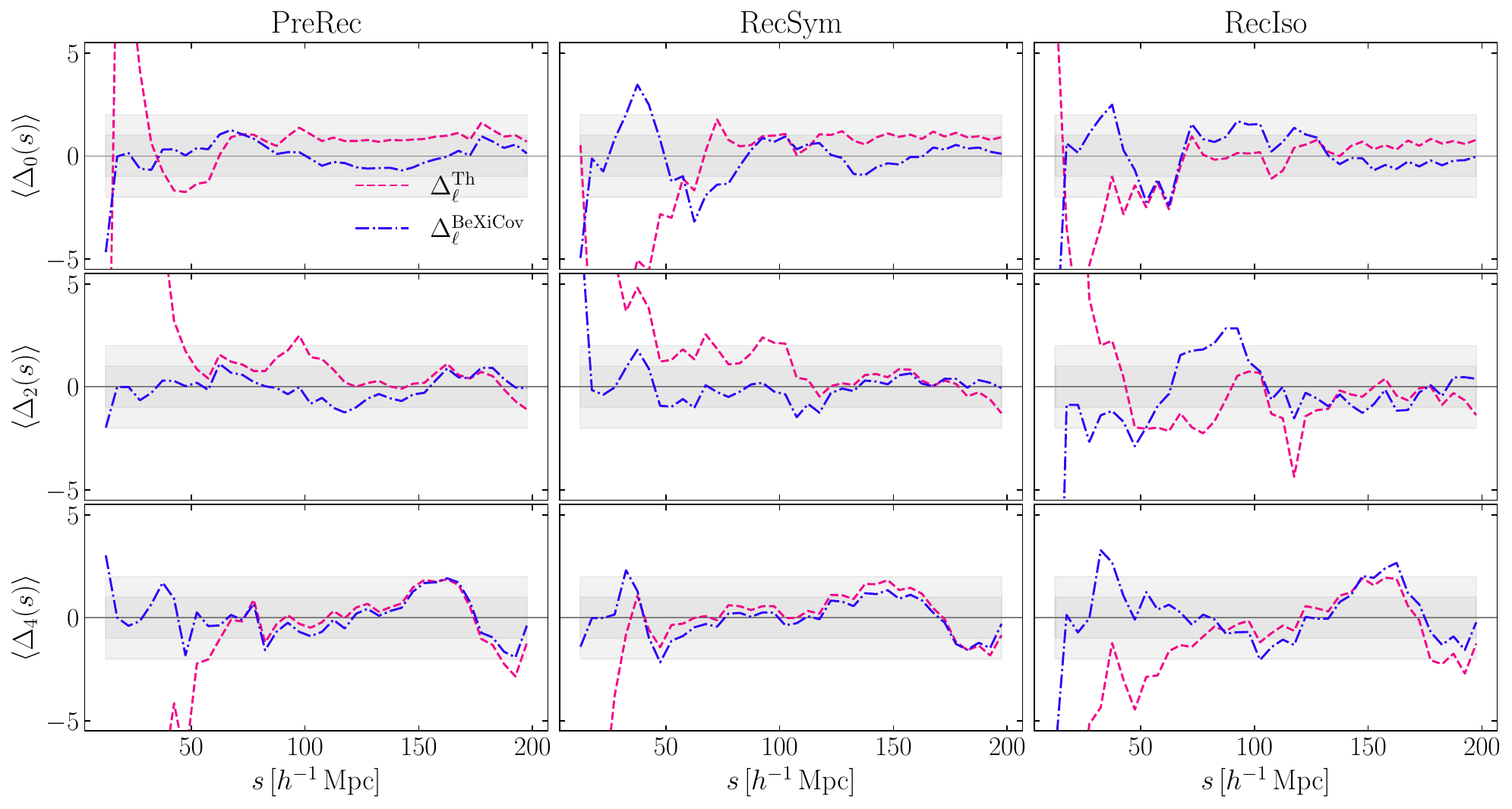}
\caption{Normalised residuals of the theoretical (magenta, dashed) and \textsc{BeXiCov} (blue, dot-dashed) models with respect to the mean \tpcf multipoles at $z=0.9$. The leftmost column shows \textsc{PreRec}, the middle \textsc{RecSym}, and \textsc{RecIso} is presented in the rightmost column. Residuals are computed as $\Delta_\ell=(\xi_\ell^{\mathrm{model}}-\xi_\ell^{\mathrm{data}})/(\sigma_\ell^{\mathrm{data}}/\sqrt{N_{\mathrm{mocks}}})$ with $N_{\mathrm{mocks}}=8$ and shaded bands indicate $|\Delta_\ell|=1$ as well as $|\Delta_\ell|=2$.}
\label{fig:cov_residuals}
\end{figure*}

We estimate the \bao parameters using a Gaussian likelihood for the \tpcf multipoles, combined with a covariance matrix calibrated on the measured clustering signal and an efficient posterior sampling engine. The first novelty is an iterative, semi-analytical covariance matrix incorporating explicit survey-window convolution. 
The second is an emulator-based Hamiltonian sampler 
that provides parameters posteriors at a fraction of the computational cost of standard methods. We detail and validate each component using \flagshipone mocks across redshift bins and reconstruction schemes.

\begin{table}
\centering
\caption{Prior distributions for the parameters entering the \tpcf model (see Eq.~\ref{eq:xi_model}).}
    \renewcommand{\arraystretch}{1.2}
    \begin{tabular}{|c|c|}
    \hline
    \rowcolor{blue!5}
    Parameter & Prior Distribution \\
    \hline
    $\alpha_\perp$ & $\mathcal{U} [0.8, 1.2]$ \\
    $\alpha_\parallel$ & $\mathcal{U}[0.8, 1.2]$ \\
    $b$ & $\mathcal{U} [0.0, 5.0]$ \\
    $f$ & $\mathcal{U} [0.0, 2.0]$ \\
    $\Sigma_\perp\, [\hMpc]$ & $\mathcal{U}[0.0, 20.0]$ \\
    $\Sigma_\parallel \, [\hMpc]$ & $\mathcal{U} [0.0, 20.0]$ \\
    $A_{\ell,i}\, [\hMpc]$ & $\mathcal{U}[-20.0, 20.0]$ \\
    \hline
    \end{tabular}
\tablefoot{$\mathcal{U}[a,b]$ denotes a uniform prior on the interval $[a,b]$. For the broadband coefficients $A_{\ell, i}$ we use the same prior for all $\ell\in\{0,2,4\}$ and $i$. For the \bao-only fit we have $i\in\{0,1,2\}$ while for the covariance we have $i\in\{0,1,2,3,4\}$.}
\label{tab:priors}
\end{table}

\subsection{Likelihood definition}\label{sec:Likelihood_definition}
We fit the data vector $\vec{d}$ containing 60 elements constructed from a concatenation of the first three even multipoles, that is, $\xi_0(s)$, $\xi_2(s)$, and $\xi_4(s)$, where $s$ consists of 20 bins spanning \(s \in [50,150]\Mpc\) with a bin width of \(5\Mpc\). The model multipoles, defined in Eq.~\eqref{eq:xi_model}, are evaluated at the same sample points $s_i$. The parameters \(\boldsymbol{\theta} = \{\alpha_\perp, \alpha_\parallel, b, f, \Sigma_\perp, \Sigma_\parallel, \{A_{\ell,i}\}\}\) are inferred from the posterior
\begin{equation}\label{eq:posterior}
\mathcal{P}(\boldsymbol{\theta}\mid\vec{d}) \; \propto\; \mathcal{L}(\vec{d}\mid\boldsymbol{\theta})\,\Pr(\boldsymbol{\theta}),
\end{equation}
where \(\Pr(\boldsymbol{\theta})\) denotes the prior density (see Table~\ref{tab:priors}). Unless stated otherwise, we report posterior medians with asymmetric 68\% credible intervals.
Assuming the distribution of the data vector to follow a Gaussian form, the likelihood reads
\begin{equation}
\label{eq:like_gauss}
    \mathcal{L}(\vec{d}\mid\boldsymbol{\theta}) =\frac{1}{\sqrt{(2\pi)^{N_{\mathrm{d}}}\,|\tens{C}|}}\, \exp\left\{-\frac{1}{2} \, \chi^2(\boldsymbol{\theta})\right\} \, ,
\end{equation}
with the $\chi^2$ being defined as
\begin{equation}
\label{eq:chi2_vec}
\chi^2(\boldsymbol{\theta})
=\left[\vec{d}-\xi^{\mathrm{model}}_\ell(s_i;\boldsymbol{\theta})\right]^{\top} \, \tens{C}^{-1} \, \left[\vec{d}-\xi^{\mathrm{model}}_\ell(s_i;\boldsymbol{\theta})\right] \, ,
\end{equation}
where $\tens{C}$ is the semi-analytical, window-convolved covariance described in Sect.~\ref{sec:covariances}, and $N_{\mathrm{d}}$ is the number of elements in the data vector.

\subsection{Covariance matrices}\label{sec:covariances}
To mitigate noise from a limited number of mocks, we estimate the \tpcf multipole covariance using a semi-analytical approach inspired by the \textsc{RascalC} formalism \citep{Philcox2019}. In our implementation, \textsc{WinCov}\footnote{\url{https://gitlab.com/veropalumbo.alfonso/windowcovariance/}} extends the original \textsc{RascalC} treatment of the survey window by re-sampling non-periodic effects with random catalogues, while \textsc{BeXiCov}\footnote{\url{https://gitlab.com/esarpa1/BeXiCov}}
replaces the simple interpolation of noisy clustering measurements with the output of a physically motivated template fit to the \tpcf multipoles, allowing for a more robust identification of stochastic fluctuations. This step is particularly important for real-data applications, where the covariance must remain flexible enough to track the clustering signal preferred by the data, including cosmology-dependent differences (see Appendix~\ref{app:fid_cov}) and residual observational systematics not included in the mocks.

We begin by estimating the Gaussian covariance of $\xi_\ell$, $\tens{C}^{\xi}_{\ell_1\ell_2}(s_i,s_j)$, under periodic boundary conditions. Following \citet{Grieb2016}, we get
\begin{equation}\label{eq:xil_cov}
\tens{C}^{\xi}_{\ell_1\ell_2}(s_i,s_j) =\frac{\mathrm{i}^{\,\ell_1+\ell_2}}{2\pi^2} \int_{0}^{\infty} k^2 \, \sigma^2_{\ell_1\ell_2}(k) \, \bar{\jmath}_{\ell_1}(k s_i)\,\bar{\jmath}_{\ell_2}(k s_j) \, \diff k \, ,
\end{equation}
where $\bar{\jmath}_\ell(ks)$ is the bin-averaged spherical Bessel function over a top-hat bin in $s$, and
\begin{equation}\label{eq:cov_sigma2}
\begin{split}
    \sigma^2_{\ell_1\ell_2}(k) = &\frac{(2\ell_1+1)(2\ell_2+1)}{V}\\
    & \times\int_{-1}^{1}\left(P^s(k,\mu)+\frac{1}{\bar n}\right)^2 \mathcal{L}_{\ell_1}(\mu)\,\mathcal{L}_{\ell_2}(\mu)\,\diff\mu \, ,
\end{split}
\end{equation}
with $V$ the survey volume and $\bar n$ the galaxy number density. The anisotropic input power spectrum follows the theoretical predictions in Sect.~\ref{sec:ZA_redshift}, meaning pre-reconstruction uses $P^\mathrm{s}_{\mathrm{obs}}(k,\mu)$ of Eq.~\eqref{eq:Pk_ZA_rsd}, while post-reconstruction employs $P^s_{\mathrm{sym}}(k,\mu)$ or $P^s_{\mathrm{iso}}(k,\mu)$ as given in  Eq.~\eqref{eq:Pk_symiso}, for \textsc{RecSym} and \textsc{RecIso}, respectively.

We initially fit the mean \tpcf multipoles (averaged over eight realisations per snapshot) using the theoretical covariance to derive a data-calibrated input spectrum. Unlike the \bao-only fit in Sect.~\ref{sec:fitting_templates}, we model the full shape over $s\in[10,200]\,\hMpc$. The broadband term in Eq.~\eqref{eq:BB} is thus extended to
\begin{equation}\label{eq:BB_ext}
\mathrm{BB}^\mathrm{ext}_\ell(s) = A_{\ell,0}
           + A_{\ell,1}\frac{s_\mathrm{ref}}{s}
           + A_{\ell,2}\frac{s_\mathrm{ref}^2}{s^2}
           + A_{\ell,3}\frac{s}{s_\mathrm{ref}}
           + A_{\ell,4}\frac{s^2}{s^2_\mathrm{ref}} \, ,
\end{equation}
in order to incorporate small- and large-scale behaviour.
From the best-fit $\xi_\ell^{\mathrm{ext,BF}}$ we obtain $P_\ell^{\mathrm{BF}}(k)$ via inverse Hankel transform and construct
\begin{equation}
\label{eq:PBF_mu}
    P^{\mathrm{BF}}(k,\mu)=\sum_{\ell=0,2,4} P^{\mathrm{BF}}_\ell(k) \, \mathcal{L}_\ell(\mu) \, .
\end{equation}
The Gaussian covariance is then updated using $P^{\mathrm{BF}}(k,\mu)$ in Eqs.~\eqref{eq:xil_cov} and \eqref{eq:cov_sigma2}, and the procedure is iterated until convergence, defined as an absolute change in $\chi^2$ of less than 1\% between successive iterations. In practice, convergence is typically achieved after two iterations for all redshifts considered.

\begin{figure}
  \centering
  \includegraphics[width=\columnwidth]{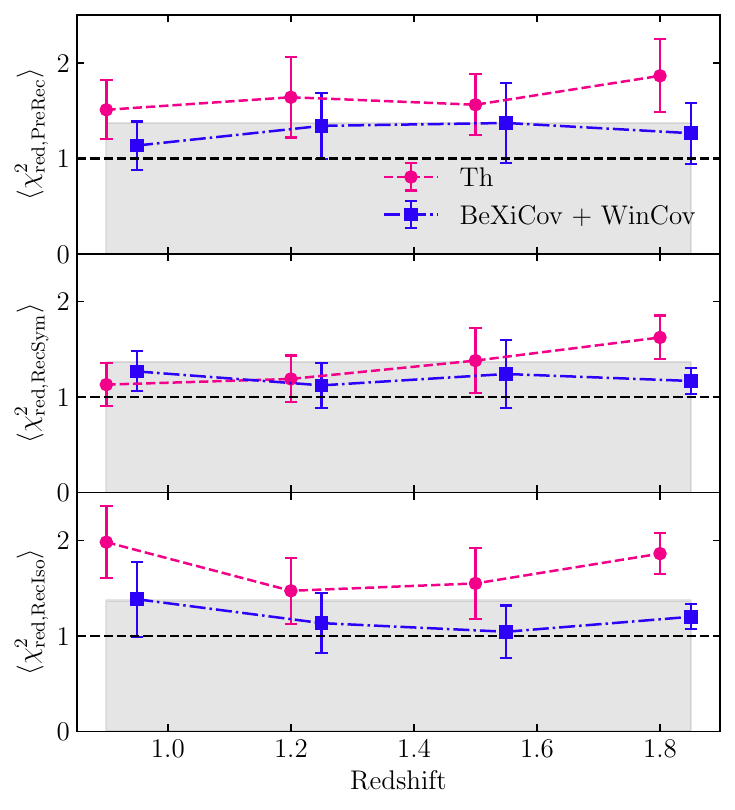}
  \caption{Mean values of the $\chired$ from fits to the \tpcf multipoles, averaged over eight realisations at $z=\{0.9,1.2,1.5,1.8\}$. Magenta circles (dashed line) denote the non-iterated theory covariance, while blue squares (dot-dashed) refer to the iterative \texttt{BeXiCov+WinCov} covariance. Panels in order from top to bottom present \textsc{PreRec}, \textsc{RecSym}, and \textsc{RecIso}, respectively. Grey bands indicate the acceptance regions quantified by a $p$-value $>0.05$. Blue dots are shifted horizontally to improve readability.}
  \label{fig:CHi2_THvsWinCov}
\end{figure}

Figure~\ref{fig:cov_residuals} shows the normalised residuals between the model predictions and the mean \tpcf multipoles for different reconstruction schemes at \( z = 0.9 \). The non-iterated theoretical covariance reproduces large-scale trends but underestimates nonlinear effects and \bao damping on small scales, especially in the quadrupole. In contrast, the iterative \textsc{BeXiCov} model achieves agreement within \( 1\sigma \) across the full separation range. This improvement is consistent across all redshift snapshots.

To account for finite-volume effects, we convolve the periodic Gaussian covariance with the survey window using the \textsc{RascalC} formalism, retaining only Gaussian contributions. The window-convolved covariance matrix of $\xi_\ell$ is expressed as
\begin{equation}
    \tens{C}^{\xi,\mathrm{Win}}_{\ell_1\ell_2}(s_1,s_2) = \tens{C}^{(2)}_{\ell_1\ell_2}(s_1,s_2) + \tens{C}^{(3)}_{\ell_1\ell_2}(s_1,s_2) + \tens{C}^{(4)}_{\ell_1\ell_2}(s_1,s_2) \, ,
\end{equation}
where $\tens{C}^{(2)}$ is the disconnected (shot-noise) component, and $\tens{C}^{(3)}$ and $\tens{C}^{(4)}$ encode the mixed and Gaussian contributions responsible for generating the off-diagonal covariance. The amplitude of all terms is regulated by the input anisotropic \tpcf\ $\xi^\mathrm{in}$. Following \citet{Philcox2019}, each $\tens{C}^{(n)}$ term is written as a summation over $n$-tuples of random-catalogue points, weighted by the measured \tpcf of the data catalogue. In our implementation, these summations use the best-fit \tpcf fitted to the data in the previous step,
\begin{equation*}
    \xi^{\mathrm{BF}}(s,\nu) = \sum_{\ell=0,2,4} \xi^{\mathrm{BF}}_\ell(s)\,\mathcal{L}_\ell(\nu) \, .
\end{equation*}
This procedure suppresses sampling noise with respect to the standard implementation.
We validate \textsc{WinCov} in Appendix~\ref{app:WinCov validation} by comparing the semi-analytical estimates against those from 1000 independent mocks not used in the fitting process.


We assess the robustness of the semi-analytical covariances by comparing the goodness-of-fit statistics from \tpcf multipole fits using our fiducial \texttt{BeXiCov+WinCov} covariance to those obtained with the initial, non-iterated theory covariance.
Figure~\ref{fig:CHi2_THvsWinCov} presents the mean $\chired$ values from fits to mock \tpcf multipoles across multiple redshifts and reconstruction schemes. The analytical covariance systematically overestimates the $\chired$, frequently falling outside the statistical acceptance region, particularly for \textsc{PreRec} and \textsc{RecIso}. In contrast, the iterative \texttt{BeXiCov+WinCov} covariance consistently yields statistically acceptable fits, indicating improved modelling fidelity. This improvement stems from a more accurate treatment of mask-induced mode couplings and the correction of theoretical inaccuracies, which persist even in sub-boxes lacking periodic boundaries. As a result, the reliability of inferred parameter constraints is significantly enhanced.

\subsection{Posterior sampling: \texttt{Bora.jl} and the \texttt{NUTS} sampler}\label{sec:Bora}
\begin{figure}
  \centering
  \includegraphics[width=0.99\columnwidth]{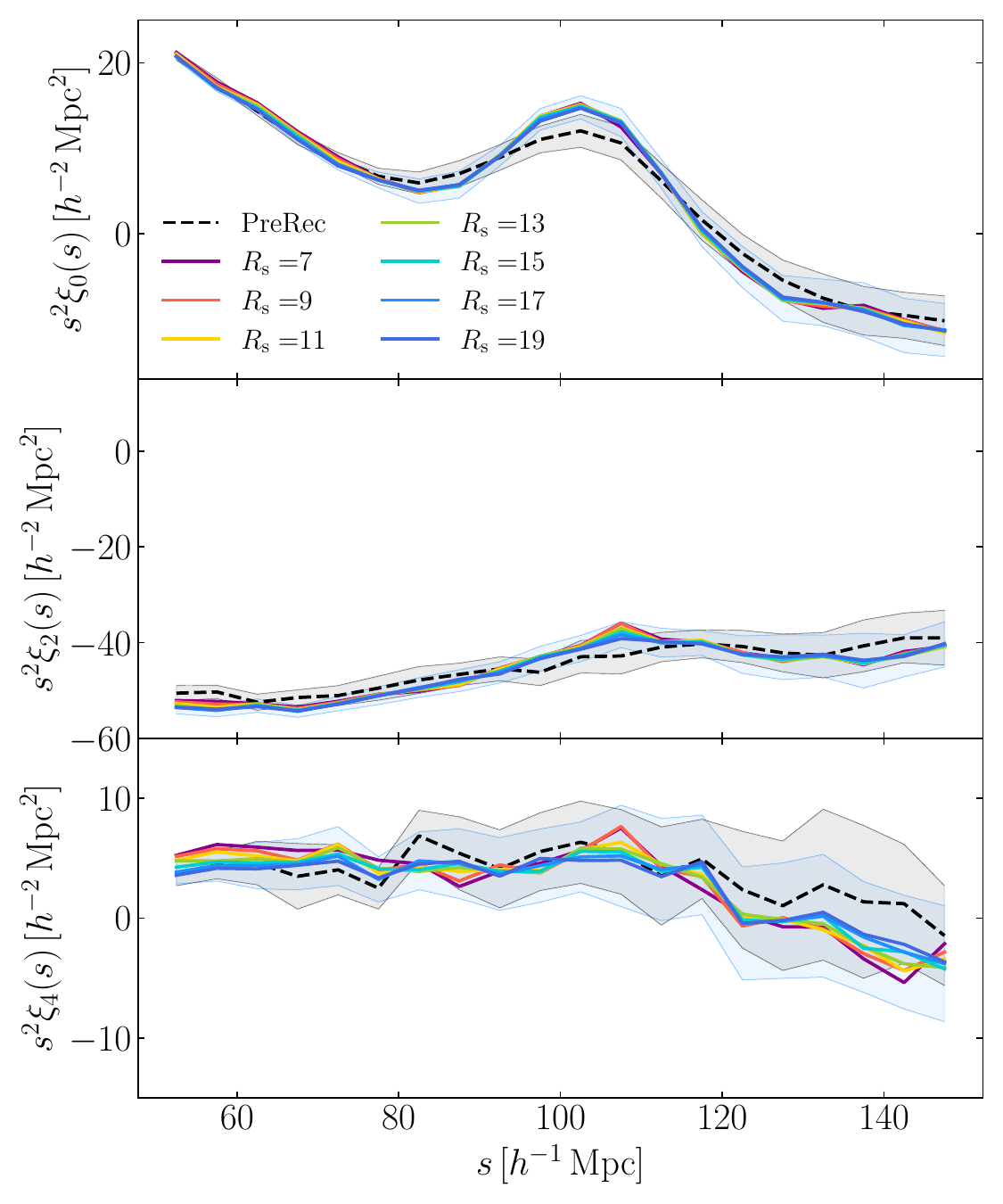}
  \caption{Impact of the smoothing scale \(R_{\mathrm{s}}\) on \textsc{RecSym} at \(z=0.9\). Coloured lines show the mean monopole (top panel), quadrupole (middle panel), and hexadecapole (bottom panel) of the reconstructed \tpcf, averaged over eight sub-boxes. Different colours correspond to different \(R_{\mathrm{s}}\) (expressed in units of $h^{-1}\,\mathrm{Mpc}$). Shaded bands, shown only for the case of \textsc{PreRec} and $R_\mathrm{s}=15h^{-1}\,\mathrm{Mpc}$ for visual clarity, indicate the standard error of the mean. The black dashed curve is the mean \textsc{PreRec} signal.}
  \label{fig:ZA_rec_sym_smoothing_xi_z0p9}
\end{figure}

\begin{figure}
  \centering
  \includegraphics[width=0.99\columnwidth]{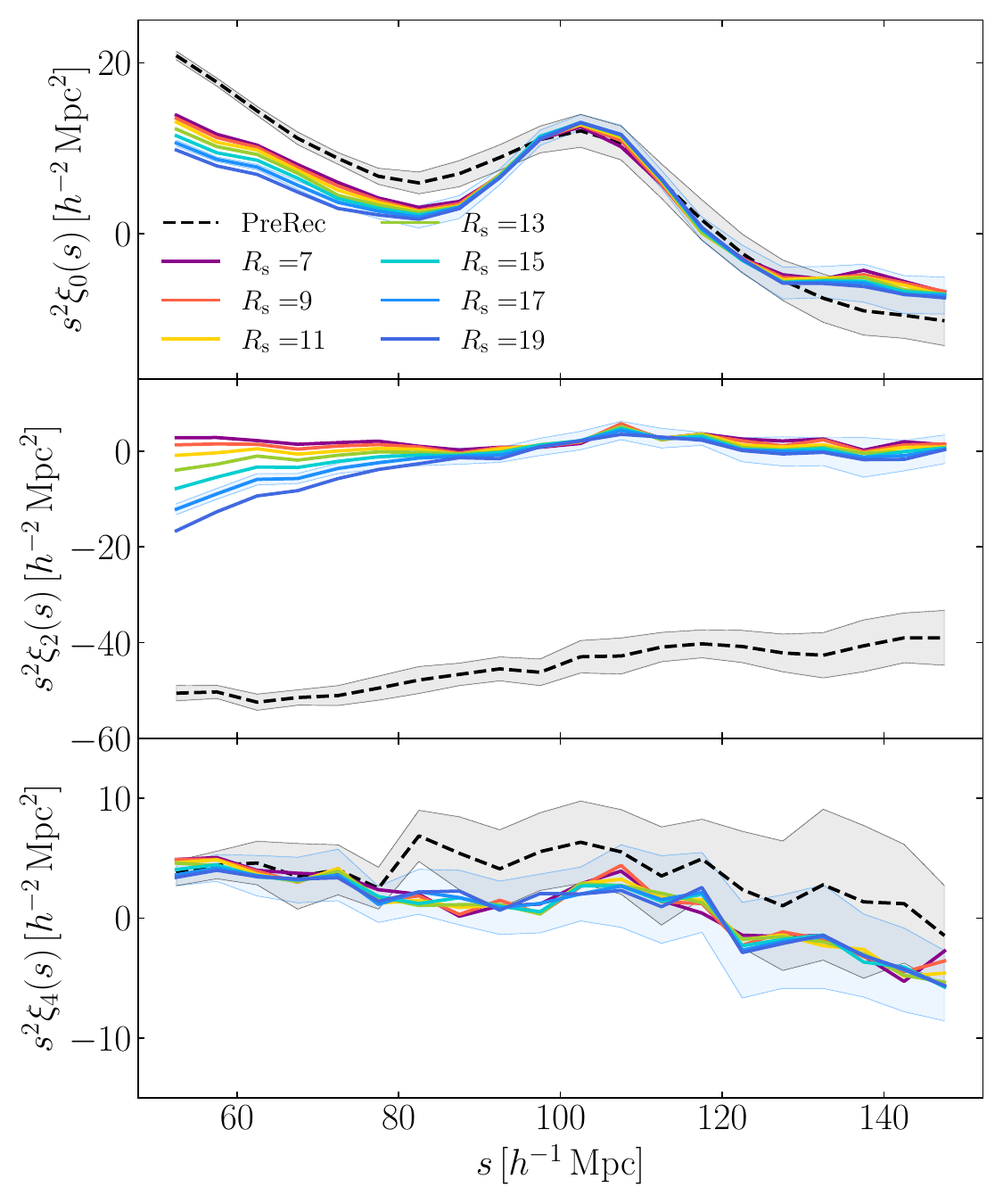}
  \caption{Same as Fig.~\ref{fig:ZA_rec_sym_smoothing_xi_z0p9}, but for \textsc{RecIso}.}
  \label{fig:ZA_rec_iso_smoothing_xi_z0p9}
\end{figure}
Sampling the posterior distribution of the parameters in the $\xi_\ell$ model -- Eq.~\eqref{eq:xi_model} -- is computationally intensive, as each evaluation requires multiple integrals and Fourier/Hankel transforms. Traditional \mcmc methods \citep{Metropolis,Hastings} typically demand of the order of $10^5\text{--}10^6$ likelihood evaluations to achieve convergence in high-dimensional parameter spaces, often incurring substantial computational costs.

To alleviate this burden, we developed \texttt{Bora.jl}, a \tpcf emulator that approximates the physical component $\xi^{\mathrm{ph}}_\ell$ of the model, enabling rapid predictions over the same pair-separation bins as the data vector. The emulator is paired with the No-U-Turn Sampler (\texttt{NUTS}; \citealt{NUTS}), a \hmc method \citep{DUANE,Radford2011} optimised for efficient exploration of complex posteriors. For each redshift bin, the emulator is trained on 10\,000 theoretical realisations of $\xi^{\mathrm{ph}}_\ell$, each corresponding to a unique parameter set $\{\alpha_\perp,\alpha_\parallel,b,f,\Sigma_\perp,\Sigma_\parallel,R_{\mathrm{s}}\}$, using an 80/20 training-validation split and a mean squared error (MSE) loss function. Training requires $\sim0.25\CPUh$. Further implementation details are provided in Appendix~\ref{app:Bora}. Once trained, \texttt{Bora.jl} enables posterior inference over the full parameter set (including broadband coefficients) using \texttt{NUTS}. This sampler performs long, gradient-informed transitions towards high-likelihood regions, thereby reducing sample autocorrelation and accelerating convergence. Its adaptive step-size control minimises manual tuning and enhances robustness across different likelihood geometries.

\section{Results and discussion}\label{sec:results}
In this section we present a comprehensive \bao analysis of the pre- and post-reconstruction clustering signals extracted from the \flagshipone mock catalogues (Sect.~\ref{sec:catalogues}). In Sect.~\ref{sec:ZA_sys_smoothing}, we determine the optimal reconstruction smoothing, quantifying its impact on the recovered \bao scale and fit reliability. Section~\ref{sec:fiducial_cosmo} tests the stability of post-reconstruction \bao constraints to variations in the fiducial cosmology, as expected when the true cosmology is unknown. Finally, Sect.~\ref{sec:BAO_constraints} reports the optimised \bao distances from the \flagshipone mocks and their propagation to cosmological parameters.

\subsection{Optimal smoothing scale}\label{sec:ZA_sys_smoothing}
\begin{figure}
\centering
\includegraphics[width=0.99\columnwidth]{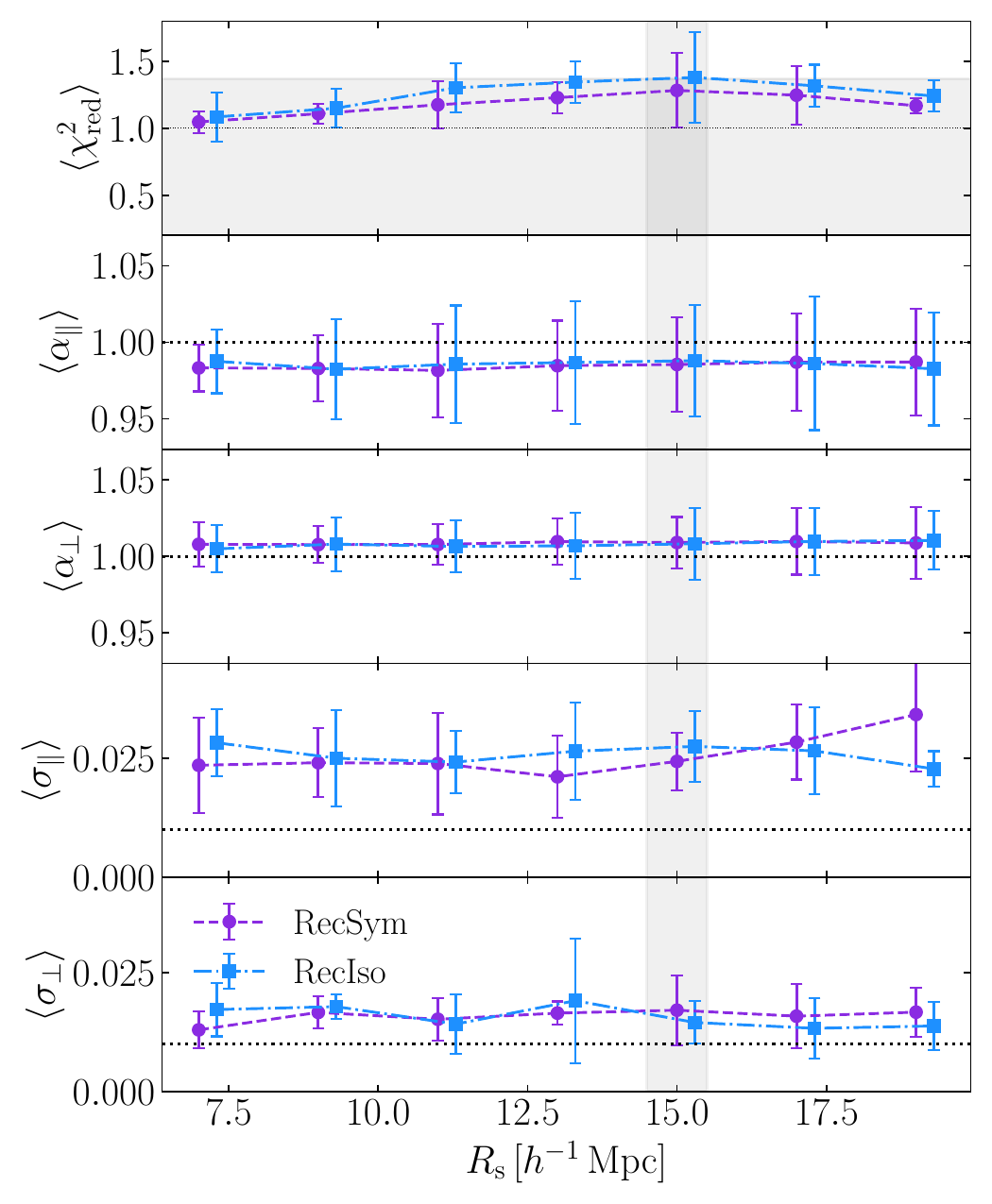}
\caption{Effect of the smoothing scale \(R_{\mathrm s}\) on post-reconstruction \bao fits at \(z=0.9\). Purple dashed and light-blue dot-dashed lines correspond to  \textsc{RecSym} and \textsc{RecIso}, respectively. From top to bottom, the panels show the mean \(\chired\), averaged over the eight realisations at fixed \(R_{\mathrm s}\), the mean values of \(\alpha_\parallel\) and \(\alpha_\perp\), and their profile-likelihood uncertainties. Error bars indicate the standard deviation across the eight realisations. The horizontal grey band in the top panel marks the acceptance region \(p>0.05\), while the vertical grey band highlights the standard choice \(R_{\mathrm s}=15\Mpc\) adopted in past surveys (see main text). Black dotted horizontal lines mark the expected AP values, while in the uncertainty panels they indicate the 1\% precision level for visual reference. Light-blue points are slightly shifted horizontally to improve readability.}
\label{fig:ZA_Smoothing_opt_z0p9}
\end{figure}

\begin{figure}
\centering
\includegraphics[width=0.99\columnwidth]{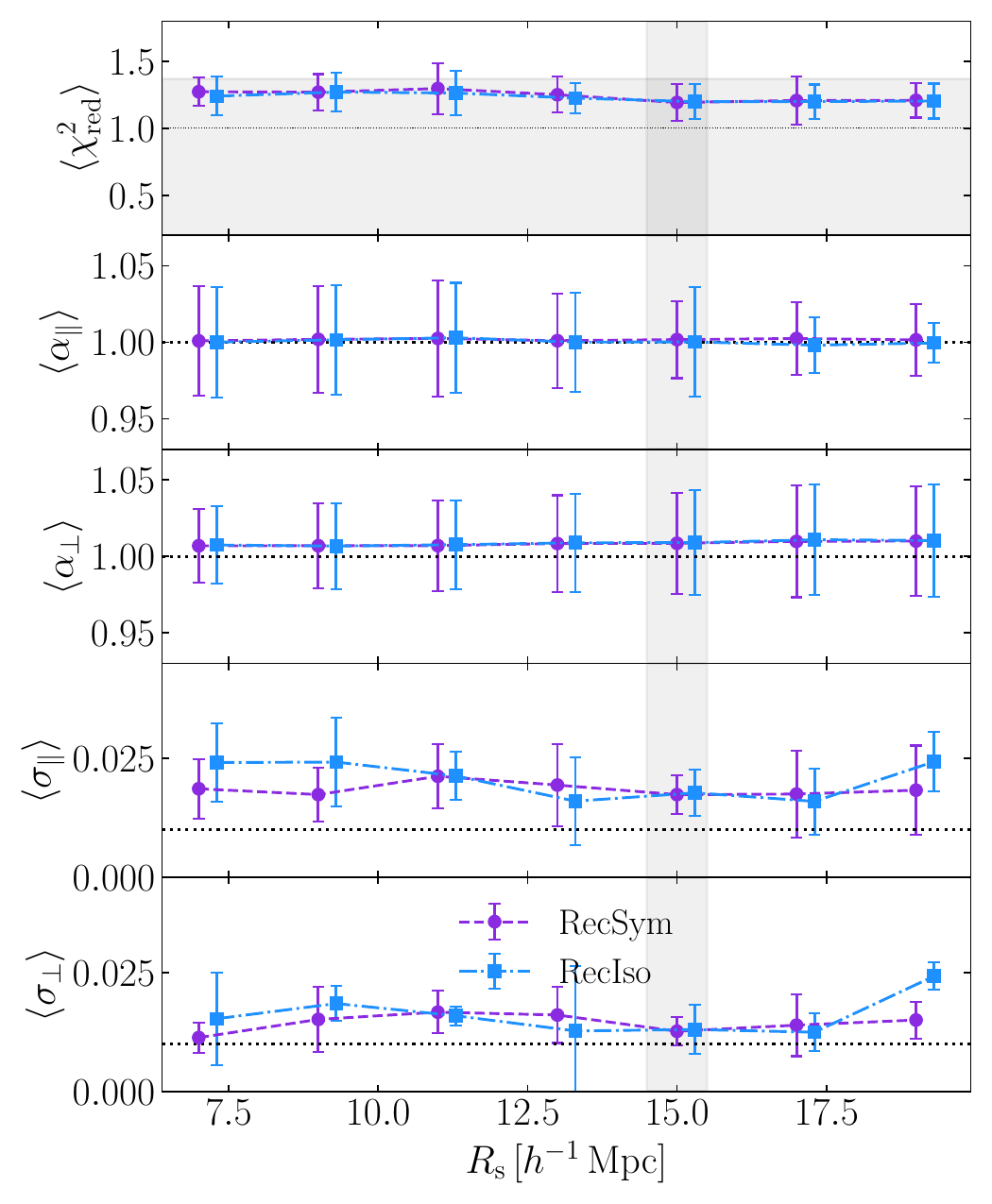}
\caption{Same as Fig.~\ref{fig:ZA_Smoothing_opt_z0p9}, but for \(z=1.8\).}
\label{fig:ZA_Smoothing_opt_z1p8}
\end{figure}

\begin{figure*}
\centering
\subfloat[\textsc{PreRec}, \(z=0.9\)]{\includegraphics[width=.33\textwidth]{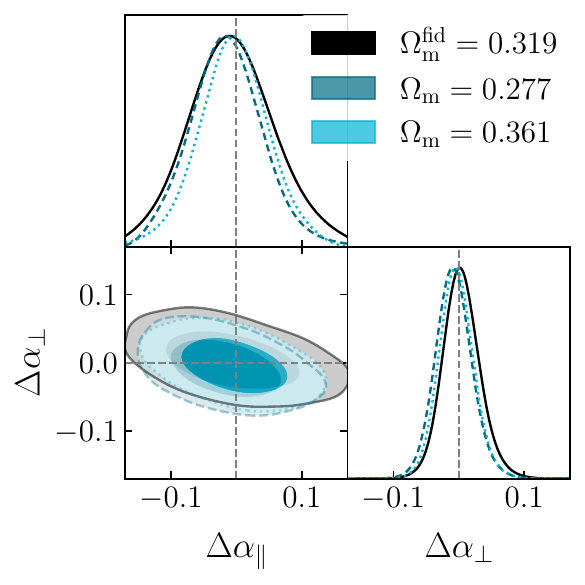}}
\subfloat[\textsc{RecSym}, \(z=0.9\)]{\includegraphics[width=.33\textwidth]{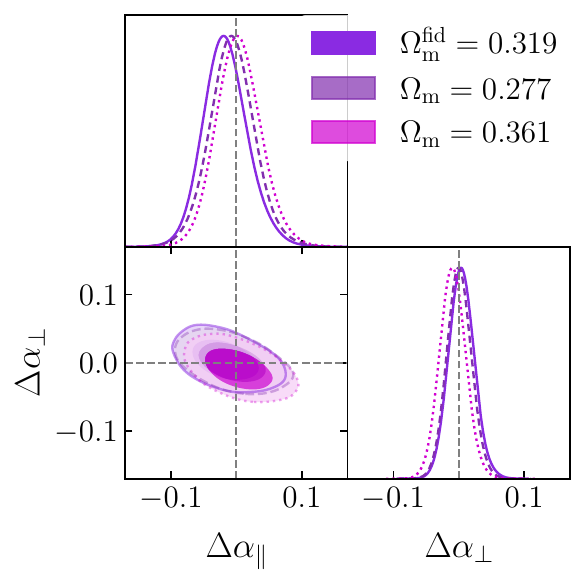}}
\subfloat[\textsc{RecIso}, \(z=0.9\)]{\includegraphics[width=.33\textwidth]{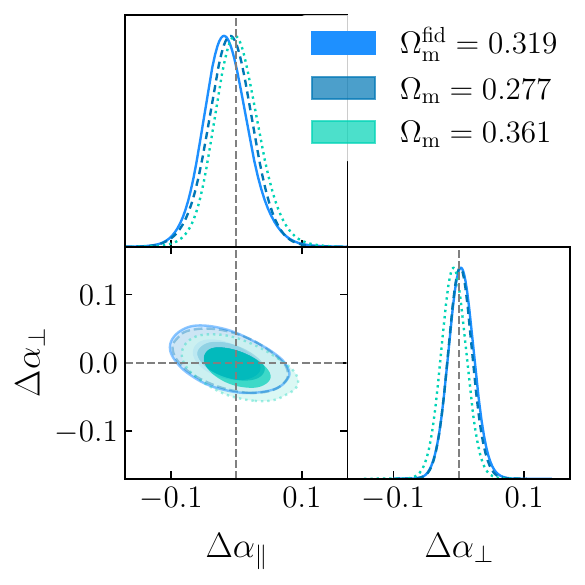}}
\caption{Impact of the fiducial cosmology on \bao constraints at \(z=0.9\). Posterior contours show the \(68\%\) and \(95\%\) credible regions in \(\Delta\alpha_\parallel\) and \(\Delta\alpha_\perp\), after subtracting the deterministic \apdist remapping. Colours refer to three fiducial cosmologies encoded also with different linestyles, where solid corresponds to $\Omega_\mathrm{m}^{\rm fid}$, dashed to $\Omega_\mathrm{m}^{\rm lower}$, and dotted to $\Omega_\mathrm{m}^{\rm upper}$.}
\label{fig:Fid_cosmo_chain09}
\end{figure*}

In reconstruction, the observed density field is smoothed before estimating displacements. In the idealised limit of a continuously sampled field, this smoothing has a known optimal value, \(R_{\mathrm{s}}^{\mathrm{opt}}\simeq 8.4\,h^{-1}\mathrm{Mpc}\), which minimises nonlinear residuals and yields the sharpest \bao features (see Sect.~\ref{sec: Recon_methods}). Real catalogues, however, are sparse and incomplete, so the smoothing scale \(R_{\mathrm{s}}\) must be validated empirically. To this end, we apply both \textsc{RecSym} and \textsc{RecIso} to the \flagshipone mock catalogue, varying \(R_{\mathrm{s}}\in[7,19]\,h^{-1}\mathrm{Mpc}\). This range contains the theoretically motivated scales \(R_{\mathrm{opt}}\) and \(R_{\mathrm{eq}}\) defined in Sect.~\ref{sec: Recon_methods}.

For each combination of redshift, reconstruction algorithm, and smoothing scale, we recompute the covariance matrix using \texttt{BeXiCov+WinCov}, calibrated at the same \(R_{\mathrm{s}}\). This ensures consistent treatment of smoothing and mask-induced mode coupling between the data, the template, and the covariance. To reduce computational cost, best-fit values and uncertainties are derived from a profile-likelihood fit using \texttt{iMinuit} \citep{James_Minuit, iminuit_v2_11_2}. While this approach may underestimate absolute errors, it provides unbiased point estimates and captures the relative variation across smoothing scales.

Figures~\ref{fig:ZA_rec_sym_smoothing_xi_z0p9} and \ref{fig:ZA_rec_iso_smoothing_xi_z0p9} present the reconstructed \tpcf multipoles for \textsc{RecSym} and \textsc{RecIso} at \( z = 0.9 \), as a function of the smoothing scale \( R_\mathrm{s} \). For \textsc{RecSym}, the monopole, quadrupole, and hexadecapole remain stable across the range of \( R_\mathrm{s} \), indicating robustness to the smoothing choice. By contrast, \textsc{RecIso} exhibits increased sensitivity, with lower \( R_\mathrm{s} \) values enhancing small-scale features, particularly in the quadrupole below \( s \lesssim 85\,h^{-1}\mathrm{Mpc} \). This behaviour reflects the differing treatment of \rsd; \textsc{RecSym} retains the large-scale distortions (Eq.~\ref{eq:Pk_ZA_rsd}), while \textsc{RecIso} removes them, leaving residual structure through the smoothing kernel (Eq.~\ref{eq:DA_iso}).

These results highlight two key points. First, \textsc{RecIso} analyses require a dedicated covariance recalibration at each \(R_{\mathrm{s}}\), because the physical clustering component entering the covariance model, \(P^s_{\mathrm{iso}}\), depends explicitly on the smoothing scale, as described in Sect.~\ref{sec:covariances}. The broadband term is then left to absorb only additional smooth, residual shape-dependent effects. Second, \textsc{RecSym} is much less sensitive to the smoothing choice, since the corresponding physical component, \(P^s_{\mathrm{sym}}\), does not retain this explicit \(R_{\mathrm{s}}\) dependence in the adopted template. This avoids repeated covariance recalibration and thereby simplifies pipeline development. These trends persist across redshifts (not shown here for brevity).

To quantify the impact on \bao fits, we focus on \(z=0.9\) and \(z=1.8\), representative of \DRone conditions. At \(z=0.9\), the field is dense (mean particle separation of \(L_{\mathrm{p}}\simeq 8\,h^{-1}\mathrm{Mpc}\)) but more nonlinear; at \(z=1.8\), the field is more linear but sparsely sampled (\(L_{\mathrm{p}}\simeq 15\,h^{-1}\mathrm{Mpc}\)). 
Figures~\ref{fig:ZA_Smoothing_opt_z0p9} and \ref{fig:ZA_Smoothing_opt_z1p8} summarise the \bao fitting results at the lowest and highest redshifts. Shown are the mean values of $\chired$, of \( \alpha_{\parallel} \) and \( \alpha_{\perp} \), as well as their profile-likelihood uncertainties, each averaged over eight realisations at fixed \( R_\mathrm{s} \)\footnote{Given the limited number of realisations, the error on the mean should not be interpreted as a precise characterisation of the underlying distribution.}. The $\chired$ values remain within the statistical acceptance band for all smoothing scales, indicating that the \bao template -- incorporating broadband terms and anisotropic damping (Eq.~\ref{eq:xi_model}) -- adequately accounts for distortions induced by reconstruction. At \( z = 0.9 \), there is a mild preference for smoothing scales around \( R_\mathrm{s} \approx 7\)--\(9\,h^{-1}\mathrm{Mpc} \), in agreement with theoretical expectations. At \( z = 1.8 \), the dependence on \( R_\mathrm{s} \) is negligible. Across the full range, the inferred scaling parameters \( \alpha_{\parallel} \) and \( \alpha_{\perp} \), as well as their uncertainties remain stable, with no significant differences between \textsc{RecSym} and \textsc{RecIso}.

In summary, at the level of precision expected for \DRone, the \bao scale is only weakly sensitive to the smoothing scale, provided \(R_{\mathrm{s}} \in [R_{\mathrm{opt}}, R_{\mathrm{eq}}]\), leaving some flexibility in the choice of this parameter. We adopt \(R_{\mathrm{s}} = 15\,h^{-1}\mathrm{Mpc}\) as our baseline for three main reasons. First, it lies closer to the equivalence scale \(R_{\rm eq}\) discussed in Sect.~\ref{sec:ZA_nonlinear}, where the reconstructed signal can be described by a single damping parameter, making it more consistent with the BAO template used to extract the AP parameters (see Sect.~\ref{sec:fitting_templates}). Second, as shown in Fig.~\ref{fig:MultiGrid_massassignment}, the grid size used for reconstruction must remain smaller than the smoothing scale, so a slightly larger \(R_{\mathrm{s}}\) allows the use of a coarser reconstruction grid and reduces the computational cost when analysing large suites of mocks. Third, this choice maintains consistency with previous BAO analyses in the literature and facilitates direct comparison with earlier results \citep{Ross2017,Bautista2018,Paillas2025}.


\subsection{Fiducial cosmology}\label{sec:fiducial_cosmo}

The choice of fiducial cosmology enters reconstruction and \bao fitting at several levels. It determines the redshift-to-distance conversion, imprinting geometric distortions in the estimated density field used to infer displacements. Furthermore, it sets the growth rate $f$, which governs the amplitude of those displacements (see Eq.~\ref{eq:Multigrd_pot} for both effects), and, through comoving mapping, it affects the effective survey volume and the clustering amplitude used to calibrate the covariance (Sect.~\ref{sec:covariances}). Because these components are not explicitly marginalised in our pipeline, a mismatch between the true and fiducial cosmology can bias the reconstructed field and the recovered \bao scale, and distort the inferred uncertainties.

To quantify this impact, we repeat the full analysis pipeline using three fiducial cosmologies. These are the \Euclid baseline value \(\Omega_{\rm m}^{\rm fid} = 0.319\), and two variants, \(\Omega_{\rm m}^{\rm low} = 0.277\) and \(\Omega_{\rm m}^{\rm high} = 0.361\), selected to span \(\pm 5\sigma\) deviations based on \Planck constraints \citep{Planck2020}, as summarised in Table~\ref{table:Fiducial_cosmo_exp}. For each combination of redshift, reconstruction type, and fiducial cosmology, we generate new mock catalogues matching the expected \DRone volume (see Sect.~\ref{sec:redshift_cat}), apply reconstruction and \tpcf measurements, and recompute the covariance matrix using \texttt{BeXiCov+WinCov} for the updated fiducial model. As shown in Appendix~\ref{app:fid_cov}, this last step is essential, covariances calibrated on an incorrect cosmology with no data anchoring (either mock-based or purely analytical with a theoretical prediction for the input power spectrum) can misestimate \bao errors by \(\sim20\%\). Our semi-analytical, data-anchored covariance reduces this sensitivity, limiting the residual miscalculation below \(\sim5\%\). 

Because the \apdist remapping deterministically shifts the \bao feature, we assess the bias via
\(\Delta\alpha_\parallel \equiv \alpha_\parallel - \alpha_\parallel^{\rm exp}\) and \(\Delta\alpha_\perp \equiv \alpha_\perp - \alpha_\perp^{\rm exp}\), where the expected scaling factors are given by
\begin{equation}\label{eq:AP_exp}
    \alpha_\perp^{\exp} = \frac{r_{\mathrm{s}}^{\rm f}}{r_{\mathrm{s}}^{\rm t}} \quad \mathrm{and} \quad
    \alpha_\parallel^{\exp} = \frac{H^{\rm t}(\bar z)}{H^{\rm f}(\bar z)} \frac{r_{\mathrm{s}}^{\rm f}}{r_{\mathrm{s}}^{\rm t}} \, .
\end{equation}
The values of \(r_{\mathrm{s}}^{\rm f}\) are estimated from the CMB-calibrated sound horizon in a model with extra relativistic species \citep[see Eq.~17 in][]{AubourgEtAl2015_r_s} and reported in Table~\ref{table:Fiducial_cosmo_exp}. In this convention, an unbiased recovery corresponds to posteriors centred at \(\Delta\alpha_\parallel = \Delta\alpha_\perp = 0\). Summary statistics for our chains as well as plots are obtained with the \texttt{GetDist}\footnote{\url{https://github.com/cmbant/getdist}} software package \citep{GetDist}.

\begin{table}[]
\centering
\caption{Expectation values for $\alpha_\parallel$, $\alpha_\perp$, $f$, and $r_\mathrm{s}$ for different fiducial cosmologies.}
\renewcommand{\arraystretch}{1.4}
\smallskip
\label{table:Fiducial_cosmo_exp}
\smallskip
\begin{tabular}{|c|cccc|}
\hline  
\rowcolor{blue!5}
 $\Omega_\mathrm{m}$ & $\alpha^\mathrm{exp}_{\parallel_{\phantom{A}_{\phantom{A}}}}$ & $\alpha^\mathrm{exp}_\perp$ & $f$ & $r_\mathrm{s} \; [\mathrm{Mpc}]$ \\
\hline
$\Omega^\mathrm{fid}_\mathrm{m} =  0.319$ & 1.000 & 1.000 & 0.862 & 147.2\\
$\Omega^\mathrm{lower}_\mathrm{m} =  0.277$ & 0.990 & 1.035 & 0.838 &  152.3 \\
$\Omega^\mathrm{upper}_\mathrm{m} =  0.361$ & 1.011 & 0.970 &  0.882 & 142.8\\
\hline
\end{tabular}
\end{table}

\begin{table}[]
\caption{Shift in the best-fit values of \apdist parameters with respect to the expectation values (compare Table~\ref{table:Fiducial_cosmo_exp}). The table compares results for analyses done pre- and  post-reconstruction (\textsc{RecSym} and \textsc{RecIso}), as well as for different fiducial cosmologies.}
\renewcommand{\arraystretch}{1.2}
\smallskip
\label{table:fid_cosmo_BF}
\smallskip
\centering
\begin{tabular}{|c|l|r|r|r|r|r|}
\hline
\rowcolor{blue!5}
 Rec. & $\Omega_\mathrm{m}^{\rm fid}$ & $\Delta\alpha_\parallel$ & $\sigma_{\alpha_\parallel}$ & $\Delta\alpha_\perp $ & $\sigma_{\alpha_\perp}$  & $\mathrm{FoB}$ \\
 \rowcolor{blue!5}
  & & & & & &\\[-12pt]   
 \hline
\rowcolor{blue!5}
$z=0.9$ & & & & &  &\\
\hline 
\textsc{PreRec} & $\Omega^{\mathrm{true}}_\mathrm{m}$ & $-0.7$  &  $6.7$ & $0.4$  &  $2.8$ & 0.15 \\

 & $\Omega^\mathrm{lower}_\mathrm{m}$ & $-1.5$  &  $5.4$ & $-0.8$  &  $2.7$ & $0.48$\\

 & $\Omega^\mathrm{upper}_\mathrm{m}$ & $-0.5$ &  $5.7$ & $-0.4$  &  $2.9$ & $0.20$ \\

\hline
\textsc{RecSym} & $\Omega^\mathrm{true}_\mathrm{m}$ &  $-1.6$  &  $3.4$ & $0.4$  &  $2.0$ & 0.47 \\

 & $\Omega^\mathrm{lower}_\mathrm{m}$ & $-0.7$ &  $3.4$ & $-0.1$  &  $1.8$ & 0.19\\

 & $\Omega^\mathrm{upper}_\mathrm{m}$ &  $0.4$  &  $3.6$ & $-0.9$  &  $2.1$ & 0.47\\

\hline
\textsc{RecIso} & $\Omega^\mathrm{true}_\mathrm{m}$  & $-1.5$  &  $3.5$ & $0.4$  &  $2.0$ & 0.42 \\

 &  $\Omega^\mathrm{lower}_\mathrm{m}$  & $-0.9$  &  $3.0$ & $0.2$  &  $1.8$  & 0.26 \\

 &  $\Omega^\mathrm{upper}_\mathrm{m}$  & $0.1$  &  $3.5$ & $-0.8$  &  $2.0$ & 0.42 \\

 \hline
 \rowcolor{blue!5}
$z=1.8$ & & & & &  &\\
\hline
\textsc{PreRec} & $\Omega^\mathrm{true}_\mathrm{m}$ & $-2.0$  &  $3.4$ & $0.6$  &  $2.0$ & 0.34 \\

 & $\Omega^\mathrm{lower}_\mathrm{m}$ & $1.2$  &  $3.1$ & $0.6$  &  $2.0$ & 0.47\\

 & $\Omega^\mathrm{upper}_\mathrm{m}$ & $2.0$ &  $3.7$ & $-0.1$  &  $2.5$ & 0.62 \\

\hline
\textsc{RecSym} & $\Omega^\mathrm{true}_\mathrm{m}$ &  $0.6$  &  $2.5$ & $-0.1$  &  $1.7$ & 0.23\\

 & $\Omega^\mathrm{lower}_\mathrm{m}$ & $0.5$ &  $2.3$ & $-0.1$  &  $1.6$ & 0.23\\

 & $\Omega^\mathrm{upper}_\mathrm{m}$ &  $1.2$  &  $2.8$ & $-0.7$  &  $2.0$ & 0.50\\

\hline
\textsc{RecIso} & $\Omega^\mathrm{true}_\mathrm{m}$  & $0.5$  &  $2.5$ & $-0.0$  &  $1.6$ & 0.24\\

 &  $\Omega^\mathrm{lower}_\mathrm{m}$  & $0.5$  &  $2.3$ & $-0.1$  &  $1.6$  & 0.22\\

 &  $\Omega^\mathrm{upper}_\mathrm{m}$  & $1.3$  &  $2.9$ & $-0.7$  &  $1.9$ & 0.55 \\

\hline
\end{tabular}
\tablefoot{For spatial compactness of the table, we reduced the number of leading zeros via multiplying the columns of $\Delta\alpha_{\parallel}$, $\sigma_{\alpha_{\parallel}}$, $\Delta\alpha_{\perp}$, and $\sigma_{\alpha_{\perp}}$ by 100.}
\end{table}

Figure~\ref{fig:Fid_cosmo_chain09} presents the \( 68\% \) and \( 95\% \) credible regions for \( \Delta\alpha_\parallel \) and \( \Delta\alpha_\perp \) at \( z = 0.9 \), derived from the mean \tpcf of the mocks under varying fiducial cosmologies. The analysis accounts for the \apdist remapping by subtracting the deterministic shift associated with each fiducial model. In all cases, the posteriors are centred near zero, indicating no detectable bias in the recovered \bao scale at the precision of \DRone. The constraints remain approximately Gaussian, with a stable correlation structure between \( \Delta\alpha_\parallel \) and \( \Delta\alpha_\perp \), and only weak sensitivity to the choice of fiducial cosmology. Best-fit values and marginalised uncertainties are reported in Table~\ref{table:fid_cosmo_BF}. The residual bias in the $(\alpha_\parallel,\alpha_\perp)$ plane is parametrised by  the Figure of Bias (FoB), defined as
\(
\mathrm{FoB} \equiv \sqrt{\Delta\boldsymbol{\alpha}^{\rm T}\,\tens{S}^{-1}\,\Delta\boldsymbol{\alpha}} \, ,
\)
where \(\Delta\boldsymbol{\alpha}\) is the vector of shifts of the best-fit AP parameters with respect to their expected values, and \(\tens{S}\) represents their covariance matrix estimated from the posterior samples.

Across all tested configurations, 
the largest $\mathrm{FoB}$ is given by $\mathrm{FoB}=0.62$, obtained for \textsc{PreRec} at $z=1.8$ with $\Omega_\mathrm{m}^{\rm upper}$. All other shifts lie below the acceptance threshold, $\mathrm{FoB}_{\rm th}=1.52/\sqrt{8}=0.54$, corresponding to the level at which the two-dimensional best-fit AP parameters, estimated from the mean of 8 mock realisations, remain indistinguishable from statistical noise.\footnote{In more detail, the acceptance threshold of $\mathrm{FoB}=1.52$ corresponds to the $68.3$ percentile of a $\chi^2$-distribution with two degrees of freedom computed via the percent-point function.} We therefore find no compelling evidence for a detectable systematic bias. Reconstruction consistently improves the parameter constraints. Defining the improvement factor as \(I_{i}\equiv \sigma^{\rm pre}_{i}/\sigma^{\rm post}_{i}\) for \(i\in\{\parallel,\perp\}\), we find that at $z=0.9$ \textsc{RecSym} yields median gains of \(I_\parallel \simeq 1.59\) and \(I_\perp \simeq 1.40\), with respective ranges \([1.58, 1.97]\) and \([1.38, 1.50]\) across fiducial cosmologies. \textsc{RecIso} instead, achieves \(I_\parallel \simeq 1.64\) and \(I_\perp \simeq 1.45\), with respective ranges [1.63, 1.91] and [1.40, 1.50]. At $z=1.8$, gains are milder -- owing to the reduced nonlinear displacement field at high redshift \((\Sigma_\mathrm{nl}\) is proportional to the growth factor) -- but more consistent across fiducial cosmologies as both \textsc{RecSym} and \textsc{RecIso} yield \(I_\parallel \simeq 1.35\), \(I_\perp \simeq 1.25\).

These trends are consistent with our numerical tests on theory data vectors (Appendix~\ref{app:fid_cov}). Increasing \(\Omega_{\rm m}\) tends to increase the predicted \bao uncertainties, making reconstruction gains slightly less pronounced. This behaviour is clearly visible at \(z=1.8\) for pre- and post-reconstruction as well as at \(z=0.9\) for post-reconstruction only, where the increased linearity of the clustering signal improves model fidelity. At \(z=0.9\) pre-reconstruction, the \bao template captures nonlinear signature less accurately and the monotonic dependence of the uncertainty on \(\Omega_{\rm m}\) is less apparent, which blurs the corresponding improvement factors.

Overall, our pipeline exhibits excellent robustness to the assumed fiducial cosmology as we observe no statistically significant \bao bias and only modest, well-understood variations in uncertainty estimates. For fixed \(R_{\rm s}\), both \textsc{RecSym} and \textsc{RecIso} yield consistent \bao constraints.

\subsection{Cosmological inference from reconstructed \bao}
\label{sec:BAO_constraints}
\begin{table}[]
\centering
\caption{Posterior constraints on the parameter set $\{\Omega_{\mathrm{m}},\,H_0 r_{\mathrm{s}}\}$ for each individual redshift bin as well as considering the joint redshift analysis. The final column reports the increase in \fom relative to the pre-reconstruction (\textsc{PreRec}) case.}
\renewcommand{\arraystretch}{1.4}
\smallskip
\label{tab:Cosmo_const}
\smallskip
\begin{tabular}{|c|c|c|c|}
\hline
\rowcolor{blue!5}
 & & &\\[-12pt]  
\rowcolor{blue!5}
Redshift & $\Omega_{\mathrm{m}}$ & $10^{-2}H_0 r_{\mathrm{s}}\ [\mathrm{km}\,\mathrm{s}^{-1}]$ & $\frac{\mathrm{FoM}^\mathrm{Rec. Type}}{\mathrm{FoM}^\mathrm{PreRec}}$ \\
\rowcolor{blue!5}
 & & &\\[-12pt]   
\hline
\rowcolor{blue!5}
\textsc{PreRec} & & & \\
\hline
0.9 & $0.42^{+0.13}_{-0.25}$ & $95^{+10}_{-12}$ & -- \\
1.2 & $0.42^{+0.11}_{-0.20}$ & $94^{+9}_{-11}$ & -- \\
1.5 & $0.34^{+0.07}_{-0.13}$ & $98^{+9}_{-9}$ & -- \\
1.8 & $0.35^{+0.06}_{-0.10}$ & $97^{+8}_{-8}$ & -- \\
Joint & $0.33^{+0.04}_{-0.05}$ & $98^{+4}_{-4}$ & -- \\
\hline
\rowcolor{blue!5}
\textsc{RecSym} & & & \\
\hline
0.9 & $0.39^{+0.08}_{-0.13}$ & $95\pm 7$ & 3.1 \\
1.2 & $0.32^{+0.10}_{-0.12}$ & $99\pm 6$ & 3.5 \\
1.5 & $0.33^{+0.06}_{-0.09}$ & $99\pm 6$ & 2.6 \\ 
1.8 & $0.32^{+0.05}_{-0.07}$ & $99\pm 6$ & 2.1 \\
Joint & $0.32^{+0.03}_{-0.03}$ & $99\pm 3$ & 2.1 \\
\hline
\rowcolor{blue!5}
\textsc{RecIso} & & & \\
\hline
0.9 & $0.39^{+0.08}_{-0.13}$ & $95\pm 7$ & 3.0 \\
1.2 & $0.32^{+0.07}_{-0.10}$ & $99\pm 7$ & 3.0 \\
1.5 & $0.32^{+0.05}_{-0.08}$ & $99\pm 6$ & 3.1 \\
1.8 & $0.32^{+0.05}_{-0.07}$ & $99\pm 6$ & 2.0 \\
Joint & $0.31^{+0.03}_{-0.03}$ & $99\pm 3$ & 2.2 \\
\hline
\end{tabular}
\end{table}

\begin{figure*}
\centering
\subfloat[\textsc{PreRec}]{\includegraphics[width=.33\textwidth]{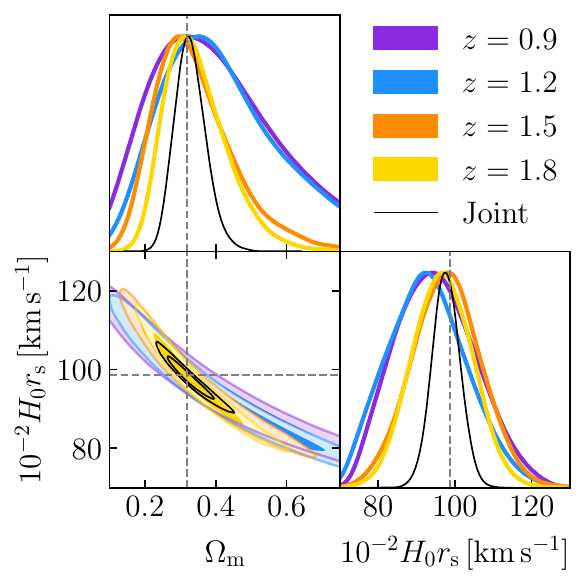}}
\hfill
\subfloat[\textsc{RecSym}]{\includegraphics[width=.33\textwidth]{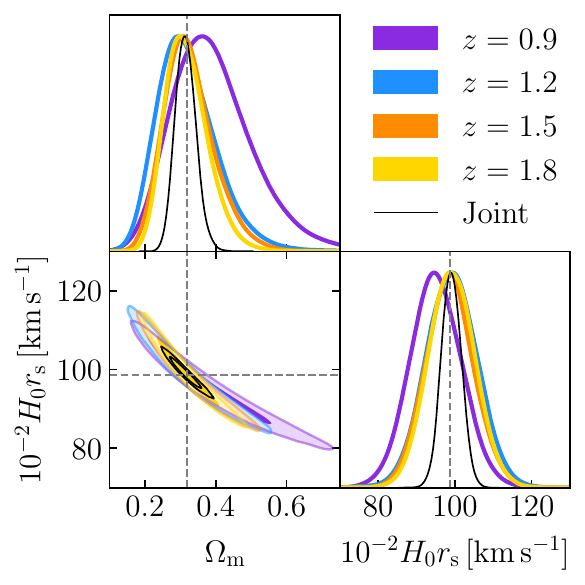}}
\hfill
\subfloat[\textsc{RecIso}]{\includegraphics[width=.33\textwidth]{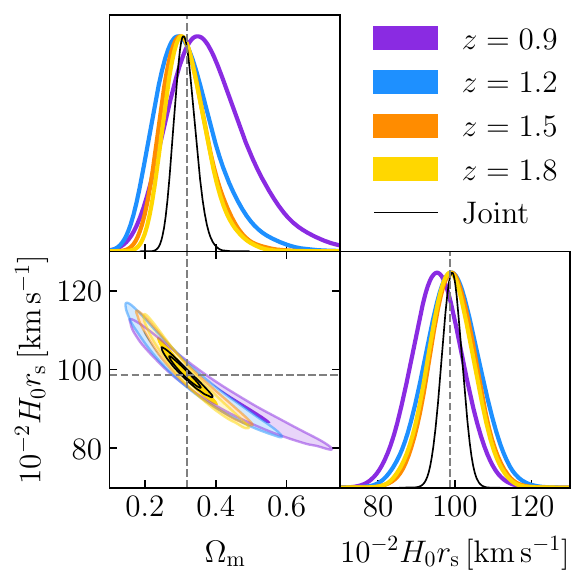}}
\caption{Posterior constraints on $\Omega_{\mathrm{m}}$ and $H_0 r_{\mathrm{s}}$ for pre-reconstruction (left), \textsc{RecSym} (middle), and \textsc{RecIso} (right). Coloured contours refer to the $68\%$ and $95\%$ credible regions for $z \in \{0.9, 1.2, 1.5, 1.8\}$ and black curves show the combined constraints (assuming independence). Grey dashed lines denote the mock fiducial values of $\Omega_{\mathrm{m}} = 0.319$ and $H_0 r_{\mathrm{s}} = 98.6\times100\,\mathrm{km\,s^{-1}}$.}
\label{fig:Cosmo_const}
\end{figure*}
We assess the cosmological implications of the reconstructed \bao measurements by translating them into constraints on the matter density and expansion rate. This provides a full end-to-end validation of the \Euclid \bao cosmology pipeline and offers a quantitative precision forecast for \DRone.

We adopt a flat $\Lambda$CDM framework and vary only the parameters most directly constrained by the \bao signal, $\left\{\Omega_{\mathrm{m}},\,H_0 r_{\mathrm{s}}\right\}$, while fixing all others to their \Euclid fiducial values. This setup follows the DESI  analysis \citep{DESI_BAO_Y1,DESI_BAO_DR2} and reflects the weak dependence of \bao-only constraints on additional cosmological parameters \citep{AubourgEtAl2015_r_s}. Parameter inference is performed directly in the \apdist parameters space, using a Gaussian likelihood built from the posterior mean and sample covariance of $(\alpha_\parallel,\,\alpha_\perp)$ obtained from the \bao template fits (Sect.~\ref{sec:fiducial_cosmo}).

For each redshift bin, we compute the mean mock signal assuming the fiducial cosmology and evaluate the likelihood from $\chi^2(\Omega_{\mathrm{m}},H_0 r_{\mathrm{s}}) = [\vec{\alpha}_\mathrm{data}-\vec{\alpha}_\mathrm{model}]^{\top} \, \tens{C}_\alpha^{-1} \, [\vec{\alpha}_\mathrm{data}-\vec{\alpha}_\mathrm{model}],$
where $\vec{\alpha}_\mathrm{data}$ denotes the posterior mean of the $\vec{\alpha}=(\alpha_\parallel,\alpha_\perp)$ vector, and $\vec{\alpha}_\mathrm{model}$ is its theoretical prediction computed at each \mcmc step.
Sampling is performed with the public code \texttt{CosmoSampler}\footnote{\url{https://gitlab.com/esarpa1/cosmosampler}}, assuming uniform in the range $\Omega_{\mathrm{m}} \in [0,1]$ and $H_0 r_{\mathrm{s}}/(100\,\mathrm{km\,s^{-1}}) \in [30,150]$.
Each redshift bin is analysed independently, and combined constraints are obtained by stacking the posteriors under the assumption of statistical independence.

Figure~\ref{fig:Cosmo_const} compares the cosmological posteriors before and after applying \textsc{RecSym} and \textsc{RecIso} reconstruction. Constraints from the individual redshift bin analyses are shown alongside the combined posterior representing the final \Euclid \bao-only result. In all cases, the recovered parameters remain consistent within $1\sigma$ of the mock fiducial values, confirming the absence of significant bias at \DRone-level precision. As expected, the constraints tighten towards higher redshift, where the \bao signal is less affected by nonlinear evolution and the modelling becomes more accurate. The contours exhibit mild non-Gaussianity and redshift-evolving degeneracy orientation, reflecting the changing radial and transverse sensitivity of the \bao signal \citep{DESI_BAO_DR2}. This geometric complementarity yields more compact, nearly Gaussian joint posteriors when stacking bins.

Both reconstruction schemes lead to substantial precision gains. As shown in Table~\ref{tab:Cosmo_const}, the \fom, which is defined as \citep[see][]{FOM} $\mathrm{FoM} = 1/\sqrt{\det \tens{S}}$, improves by a factor $\sim3$ in each redshift bin compared to pre-reconstruction. The only exception is the bin at $z = 1.8$, where the improvement reduces to a factor of 2, owing to the signal already being close to linear, thus limiting the impact of the reconstruction. Increasing the \fom by a factor of 3 corresponds to an approximate threefold increase in effective survey volume. Results are statistically consistent across reconstruction types, confirming their robustness over redshift and analysis choices. In the joint case, the improvement reduces to $\sim2$, as most of the gain stems from combining bins with complementary degeneracies.

At last, the reconstructed \Euclid-like \bao measurements deliver unbiased cosmological constraints with significantly enhanced precision, in line with previous SDSS-based analyses \citep{Padmanabhan2012}. By combining four tomographic bins at $z \in \{0.9, 1.2, 1.5, 1.8\}$, we forecast $\sim$10\% relative precision on $\Omega_{\mathrm{m}}$ and $\sim$3\% on $H_0 r_{\mathrm{s}}$, independent of the reconstruction choice. These results demonstrate the accuracy and scientific readiness of the \Euclid \bao pipeline ahead of \DRone.

\section{Conclusions}
\label{sec:Discussion_Conclusions}

We have presented the first end-to-end execution of the \Euclid \bao analysis pipeline, encompassing the full chain from \bao reconstruction and \tpcf measurement to cosmological inference. Compared to current state-of-the-art implementations, our approach enhances both computational efficiency and robustness by combining physically motivated modelling with modern inference techniques. Parameter fitting is performed using an emulator-based model evaluator (\texttt{Bora.jl}) and a \hmc sampler (\texttt{NUTS}), yielding a speed-up of approximately a factor of 500 over standard \mcmc methods without compromising accuracy. In parallel, our semi-analytical covariance estimator (\texttt{BeXiCov+WinCov}) eliminates the need for extensive mock realisations -- requiring only eight -- and reduces bias in \bao uncertainty estimates from $\sim 20\%$ to below $5\%$, even for cosmologies that differ by five times the Planck CMB uncertainties from the fiducial model.

Motivated by theoretical differences between the two standard Zeldovich-based reconstruction schemes, we assessed the performance of \textsc{RecSym} and \textsc{RecIso} using $8\times4$ mock galaxy catalogues derived from the \Euclid-like \flagshipone simulation. These catalogues reproduce the expected nonlinear galaxy clustering for \Euclid \DRone across four redshift bins centred at $z \in \{0.9,1.2,1.5,1.8\}$. We first tested the sensitivity of the reconstructed \bao signal to the smoothing scale $R_{\mathrm s}$ used in the displacement estimation. Within the range defined by the theory-motivated optimum $R_\mathrm{opt}$ and the empirical threshold $R_\mathrm {eq}$ -- where the \tpcf retains pre-reconstruction scale dependence -- we observed no significant bias or degradation in the recovered \bao scale or its uncertainty, and both reconstruction schemes performed consistently. We then examined the impact of the fiducial cosmology, varying $\Omega_{\mathrm m}$ by up to 5 times the standard deviation of Planck \citep{Planck2020} around the true value of \flagshipone. Across all variations, the recovered \bao scale and its uncertainty remained stable, confirming the robustness of both the reconstruction and inference stages of the pipeline.

Finally, we derived the cosmological constraints achievable from a \bao-only analysis of \DRone-quality data. Focusing on the pair $\{\Omega_{\mathrm m},\,H_0 r_{\mathrm s}\}$, we found that \Euclid-like \bao measurements yield unbiased results, with reconstruction improving the \fom of single-bin constraints by a factor of $\sim 3$, equivalent to tripling the effective survey volume. When combining all four redshift bins, we forecast $\sim 10\%$ relative precision on $\Omega_{\mathrm m}$ and $\sim 3\%$ on $H_0 r_{\mathrm s}$, independently of the reconstruction scheme. These results validate the accuracy and scientific maturity of the \Euclid \bao pipeline in preparation for \DRone, in excellent agreement with legacy SDSS benchmarks \citep{Padmanabhan2012}.


Our results support the use of relatively large smoothing scales. To be effective, $R_{\mathrm s}$ must exceed the grid-cell size used for density field interpolation, since the computational cost of reconstruction scales as $N_{\mathrm{cell}}^3$. A larger $R_{\mathrm s}$ allows the use of coarser grids and faster computation, provided that it does not excessively smooth the survey mask or blend regions with different sampling densities. As for the reconstruction scheme, we observe no significant numerical differences between \textsc{RecSym} and \textsc{RecIso}. However, \textsc{RecIso} introduces a scale-dependent residual that requires recomputation of the covariance matrix for each $R_{\mathrm s}$ value. We therefore recommend using \textsc{RecSym} during testing and validation phases to minimise computational overhead.

The work presented here establishes the methodological foundation of the \Euclid \DRone \bao analysis. By delivering a fast, accurate, and bias-free reconstruction-to-inference pipeline, this study ensures that the first cosmological results from \Euclid will be derived from a robust and fully validated framework. Looking beyond \DRone, the techniques developed here -- including emulator-based modelling, Hamiltonian sampling, and semi-analytical covariance estimation -- provide a scalable foundation for high-precision cosmological inference from future \Euclid data releases.
While this work focuses on methodological validation with idealised \Euclid-like mocks, future analyses based on more realistic \DRone simulations will include redshift errors, survey geometry, and model refinements to account for these systematics.

\begin{acknowledgements}

\AckEC
The majority of the analysis carried out in this manuscript has been produced by a joint effort among several Euclid members and work centers. This research made use of matplotlib, a Python library for publication quality graphics \citep{Hunter:2007_matplotlib}.

\end{acknowledgements}

\bibliography{Euclid, AandA}

\begin{appendix}

\section{Reconstructed power spectrum for biased tracers}
This appendix derives the pre- and post-reconstruction power spectra for biased tracers within the Zeldovich approximation. While related treatments exist in the literature -- for example \citet{Chen2019} -- a complete derivation including bias terms is not explicitly presented elsewhere. We therefore provide it here for completeness. We begin in real space and adopt the large-separation (leading-order) expansion throughout. Einstein summation over repeated Latin indices is implied and for brevity of the upcoming expressions we often use 
\begin{equation}
    \mathcal{F}\left[f(q)\right] \equiv \int \diff^3 q \, \mathrm{e}^{-\mathrm{i}\vec{k}\cdot \vec{q}} f(q)
\end{equation}
to denote Fourier transformations from $q$- to $k$-space.
Furthermore, when the argument of the smoothing kernel $\mathcal{S}$ is evident from the context of the remaining formula we will omit it.

\subsection{Real space}\label{app:recon_real}

We model the observed tracer overdensity as
\begin{equation}\label{eq:delta_obs_lag}
\delta^\mathrm{g}_{\mathrm{obs}}(\vec{k})=\int \diff^3q \, \left[1+b_\mathrm{L}\,\delta_{\mathrm{lin}}(\vec{q})\right] \, \mathrm{exp}\left\{-\mathrm{i}\,\vec{k}\cdot\left[\vec{q}+\vec{\psi}_{\mathrm{obs}}(\vec{q})\right]\right\} \, ,
\end{equation}
which extends the computations of Sect.~\ref{sec:ZA_nonlinear} to include a non-zero Lagrangian linear bias \(b_\mathrm{L}\), as discussed in the first-order limit by \citet{Chen2019}. As before, \(\vec{\psi}_{\mathrm{obs}}\) denotes the Zeldovich displacement (see Eq.~\ref{eq:ZA_disp}), treated as a linear functional of the Gaussian field \(\delta_{\mathrm{lin}}\).
By definition, the galaxy power spectrum reads
\begin{equation}\label{eq:P_obs_initial}
P^\mathrm{g}(k)= \mathcal{F}\left[\left\langle \left(1+b_\mathrm{L}\delta_1\right) \left(1+b_\mathrm{L}\delta_2\right)\,\mathrm{e}^{-\mathrm{i}\vec{k}\cdot\Delta\vec{\psi}}\right\rangle \right]\, ,
\end{equation}
with \(\delta_i\equiv\delta_{\mathrm{lin}}(\vec{q}_i)\), \(\vec{q}\equiv\vec{q}_2-\vec{q}_1\), and \(\Delta\vec{\psi}\equiv\vec{\psi}_{\mathrm{obs}}(\vec{q}_2)-\vec{\psi}_{\mathrm{obs}}(\vec{q}_1)\).

To expose the dependence on $\Plin(k)$, we first expand the product of the two expressions in square brackets and evaluate its product with the displacement exponential $\mathrm{e}^{-\mathrm{i}\vec{k}\cdot\Delta\vec{\psi}}$ term by term.
Using the cumulant expansion for Gaussian fields we get
\begin{equation}
\left\langle \mathrm{exp}\left(-\mathrm{i}\,\vec{k}\cdot\Delta\vec{\psi}\right)\right\rangle=\mathrm{exp}\left[-k_i k_j\,A_{ij}(\vec{q})/2\right] \, ,
\end{equation}
with the 2-point correlator of the displacement field difference defined as $A_{ij}(\vec{q})\equiv\langle \Delta\psi_{\mathrm{obs},i}\,\Delta\psi_{\mathrm{obs},j}\rangle$, where $i,j$ label components of a vector and the moment is identical to the cumulant due to a vanishing mean of $\Delta \psi_{\mathrm{obs}}$.
For the mixed terms we use the identity for a jointly Gaussian scalar $X$ and vector $\vec{Y}$ given by
\begin{equation}
    \left\langle X\,\mathrm{exp}\left(\mathrm{i} \, \boldsymbol{\lambda}\cdot\vec{Y}\right)\right\rangle=\mathrm{i} \, \lambda_j\,\langle X Y_j\rangle \, \mathrm{exp}\left(-\lambda_k\lambda_l\,\langle Y_k Y_l\rangle/2\right) \, .
\end{equation}
Using $X=\delta_1$, $\vec{Y}=\Delta\vec{\psi}$, and $\boldsymbol{\lambda}=-\vec{k}$, we obtain
\begin{equation}
\begin{split}
    \left\langle \delta_1\,\mathrm{exp}\left(-\mathrm{i}\,\vec{k}\cdot\Delta\vec{\psi}\right)\right\rangle
        &=-\mathrm{i}\,k_i\,\left\langle \delta_1\,\Delta\psi_{\mathrm{obs},i}\right\rangle \, \mathrm{exp}\left[-k_i k_j \,A_{ij}(\vec{q})/2\right] \\
        &\equiv -\mathrm{i}\,k_i\,U_i(\vec{q})\,\mathrm{exp}\left[-k_i k_j\, A_{ij}(\vec{q})/2\right] \, ,
\end{split}
\end{equation}
where we defined
\begin{equation}
\begin{split}
    U_i(\vec{q})&\equiv\left\langle \delta_1\,\Delta\psi_{\mathrm{obs},i}\right\rangle \\
    & =\left\langle \delta_{\mathrm{lin}}(\vec{q}_1)\,[\psi_{\mathrm{obs},i}(\vec{q}_2)-\psi_{\mathrm{obs},i}(\vec{q}_1)]\right\rangle \\
    &= \left\langle \delta_{\mathrm{lin}}(\vec{q}_1)\,\psi_{\mathrm{obs},i}(\vec{q}_2)\right\rangle \, .
\end{split}
\end{equation}
The zero-lag cross term (i.e., a moment of two fields evaluated at the same point) vanishes due to the statistical homogeneity of the fields and the odd parity of the cross-correlation at zero lag.
By the same argument and translational invariance, the term involving only $\delta_2$ becomes
\begin{equation} 
    \left\langle \delta_2\,\mathrm{exp}\left(-\mathrm{i}\,\vec{k}\cdot\Delta\vec{\psi}\right)\right\rangle=-\mathrm{i}\,k_i\,U_i(\vec{q})\,\mathrm{exp}\left[-k_i k_j\, A_{ij}(\vec{q})/2\right] \, .
\end{equation}
Finally for the term involving the product of $\delta_1$ with $\delta_2$, we obtain
\begin{equation}\label{eq:delta_cross}
    \left\langle \delta_1\delta_2\,\mathrm{exp}\left(-\mathrm{i}\,\vec{k}\cdot\Delta\vec{\psi}\right)\right\rangle=\xi_{\mathrm{lin}}(\vec{q})\,\mathrm{exp}\left[-k_i k_j \, A_{ij}(\vec{q})/2\right],
\end{equation}
where \(\xi_\mathrm{lin}\) is the \tpcf of \(\delta_\mathrm{lin}\).

It is convenient to write the 2-point correlator of the displacement field as
\begin{equation}
    A_{ij}(\vec{q}) = 2\left[A_{ij}-\xi_{ij}(\vec{q})\right] \, ,
\end{equation}
with
\begin{align}
    & A_{ij} \equiv \left\langle \psi_{\mathrm{obs},i}(0) \, \psi_{\mathrm{obs},j}(0) \right\rangle \\
    \mathrm{and} \quad & \xi_{ij}(\vec{q}) \equiv \left \langle \psi_{\mathrm{obs},i}(0) \, \psi_{\mathrm{obs},j}(\vec{q}) \right\rangle \, ,
\end{align}
such that the exponential of the full $A_{ij}(\vec{q})$ factorises into a product involving $A_{ij}$ and $\xi_{ij}$.
Here, $A_{ij} $ is the zero-lag term and $\xi_{ij}(\vec{q})$ is the displacement correlation function.
In the large-separation limit, which is given by \(\xi_{ij}(\vec{q})\ll A_{ij}\), we can expand
\begin{equation}
    \mathrm{exp}\left[\,k_i k_j \xi_{ij}(\vec{q})\right] \simeq 1+k_i k_j \xi_{ij}(\vec{q}) + \mathcal{O} \left(\xi^2\right) \, .
\end{equation}
Inserting Eq.~\eqref{eq:delta_cross} into Eq.~\eqref{eq:P_obs_initial}, we obtain
\begin{equation}
\begin{split}
    P^\mathrm{g}(k) = &\mathrm{exp}\left(-k_i k_j A_{ij}\right) \\
    & \times \mathcal{F}\left[b_\mathrm{L}^2\,\xi_{\mathrm{lin}}(q) - 2 \mathrm{i}\,  b_\mathrm{L}\,k_i U_i(\vec{q}) + k_i k_j \,\xi_{ij}(\vec{q}) \right] \, ,
\end{split}
\end{equation}
where we neglected the $\delta_{\mathrm{D}}(\vec{k})$ contribution (Dirac delta) originating from the constant zeroth-order expansion piece.
To further simplify the expression we can use the following Fourier transforms
\begin{align} 
    &\mathcal{F}\left[\xi_{\mathrm{lin}}(q)\right] = \Plin(k) \, , \label{eq:F1} \\
    &\mathcal{F}\left[k_i U_i(\vec{q})\right] = \mathrm{i}\,\Plin(k) \, , \label{eq:F2}\\
    \mathrm{and} \quad &\mathcal{F}\left[ k_i k_j \, \xi_{ij}(\vec{q})\right] = \Plin(k) \, ,\label{eq:F3}
\end{align}
such that we obtain for the galaxy power spectrum in real space (matching Eq.~\ref{eq:ZA_pk})
\begin{equation}\label{eq:Pk_obs_real}
\begin{split}
    P^\mathrm{g}(k) &= \left( 1+2b_\mathrm{L}+b_\mathrm{L}^2 \right) \, \Plin(k)\,\mathrm{exp}\left(-k_i k_j A_{ij}\right) \\
    & = b^2\,\Plin(k)\,\mathrm{exp}\left(-k^2\,\Sigma^2/2\right) \, .
\end{split}
\end{equation}
Here $b$ is the Eulerian linear bias, related to the Lagrangian one via \(b\equiv 1+b_\mathrm{L}\). The displacement field variance $\Sigma^2$ is defined as $\Sigma^2 \equiv 2/3\,\langle |\vec{\psi}_{\mathrm{obs}}|^2\rangle$, where the explicit expression can be found in Eq.~\eqref{eq:ZA_sigma}.

Having derived the form of the observed clustering signal, we now derive that of the reconstructed field. We first consider the displaced data catalogue distribution. As discussed in Sect.~\ref{sec:ZA_nonlinear}, the displaced field $\vec{\psi}^{\mathrm{d}}_{\mathrm{obs}}$ is generated by the filtered displacement that satisfies in Fourier space \(\hat{\vec{\psi}}^{\mathrm{d}}_{\mathrm{obs}}=(1-\mathcal{S})\,\hat{\vec{\psi}}_{\mathrm{obs}}\). Proceeding as in Eqs.~\eqref{eq:P_obs_initial}--\eqref{eq:delta_cross}, $\delta_{\mathrm{d}}$ becomes
\begin{equation}
    \delta_{\mathrm{d}}(\vec{k})=\int \mathrm{d}^3q \, \left[1+b_\mathrm{L}\,\delta_{\mathrm{lin}}(\vec{q})\right] \mathrm{exp}\left\{-\mathrm{i}\,\vec{k}\cdot \left[\vec{q}+\vec{\psi}^{\mathrm{d}}_{\mathrm{obs}}(\vec{q})\right]\right\} \, ,
\end{equation}
and its power spectrum is given by 
\begin{equation}
    P_{\mathrm{dd}}(k)=\mathcal{F}\left[\left\langle \left(1+b_\mathrm{L}\delta_1\right)\left(1+b_\mathrm{L}\delta_2\right) \, \mathrm{exp}\left\{-\mathrm{i}\vec{k}\cdot\Delta\vec{\psi}^\mathrm{d}\right\}\right\rangle \right] \, ,
\end{equation}
with $\Delta\vec{\psi}^\mathrm{d}\equiv \vec{\psi}^{\mathrm{d}}_{\mathrm{obs}}(\vec{q}_2) - \vec{\psi}^{\mathrm{d}}_{\mathrm{obs}}(\vec{q}_1)$.
By the same steps as for Eq.~\eqref{eq:P_obs_initial} we obtain
\begin{equation}
\begin{split}
    P_{\mathrm{dd}}(k) =  \mathrm{exp}\left(-k^2\Sigma_\mathrm{dd}^2/2\right) \,  \mathcal{F} \Big[&b_\mathrm{L}^2\,\xi_{\mathrm{lin}}(q) -2 \mathrm{i} \, b_\mathrm{L}\,k_i U^\mathrm{d}_i(\vec{q}) \\
    & +k_i k_j \, \xi^\mathrm{d}_{ij}(\vec{q})\Big] \, ,
\end{split}
\end{equation}
with the correlators for the shifted catalogue are related to $\Plin$ via
\begin{align}
    &\mathcal{F}\left[k_i U^{\mathrm{d}}_i(\vec{q})\right] = \mathrm{i}\,(1-\mathcal{S}) \, \Plin(k) \, ,\\
    &\mathcal{F}\left[ k_i k_j \, \xi^{\mathrm{d}}_{ij}(\vec{q})\right] = (1-\mathcal{S})^2 \, \Plin(k) \, ,
\end{align}
and the damping $\Sigma_{\mathrm{dd}}$ is defined in Eq.~\eqref{eq:sigma_dd}.
Using the expressions in Eqs.~\eqref{eq:F1}--\eqref{eq:F3}, while suppressing the disconnected $\delta_{\mathrm{D}}(\vec{k})$ contribution originating from the constant term, we have
\begin{equation}\label{eq:pk_d_real_fixed}
    P_{\mathrm{dd}}(k) = \left(b - \mathcal{S}\right)^2 \, \Plin(k) \, \mathrm{exp}\left(-k^2\,\Sigma_{\mathrm{dd}}^2/2\right) \, .
\end{equation}

We continue with the shifted random field. We define the overdensity field by displacing a uniform distribution with $\vec{\psi}_{\mathrm{obs}}^\mathrm{s}$ that is related to original displacement field in Fourier space by \(\hat{\vec{\psi}}^{\mathrm{s}}_{\mathrm{obs}} = -\mathcal{S}\,\hat{\vec{\psi}}_{\mathrm{obs}}\) such that
\begin{equation}
    \delta_{\mathrm{s}}(\vec{k})=\int \mathrm{d}^3q \, \mathrm{exp}\left\{-\mathrm{i}\,\vec{k}\cdot \left[\vec{q}+\vec{\psi}^{\mathrm{s}}_{\mathrm{obs}}(\vec{q})\right]\right\} \, ,
\end{equation}
where again the $\delta_{\mathrm{D}}(\vec{k})$ contribution is neglected. Repeating the steps used for \(P_{\mathrm{dd}}\) with \(b_\mathrm{L}=0\) and replacing \((1-\mathcal{S})\) with \(-\mathcal{S}\) we can write the auto power spectrum of the shifted random field as 
\begin{equation}\label{eq:pk_s_real}
    P_{\mathrm{ss}}(k) = \mathcal{S}^2\,\Plin(k)\,\mathrm{exp}\left(-k^2\,\Sigma_{\mathrm{ss}}^2/2\right) \, ,
\end{equation}
where the damping $\Sigma_{\mathrm{ss}}$ is given in Eq.~\eqref{eq:sigma_ss}.

For the cross spectrum between the displaced galaxies and the shifted random catalogue, we can write
\begin{equation}
    P_{\mathrm{ds}}(k)= \mathcal{F}\left[ \left\langle \left(1+b_\mathrm{L}\delta_2\right) \, \mathrm{exp}\left\{-\mathrm{i}\,\vec{k}\cdot\Delta\vec{\psi}^\mathrm{ds}\right\}\right\rangle \right] \, ,
\end{equation}
with $\Delta\vec{\psi}^\mathrm{ds}\equiv \vec{\psi}^{\mathrm{d}}_{\mathrm{obs}}(\vec{q}_2) - \vec{\psi}^{\mathrm{s}}_{\mathrm{obs}}(\vec{q}_1)$.
As in Appendix~\ref{app:recon_real}, the large-separation expansion yields
\begin{equation}\label{eq:Pk_ds_implicit}
    P_{\mathrm{ds}}(k) =\mathrm{exp}\left(-k^2\Sigma_{\mathrm{ds}}/2\right) \,  \mathcal{F} \left[-\,\mathrm{i}\,b_\mathrm{L}\,k_i U^{\mathrm{ds}}_{i}(\vec{q})+k_i k_j \,  \xi^{\mathrm{ds}}_{ij}(\vec{q})\right] \, .
\end{equation}
We note that the cross correlator $U^{\mathrm{ds}}_i(\vec{q})$ does only contain the correlation between the linear density contrast and $\vec{\psi}^{\mathrm{s}}_{\mathrm{obs}}$. This is because the bias term -- ($1+b_{\mathrm{L}}\delta_i$) -- comes from the shifted galaxy catalogue but can only correlate with the displacement field at a different position (i.e. the $U_i$ correlator does not contain a zero-lag contribution), that is, the shifted random field.
Therefore, analogous to the expressions in Eqs. \eqref{eq:F2} and \eqref{eq:F3} the cross correlators are related to $\Plin$ via
\begin{align}
    &\mathcal{F}\left[k_i  U^\mathrm{ds}_{i}(\vec{q})\right] = -\mathrm{i} \, \mathcal{S}\,\Plin(k)\\
    &\mathcal{F}\left[k_i k_j \xi^\mathrm{ds}_{ij}(\vec{q})\right] = -\mathcal{S} \, \left(1-\mathcal{S}\right)\,\Plin(k) \, ,
\end{align}
such that we get for the cross power spectrum
\begin{equation}\label{eq:pk_ds_real}
    P_{\mathrm{ds}}(k) =-\mathcal{S}\,\left(b-\mathcal{S}\right) \, \Plin(k) \, \mathrm{exp}\left(-k^2\,\Sigma_{\mathrm{ds}}^2/2\right) \, ,
\end{equation}
and the cross damping $\Sigma_{\mathrm{ds}}$ can be found in Eq.~\eqref{eq:sigma_ds}.
Combining Eqs.~\eqref{eq:pk_d_real_fixed}, \eqref{eq:pk_s_real}, and \eqref{eq:pk_ds_real}, the reconstructed spectrum for the field \(\delta_{\mathrm{rec}}=\delta_{\mathrm{d}}-\delta_\mathrm{s}\) takes the form
\begin{equation}
\begin{split}
    P_{\mathrm{rec}}(k) = &\left[(b-\mathcal{S})^2 \, \mathrm{exp}\left(-k^2\Sigma_{\mathrm{dd}}^2/2\right) +\mathcal{S}^2\,\mathrm{exp}\left(-k^2\Sigma_{\mathrm{ss}}^2/2\right) \right. \\
    &\left. +2 \, \mathcal{S}\,(b-\mathcal{S})\,\mathrm{exp}\left(-k^2\Sigma_{\mathrm{ds}}^2/2\right)\right]\,\Plin(k) \, ,
\end{split}
\end{equation}
which reduces to a common prefactor of \(b^2\) in the limit where all the damping scales $\Sigma_i$ are equal.

\subsection{Redshift-space}\label{app:recon_redshift}
We conclude by modelling the pre- and post-reconstruction clustering in redshift space. We focus on the \textsc{RecSym} implementation, while the \textsc{RecIso} expressions follow by setting \(f\rightarrow0\) in the terms related to the random catalogue, namely  \(P_{\mathrm{ds}}\) and \(P_{\mathrm{ss}}\), but keeping the terms describing the shifted galaxies unchanged.
Redshift-space displacements are obtained by the Kaiser mapping.
At leading order this implies for the relevant correlators in Fourier space
\begin{align}
    & A^s_{ij}(\vec{k}) \equiv \left(1+f\mu^2\right)^2 A_{ij}(\vec{k}) \label{eq:Aij_s-space} \, , \\
    & \xi^s_{ij}(\vec{k}) \equiv  \left(1+f\mu^2\right)^2 \xi_{ij}(\vec{k}) \label{eq:xiij_s-space} \, , \\
    \mathrm{and}  \quad & U^s_{i}(\vec{k}) \equiv  \left(1+f\mu^2\right) U_{i}(\vec{k}) \label{eq:Ui_s-space} \, .
\end{align}
By plugging Eqs.~\eqref{eq:Aij_s-space}--\eqref{eq:Ui_s-space} into Eq.~\eqref{eq:Pk_obs_real} we obtain for the redshift-space power spectrum of galaxies
\begin{equation}\label{eq:Pobs_s}
    P_{\mathrm{obs}}^{s}(k,\mu)=\left(b+f\mu^2\right)^{2} \, \Plin(k) \, \mathcal{E}(k, \mu) \, ,
\end{equation}
with the anisotropic Zeldovich damping
\begin{equation}
    \mathcal{E}(k,\mu) = \mathrm{exp}\left\{-k^2 \left[\left(1 - \mu^2\right)\Sigma^2 + (1 + f)^2 \mu^2 \Sigma^2\right]/2 \right\} \, .
\end{equation}

Similarly, for the reconstructed clustering signal we can transform  Eqs.~\eqref{eq:pk_d_real_fixed}, \eqref{eq:pk_s_real}, and \eqref{eq:pk_ds_real} into
\begin{align}
      & P_\mathrm{dd}^s(k,\mu) = \left[(b-1)+(1-\mathcal{S}) \, \left(1+f\mu^2\right)\right]^2 \, \Plin(k) \,  \mathcal{E}_{\mathrm{dd}} \, , \\
      & P_\mathrm{ss}^s(k,\mu) = \mathcal{S}^2\left(1+f\mu^2\right)^2 \, \Plin(k) \, \mathcal{E}_{\mathrm{ss}} \, ,
\end{align}
and 
\begin{equation}
\begin{split}
      P_\mathrm{ds}^s(k,\mu) =& - \left[(b-1)+(1-\mathcal{S}) \, \left(1+f\mu^2\right) \right] \\
      & \times \mathcal{S} \, \left(1+f\mu^2\right) \, \Plin(k) \, \mathcal{E}_{\mathrm{ds}} \, ,
\end{split}
\end{equation}
that we can use to construct the reconstructed power spectrum in Eq.~\eqref{eq:Pk_symiso}. For the anisotropic damping functions we omitted the explicit dependency on $k$ and $\mu$ to improve readability and used the same notation for $\mathcal{E}_i$ as in Sect.~\ref{sec:ZA_redshift}.

\section{2PCF measurements}\label{app:2PCF}
\begin{figure*}
\subfloat[Pre-reconstruction]{\label{sfig:a}\includegraphics[width=1\textwidth]{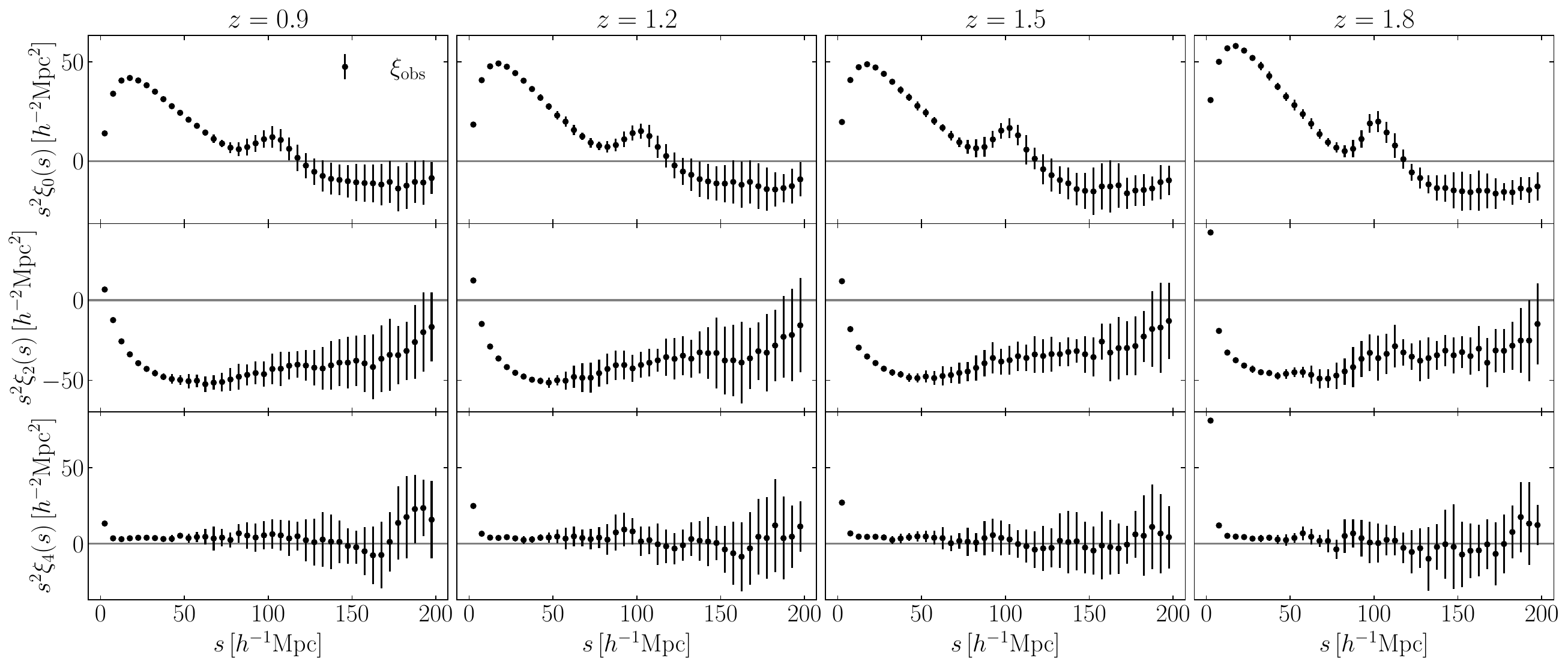}}
\\
\subfloat[Post-reconstruction]{\label{sfig:n}\includegraphics[width=1\textwidth]{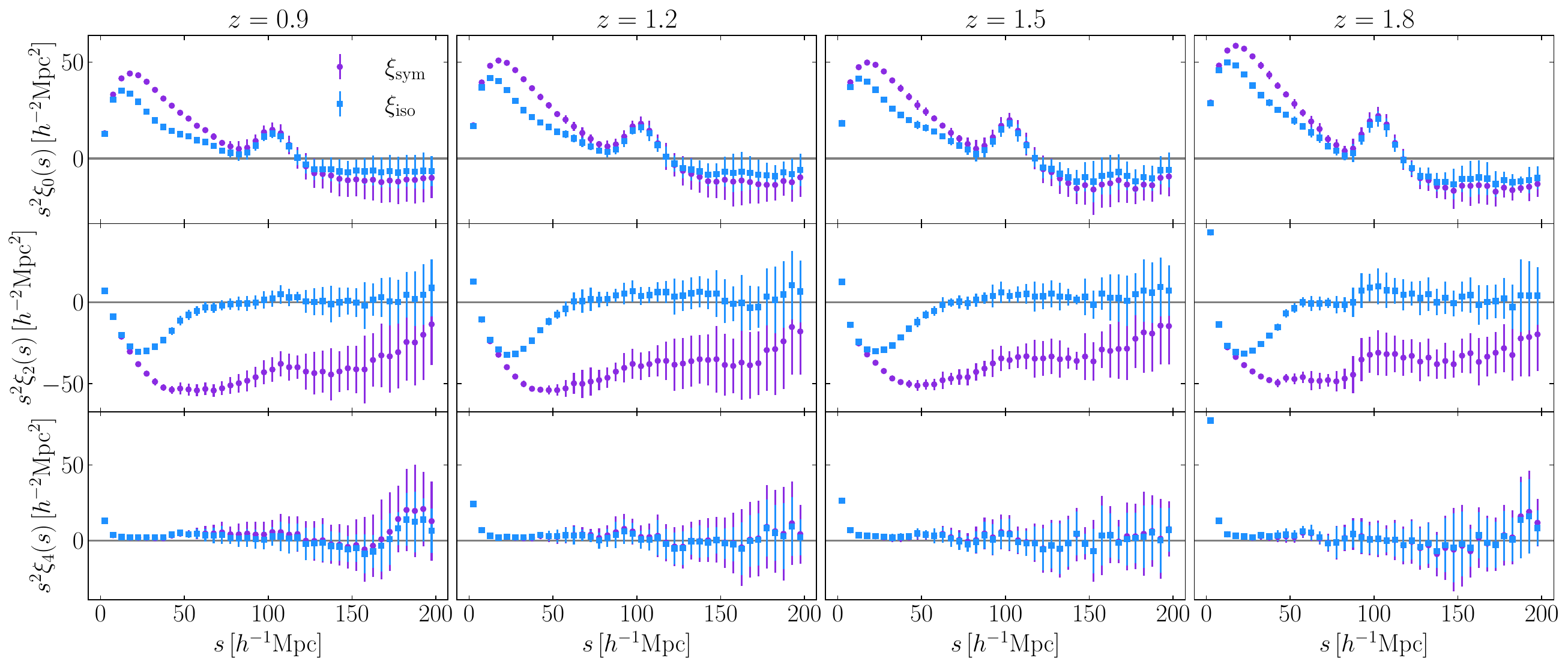}}\\
\caption{Mean \tpcf multipoles averaged over eight \flagshipone mocks at different redshifts, shown in separate columns. The top panel presents the pre-reconstruction multipoles, while the bottom panel shows the post-reconstruction results. Each plot is divided into three rows corresponding to the monopole, quadrupole, and hexadecapole, in order from top to bottom. Purple circles represent the \tpcf multipoles obtained by running the \textsc{RecSym} reconstruction method, with \(f = f^\mathrm{f}\), \(b = b^\mathrm{f}\), \(N_\mathrm{cell} = 128\), and \(R_\mathrm{s} = 15h^{-1}\mathrm{Mpc}\). The blue squares indicate the \tpcf multipoles recovered using the \textsc{RecIso} method with identical input parameters as for \textsc{RecSym}. Errorbars are the standard deviation of the measurements over the eight mock catalogues.}
\label{fig:2PCF_measurements}
\end{figure*}

Figure~\ref{fig:2PCF_measurements} shows the mean pre- and post-reconstruction
multipoles of the anisotropic two-point correlation function measured from
the eight \flagshipone mock catalogues using the \Euclid official estimator \citep{EP-delaTorre}.
Error bars indicate the standard deviation across mock realisations, providing a direct estimate of the statistical uncertainty at each redshift.
The pre-reconstruction monopole, quadrupole, and hexadecapole exhibit the expected smoothing of the acoustic feature.
Post-reconstruction, \textsc{RecSym} sharpens the \bao peak while preserving the excess monopole amplitude and a non-vanishing quadrupole induced by \rsd.
By construction, \textsc{RecIso} yields a lower small-scale clustering amplitude and drives the quadrupole towards zero for $s\gtrsim 50\,\hMpc$.
The change of slope around $s\sim 50\,\hMpc$ in both monopole and quadrupole traces the scale dependence introduced by the smoothing filter, (see Eq.~\ref{eq:D_iso}).

\section{\texttt{WinCov} validation}\label{app:WinCov validation}

\begin{figure*}
    \begin{minipage}{0.34\textwidth}%
    \begin{subfigure}{\linewidth}%
        \includegraphics[width=\textwidth]{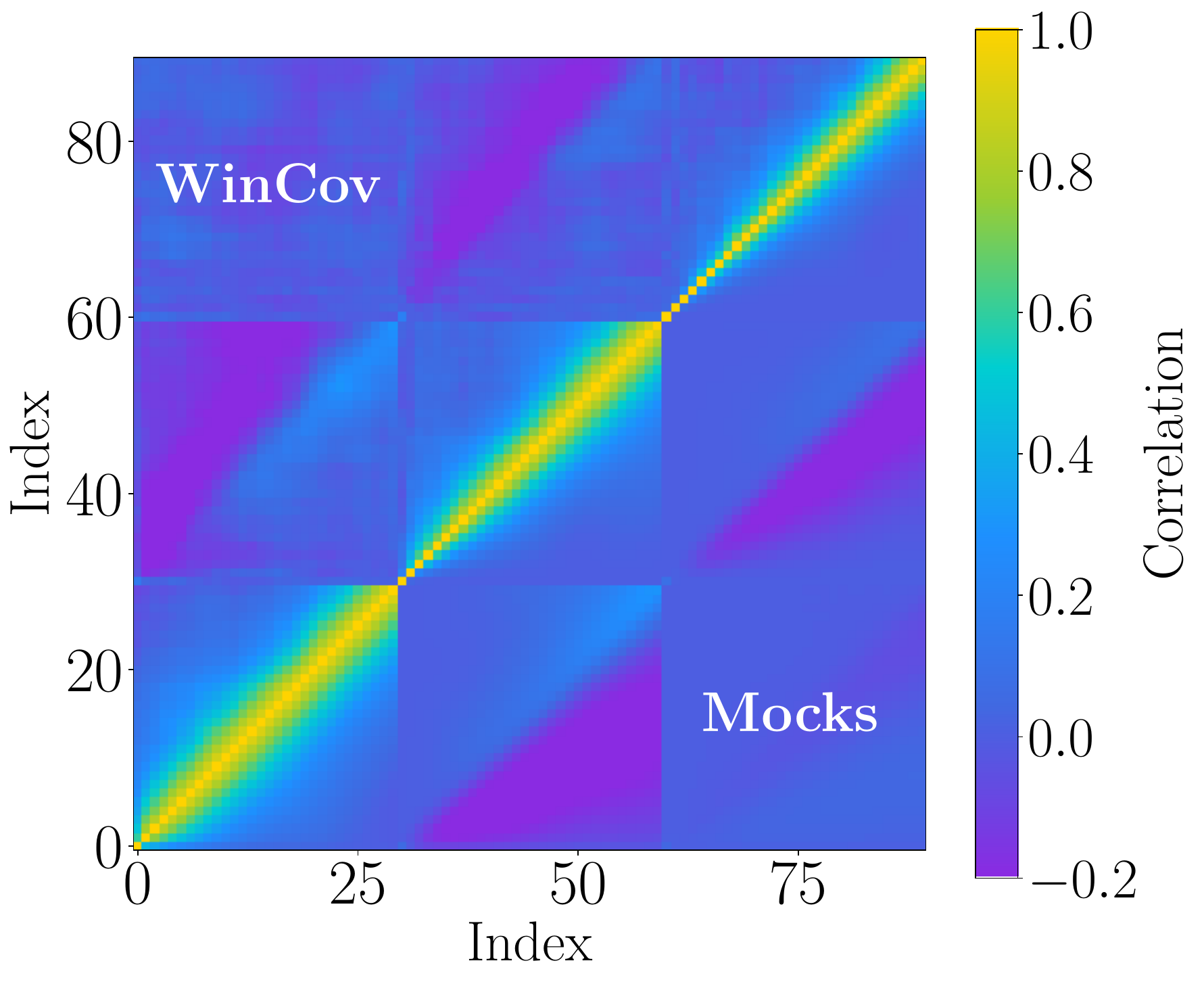}
        \caption{Correlation matrices.}
        \label{fig:wincov-corr}
    \end{subfigure}
    \end{minipage}%
    \hspace{0.5mm}
    \begin{minipage}{0.66\textwidth}%
    \begin{subfigure}{\linewidth}%
        \includegraphics[width=\textwidth]{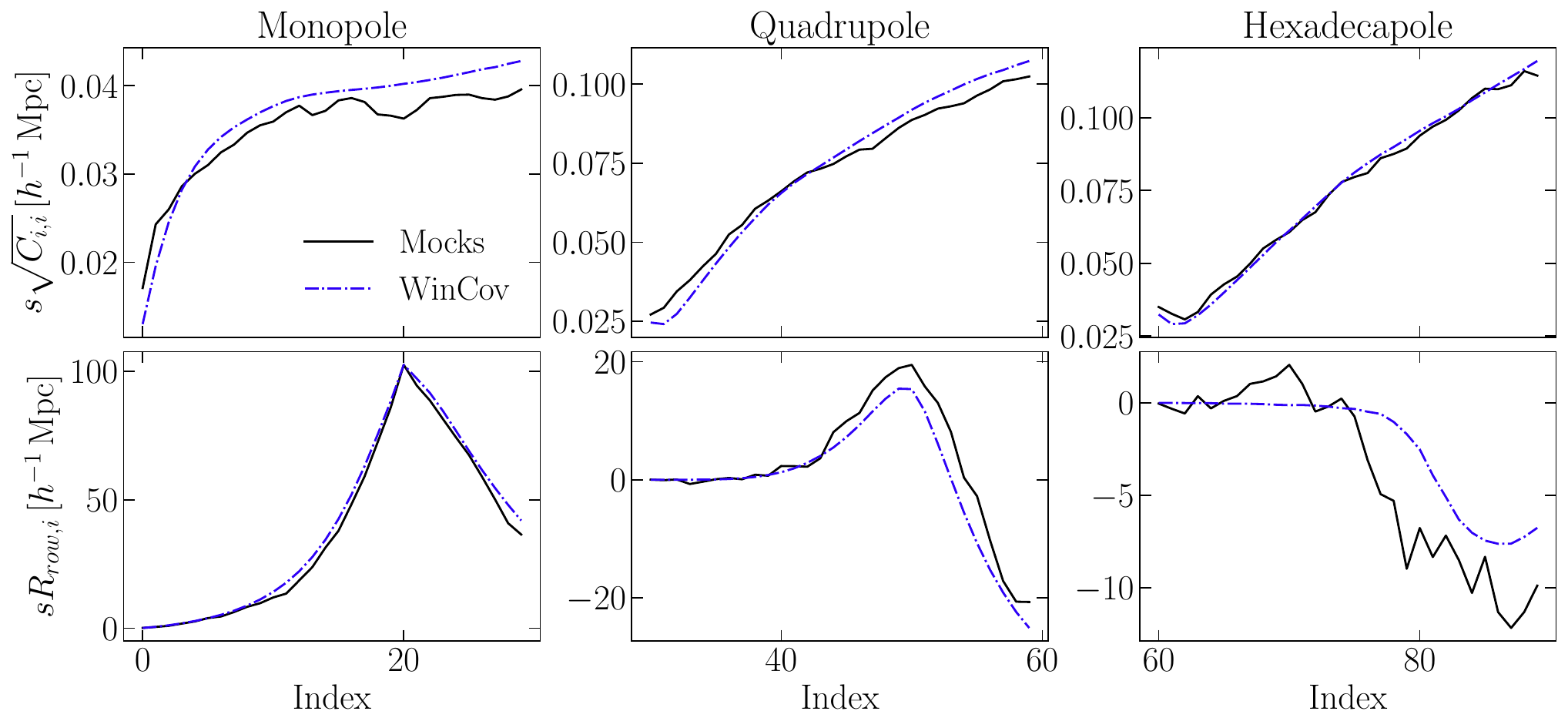}
        \caption{Diagonal and off-diagonal comparisons.}
        \label{fig:wincov-line}
    \end{subfigure}
    \end{minipage}%
    \caption{
    Validation of the \texttt{WinCov} covariance estimates. \textit{Left panel}: Correlation matrix from \texttt{WinCov} (top-left triangular matrix) compared to the numerical estimate from 1000 ELM (bottom-right triangular matrix). \textit{Top-right panels}: Square root of the diagonal elements of the \texttt{WinCov} covariance (blue) versus the mock-based estimate (black dashed). \textit{Bottom-right panel}: 20th row of the correlation matrix for the joint multipole vector -- a concatenation of the first three even multipoles -- sampled in 30 bins of \SI{5}{\per\h\mega\parsec} over the range $0\Mpc$--$150\Mpc$. The label `index' refers to the pair-separation bin, repeated for each multipole order $\ell \in\{ 0, 2, 4\}$, such that index $\in [0, 89]$
    }\label{fig:wincov}
\end{figure*}

We validate here the performance of the \texttt{WinCov} implementation of the \texttt{RascalC} method, used to generate the semi-analytical covariance matrices for the \tpcf multipoles introduced in Sect.~\ref{sec:covariances}. Extensive tests have been performed within the \Euclid framework, covering complex survey geometries, varying number densities, and different levels of redshift-sample purity \citep[see, e.g.,][]{EP-Risso}. In this appendix, we focus on a representative comparison against numerical covariance estimates derived from 1000 ELM \citep{EP-Monaco1, EP-Risso}, designed to reproduce the observed \DRone lightcone of \Euclid.

The test corresponds to a \Euclid-like configuration in terms of volume and area, covering the redshift range $0.9 < z < 1.1$, and includes a realistic contamination fraction of 20\% redshift interlopers, primarily caused by catastrophic redshift errors. These mocks achieve greater realism than the \flagshipone catalogues used in the main analysis, owing to their detailed treatment of the survey window and interloper contamination. They thus offer a more stringent validation test, extending the applicability of our approach beyond the scope of this work and towards the forthcoming \DRone analysis.\footnote{We recall that the ELM were not used in the main paper because their clustering signal is generated through a quasi-linear approximation, which is insufficiently accurate to test the performance of \bao reconstruction in recovering nonlinear features.}

Figure~\ref{fig:wincov} compares the mock-based and semi-analytical covariance estimates.
The correlation matrices (leftmost panel) show excellent agreement, with \texttt{WinCov} exhibiting significantly reduced noise, as expected from its analytical formulation. The \texttt{WinCov} model accurately reproduces both the diagonal and off-diagonal trends of the mock-derived uncertainties -- as shown in the six panels to the right -- confirming its capability to capture the amplitude and scale dependence of the statistical errors while dramatically reducing computational cost.
In conclusion, \texttt{WinCov} extends \texttt{RascalC}, which has already been successfully applied for efficient covariance estimation in the DESI survey \citep{Rashkovetskyi2023, Rashkovetskyi2025}, to the specific requirements of the \Euclid spectroscopic survey, ensuring accuracy and scalability across the wide range of redshifts and sky coverage expected for \DRone.

\section{\texttt{Bora.jl} -- a fast emulator for the \Euclid clustering signal}\label{app:Bora}

\texttt{Bora.jl} is a neural network (NN) emulator developed in \texttt{Julia} to reproduce the predictions of the \tpcf template fitting model adopted for \Euclid, described in Sect.~\ref{sec:fitting_templates}. NN emulators have become a standard tool in cosmology, providing fast and accurate surrogates for large-scale structure simulations and clustering models that would otherwise be computationally prohibitive~\citep{Fendt:2006uh, SpurioMancini:2021ppk, Arico:2021izc, Eggemeier:2022anw, Bonici:2022xlo}.
Building on this concept, \texttt{Bora.jl} extends the differentiable backend developed for \texttt{Capse.jl}~\citep{Bonici:2023xjk} and \texttt{Effort.jl}~\citep{Bonici:2025ltp} to specifically emulate the \Euclid \tpcf template with high precision. 

\subsection{Architecture, training, and validation}

\texttt{Bora.jl} predicts the physical component \(\xi^\mathrm{ph}\) of the \tpcf multipoles template, presented in Eq.~\eqref{eq:xi_model}, evaluated on a fixed $s$-grid of 40 evenly spaced points ranging from $0$ to $200\,\hMpc$ corresponding to the data separation bins. The architecture of \texttt{Bora.jl} is that of a standard multi-layer perceptron, consisting in an input layer with six neurons, an output layer with 40 neurons (each of them corresponding to one of the points of the $r$-grid), and five hidden layers, each containing 64 neurons. The used activation function is the hyperbolic tangent.

Before sent into training the \texttt{Bora.jl}, the data are preprocessed based on the max-min normalisation; a common data preprocessing technique used to scale the values of a dataset so that they fall within a specific range, typically between 0 and 1. It is also known as feature scaling or min-max scaling.
To perform max-min normalisation, the minimum value of a feature is subtracted from each value in the feature, and the result is divided by the range of the feature (i.e., the difference between the maximum and minimum values). This transformation ensures that the minimum value in the feature is scaled to 0 and the maximum value is scaled to 1. Values in between are scaled proportionally based on their relative position within the range of the feature.

Max-min normalisation can be useful when working with features that have different scales, as it allows for a more meaningful comparison between features. Additionally, it can help to prevent numerical instability during training, as scaling the data to a similar range can reduce the influence of large values on the optimisation process. This preprocessing step helps to improve the performance of the NN by normalising the input and output features to a similar scale, reducing the potential for numerical instability during training. Overall, the \texttt{Bora.jl} architecture with the preprocessing step is capable of producing predictions in microseconds.

For each considered redshift bin, we train \texttt{Bora.jl} using 10\,000 theoretical predictions for \(\xi^\mathrm{ph}_\ell\), each corresponding to a unique set of physical parameters \( \{\alpha_\perp, \alpha_\parallel, b, f, \Sigma_\perp, \Sigma_\parallel, R_\mathrm{s}\} \).
The dataset was randomly split into training and validation sets, with 80\% of the data used for training and 20\% used for validation. The training process is performed by minimising the loss function, set here to the mean square error between \texttt{Bora.jl} predictions and the analytical model, using the ADAM optimizer~\citep{kingma2015adam}. The full procedure is completed in $0.25\CPUh$.  

Once trained, we evaluated the performance of \texttt{Bora.jl} on the validation set. We compared the predictions made by \texttt{Bora.jl} with the predictions made by the standard template fitting method. We found that \texttt{Bora.jl} achieved comparable accuracy to the standard method, with a precision such that for most of the points in the test dataset, the emulation error is smaller than 0.05 times the expected error on the measured quantities, as computed from the covariance matrix used in this paper (Sect.~\ref{sec:covariances}).

\subsection{Posterior sampling and validation}\label{sec:chains_validation}

\begin{figure}
\centering
\includegraphics[width=1\columnwidth]{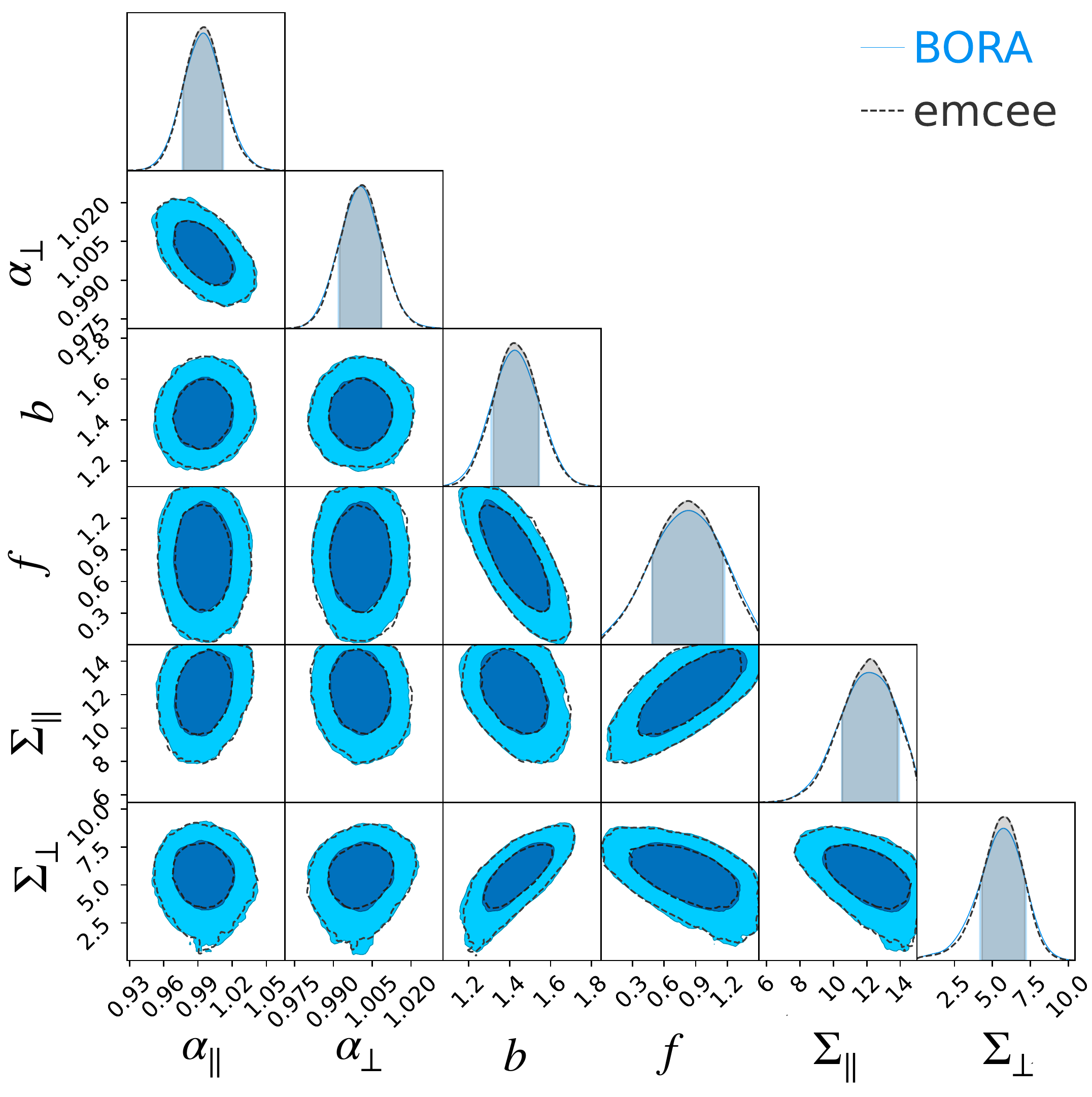}
\caption{Comparison of posterior contours from fits to the mean pre-reconstruction \tpcf multipoles at $z=0.9$, averaged over eight sub-boxes. Blue filled contours denote \texttt{Bora.jl}+\texttt{NUTS} while black dashed represents \texttt{emcee} paired with the direct model. The posterior contours represent the 68\% and 95\% credible regions.}
\label{fig:BORA_vs_emcee_prerec09}
\end{figure}

To sample the posterior we adopt the No-U-Turn Sampler (hereafter \texttt{NUTS}), developed by \citet{NUTS}, which is an extension of Hamiltonian Monte Carlo (\hmc; \citealt{2017arXiv170102434B}). Unlike traditional random-walk \mcmc methods, Hamiltonian samplers exploit gradient information to perform longer and more directed moves in parameter space, thereby reducing autocorrelation and improving sampling efficiency. \texttt{NUTS} is particularly convenient because it adaptively determines the trajectory length and step size, avoiding the need for extensive manual tuning. Since such samplers are still not commonly used in \bao analyses, we validate this choice below by comparing the resulting posteriors with those obtained from a standard \texttt{emcee} analysis.

For this comparison, we analyse the mean \textsc{PreRec} \tpcf multipoles at \(z=0.9\), averaged over eight sub-box realisations. We compare parameter constraints obtained with the \texttt{Bora.jl + NUTS} pipeline to those derived from the standard \mcmc approach implemented in \texttt{emcee} \citep{ForemanMackey2013}, which evaluates the model directly at each step without relying on the emulator. Both approaches adopt the iterative covariance matrix introduced in Sect.~\ref{sec:covariances}, normalised by the number of independent mocks.
Figure~\ref{fig:BORA_vs_emcee_prerec09} shows the marginalised posterior distributions for the physical parameters obtained from fitting the mean \textsc{PreRec} \tpcf at \( z = 0.9 \). The results from \texttt{emcee}, based on 80 walkers with 10\,000 steps each (totalling $288\CPUh$), are in excellent agreement with those from the emulator-based \texttt{Bora.jl + NUTS} pipeline, which requires only 8 chains with 1500 post-warmup draws and completes in just $0.5\CPUh$. This validates the accuracy of the emulator-based inference while demonstrating a speed-up by a factor of more than 500. To ensure robust convergence for single-realisation fits, which may exhibit noisier likelihood surfaces, we conservatively increase the number of \texttt{NUTS} samples to 6000 post-warmup draws per chain in subsequent analyses.

\section{Fiducial covariance cosmology}\label{app:fid_cov}

\begin{figure}
\centering
\includegraphics[width=.9\columnwidth]{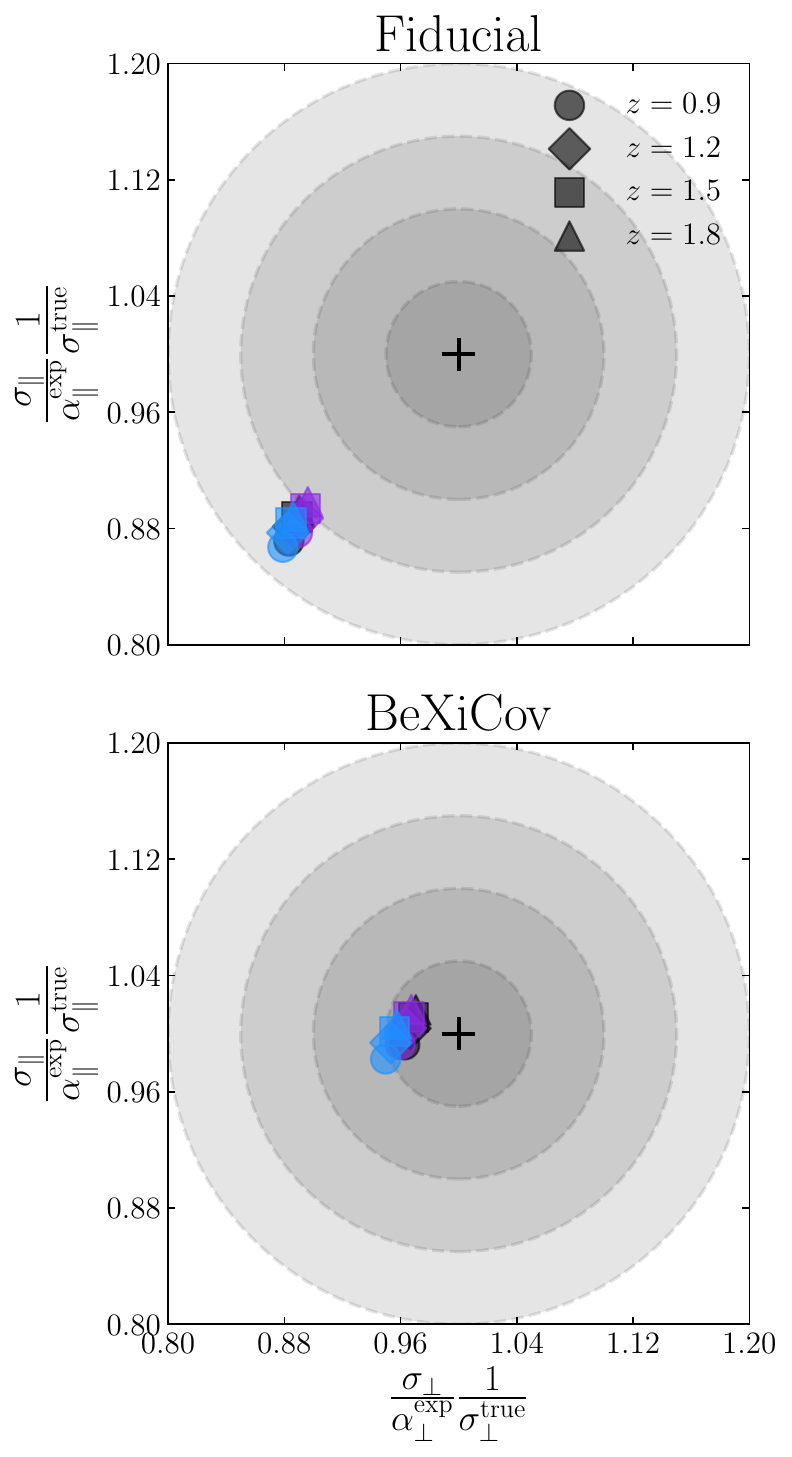}
\caption{Effect of cosmology-covariance mismatch on the inferred relative uncertainties of the \bao scale parameters, $\sigma_{\parallel,\perp}/\alpha^\mathrm{exp}_{\parallel,\perp}$. The upper panel shows results obtained using a mock-based covariance matrix, while the lower panel corresponds to the fiducial semi-analytical model introduced in this work. Black markers denote pre-reconstruction results, while purple and blue refer to the \textsc{RecSym} and \textsc{RecIso} methods, respectively. Different symbols indicate the redshift bins and shaded bands at 5\%, 10\%, 15\%, and 20\% provide visual reference.}
\label{fig:Fiducial_covariance}
\end{figure}

In this appendix, we assess the sensitivity of \bao\ constraints to the fiducial cosmology assumed in computing the covariance matrix of the \tpcf\ multipoles, adopting the same \(\Omega_{\mathrm{m}}\) variations used in Sect.~\ref{sec:fiducial_cosmo} to keep this test directly consistent with the main \bao\ analysis. As discussed in Sect.~\ref{sec:BAO_constraints}, we restrict this study to variations in \(\Omega_{\mathrm{m}}\), since within the \bao\ framework considered here it is the parameter to which the \apdist\ remapping is most directly sensitive. We do not consider departures more extreme than the \(5\sigma\) \Planck\ range explored in Sect.~\ref{sec:fiducial_cosmo}, since such cases would no longer probe the covariance response within the standard analysis setup, but rather the validity of the fiducial mapping itself, and would in practice require updating the fiducial cosmology and repeating the analysis.

To this end, we construct two synthetic data vectors that encode the clustering signal of the same underlying Universe, as measured under different fiducial cosmologies. For a first case, we use $\xi_\ell^\mathrm{true}$, obtained from the matter power spectrum evaluated at the true cosmology of the mocks ($\Omega_{\mathrm{m}} = 0.319$) and analysed under the same assumption, such that no \apdist distortion is present. The second case is described by $\xi_\ell^\mathrm{AP}$, constructed using the same model but evaluated under a fiducial matter density of $\Omega_{\mathrm{m}} = 0.277$, corresponding to a $-5\sigma$ deviation from the \Planck\ mean (see Sect.~\ref{sec:fiducial_cosmo}), and therefore subject to an artificial \apdist distortion. The vector $\xi_\ell^\mathrm{AP}$ mimics a realistic scenario with a mismatched fiducial cosmology, whereas $\xi_\ell^\mathrm{true}$ serves as a baseline. As both vectors share identical input physics, any differences in the inferred parameter uncertainties arise solely from the modelling of the covariance.

We generate both clustering signals using Eq.~\eqref{eq:2PCF_mult}, where we evaluate the real-space power spectrum (Eq.~\ref{eq:Pk_real_nl_model}) and \rsd (Eq.~\ref{eq:Pkmu_nl_model}) at the true cosmology. For simplicity, we fix the galaxy bias to the fiducial value of the mocks and the damping terms to their theoretical predictions, set by $\Sigma_\perp = \Sigma_\mathrm{ref}$, and $\Sigma_\parallel = (1+f^\mathrm{t})\,\Sigma_\mathrm{ref}$, where $\Sigma_\mathrm{ref}$ corresponds to the pre-reconstruction damping and to $\Sigma_\mathrm{eq}$ post-reconstruction (Sect.~\ref{sec: Recon_methods}). The broad-band terms are set to zero.
The fiducial cosmology enters through the \apdist parameters computed in the plane-parallel approximation. As the redshift-to-distance mapping does not affect $r_\mathrm{s}$, the \apdist parameters are simply given by $\alpha_\perp = 1$ and $\alpha_\parallel = H^\mathrm{f}/H^\mathrm{t}$, where we invert \(H^\mathrm{f}\) and \(H^\mathrm{t}\) relative to the \bao fit convention, since the model is now evaluated at the true rather than the fiducial cosmology.

We consider three standard strategies for constructing the covariance matrix needed for \bao modelling. The first utilises a numerical covariance from mock catalogues generated at the fiducial cosmology, the second an analytical covariance calibrated to the fiducial model, and the third a data-calibrated semi-analytical covariance. In the case of a purely theoretical data vector and under the assumption of periodic boundary conditions, the first and second options are equivalent. In contrast, the third strategy corresponds to the \texttt{BeXiCov} implementation introduced in Sect.~\ref{sec:covariances}, without requiring the survey window sampling performed by \texttt{WinCov}.

To compare their impact on the estimated \bao scale, we compute $\tens{C}^\mathrm{fid}$ using the Gaussian model of \citet{Grieb2016} with the first and second methods, $\tens{C}^\mathrm{BeXiCov}$ using the third method, and a reference $\tens{C}^\mathrm{true}$, equivalent to $\tens{C}^\mathrm{fid}$ but evaluated at the true cosmology. For consistency with the analysis of Sect.~\ref{sec:fiducial_cosmo}, all covariances are calibrated on the \DRone volume and number density expected for the true cosmology at $\bar{z}$.
We finally fit $\xi_\ell^\mathrm{AP}$ using both $\tens{C}^\mathrm{fid}$ and $\tens{C}^\mathrm{BeXiCov}$, and compare the constraints to those obtained from $\xi_\ell^\mathrm{true}$ with $\tens{C}^\mathrm{true}$.

Figure~\ref{fig:Fiducial_covariance} shows the variation in the rescaled \bao uncertainties -- 
$\sigma_{\parallel}/\alpha^\mathrm{exp}_\parallel$ and $\sigma_{\perp}/\alpha^\mathrm{exp}_\perp$ -- 
with respect to the benchmark values $\sigma_{\parallel}^\mathrm{true}$ and $\sigma_{\perp}^\mathrm{true}$, 
when using $\tens{C}^\mathrm{fid}$ (top) and $\tens{C}^\mathrm{BeXiCov}$ (bottom). 
In the case of $\tens{C}^\mathrm{fid}$, we systematically overestimate the error by 15--20\%. Using $\tens{C}^\mathrm{BeXiCov}$, these biases are reduced to below 5\% across all cases. Once again, \textsc{RecSym} and \textsc{RecIso} yield perfectly consistent results.
This confirms that a data-calibrated covariance model -- responsive to the observed clustering amplitude -- is essential to guarantee the robustness of \bao constraints, thus mitigating the risk of spurious cosmological tension arising from mismatched model assumptions.

\end{appendix}

\end{document}